%% file: Barmak_MHD_2phase_arxiv.tex
\documentclass[pdflatex,sn-mathphys-num]{sn-jnl}

\usepackage[]{graphicx,color}%
\usepackage{multirow}%
\usepackage{amsmath,amssymb,amsfonts}%
\usepackage{amsthm}%
\usepackage{mathrsfs}%
\usepackage[title]{appendix}%
\usepackage{xcolor}%
\usepackage{textcomp}%
\usepackage{manyfoot}%
\usepackage{booktabs}%

\usepackage{bm}							
\usepackage[position=t,caption=false,justification=centering]{subfig}
\usepackage{hyperref}
\usepackage{mathtools}
\usepackage[makeroom]{cancel}
\usepackage{stmaryrd} 
\usepackage{tabularx}
\usepackage{multirow}
\renewcommand{\vec}[1]{{\mbox{\boldmath $ #1 $}}}
\allowdisplaybreaks
\newcommand{\Rey}{Re} 
\newcommand{\Ha}{Ha} 
\newcommand{\Pran}{Pr} 
\newcommand{\Po}{Po} 
\graphicspath{{./figures/}}

\begin{document}

\title[Two-phase stratified MHD flows in wide rectangular ducts...]{Two-phase stratified MHD flows in wide rectangular ducts: analytical and numerical solutions}

\author*[1,2]{\fnm{Ilya} \sur{Barmak}}\email{ilyab@tauex.tau.ac.il}

\author[2]{\fnm{Subham} \sur{Pal}}\email{subhampal@tauex.tau.ac.il}

\author[2]{\fnm{Alexander} \sur{Gelfgat}}\email{gelfgat@tauex.tau.ac.il}

\author[2]{\fnm{Neima} \sur{Brauner}}\email{brauner@tauex.tau.ac.il}

\affil[1]{\orgname{Soreq NRC}, \orgaddress{\city{Yavne}, \postcode{8180000}, \country{Israel}}}

\affil[2]{\orgdiv{School of Mechanical Engineering}, \orgname{Tel Aviv University}, \orgaddress{\city{Tel Aviv}, \postcode{6997801}, \country{Israel}}}


\abstract{This study explores the effects of a non-conductive gas layer flowing concurrently with a conductive liquid on the two-phase flow characteristics in wide horizontal ducts under a constant vertical magnetic field. To this end, analytical solutions for the velocity profile and induced magnetic field are presented for laminar gas-liquid stratified magnetohydrodynamic (MHD) flow between two infinite plates of various conductivities. The contributions of the Lorentz force and wall shear stresses to the pressure gradient are examined. To the best of our knowledge, it is shown for the first time that, unlike the single-phase Hartmann flow, the velocity profiles in two-phase flow differ significantly depending on whether the bottom wall is conducting or insulating. In the case of an insulating bottom wall, the gas lubrication effect and potential pumping power savings are significantly greater, regardless of the magnetic Reynolds number. This conclusion also holds for gas-liquid MHD flows in rectangular ducts with finite width-to-height aspect ratios. To assess the applicability of the Two-Plate (TP) model to wide ducts, numerical solutions of the two-dimensional problem are used to investigate the influence of side walls on the two-phase flow characteristics, considering various combinations of bottom and side wall conductivities. In all cases, the results for high aspect ratios converge to the analytical solution obtained from the TP model with the same bottom wall conductivity. However, the influence of insulating side walls remains significant even at large aspect ratios when the bottom wall is conducting. Unexpectedly, in such cases, the change in the induced magnetic field due to the presence of side walls has a dramatic effect on the velocity profile, leading to a reduced pressure gradient compared to that predicted by the TP model.}


\maketitle

\section{Introduction} \label{Sec: Introduction}

Flows of electrically conducting fluids in electromagnetic fields are present in a wide range of physical phenomena and technological applications, such as power generation, nuclear reactor cooling, microfluidic pumps, drug delivery, and plasma propulsion engines (e.g., \cite{Branover78,Davidson01,Molokov07}). Conducting fluids, including plasmas, liquid metals, and electrolytes, can interact with and be influenced by magnetic fields. These interactions are extensively studied in the field of magnetohydrodynamics (MHD), which combines the principles of fluid dynamics and electromagnetism to describe the behavior of electrically conducting fluids in the presence of magnetic fields \cite[e.g., ][]{Branover78,Davidson01,Molokov07,Kadid11,Kundu22}. Numerous practical challenges involve two-phase flow systems, particularly within the nuclear and petroleum industries, geophysics, and MHD power generation. Examples include two-phase liquid metal magnetohydrodynamics generators \cite{Wang22}, magnetic field-driven micropumps \cite{Yi02,Weston10}, microchannel networks of lab-on-a-chip devices \cite{Haim03,Hussameddine08}, where the presence of a second non-conductive fluid may enhance the mobility of the conducting fluid.

Much of the research on MHD focused on single-phase flows and have been extensively documented in the literature. Hartmann \cite{Hartman37a} and Hartmann and Lazarus \cite{Hartmann37b} provided an analytical solution for the one-dimensional (1D) velocity profile of Poiseuille flow involving an isothermal, electrically conducting fluid flowing through a channel formed by two infinite plates under an external uniform wall-normal magnetic field. The analysis demonstrated that the axial velocity profile depends solely on the Hartmann number, $\displaystyle\Ha=B_0 H \sqrt{\sigma_e/\eta}$, where $H$ is the channel height, $B_0$ is the strength of the external transverse magnetic field, and $\eta,\sigma_e$ are the fluid viscosity and electric conductivity, respectively. It was shown that the magnetic field flattens the velocity profile in the core region and at high Hartmann numbers results in thin boundary layers, commonly referred to in the literature as the Hartmann boundary layers. Hartmann's solution \cite{Hartman37a} does not refer to the induced magnetic field and is, in fact, valid for ideally electrically conductive plates \cite[e.g., ][]{Muller01}. Referring to electrically insulating plates, Shercliff \cite{Shercliff53} presented a solution for the velocity profiles that accounts for the effect of the induced magnetic field. Also in this case, the velocity profile depends solely on the Hartmann number, and is the same as that obtained for conducting walls when scaled by the maximal velocity. However, as indicated in \cite{Muller01}, for a given Hartmann number and pressure gradient, the flow rate is larger in a channel with electrically insulating walls.

In rectangular ducts, the analytical solution for 2D axial velocity field is an infinite series of exponential functions. To address the slow convergence rate of this series, Shercliff \cite{Shercliff53,Shercliff62} employed approximate methods to analyze the flow field's characteristics. These solutions indicate that, for large Hartmann numbers, the solution obtained in the simple two-plate geometry is valid, except near the sidewalls, where thin boundary layers, referred to in the literature as Shercliff boundary layer (B. L.), are present. The effects of electrical conductivity of the duct walls on the boundary conditions and the solution for the velocity and induced magnetic field profiles were later studied \cite{Hunt65}. Analytical solutions were obtained for (i) ducts with perfectly conducting walls perpendicular to the imposed magnetic field and thin walls of arbitrary conductivity parallel to the field and for (ii) ducts with non-conducting walls parallel to the magnetic field and thin walls of arbitrary conductivity perpendicular to the field. Numerous additional theoretical studies focusing on single-phase laminar MHD flows in channels and ducts subjected to transverse magnetic fields are available in the literature. These works have been reviewed in various articles and book chapters \cite[e.g., ][]{Branover78,Muller01,Hunt03}. By solving numerically the problem equations, Krasnov et al. \cite{Krasnov10} and Boeck and Krasnov \cite{Boeck14} confirmed that for the same $\Ha$, the velocity profiles in non-conducting rectangular ducts are similar to those obtained by the Two-Plate model, when scaled by their maximal velocity.

In two-phase flow systems, the presence of a second phase can significantly alter the velocity profile and the associated pressure gradient. Due to the density difference between the phases, stratified flow configurations are commonly observed in both horizontal and inclined conduits. However, far fewer studies have focused on the analysis of two-phase MHD flows, and most of these have been limited to the simple geometry of two infinite plates. Shail \cite{Shail73} investigated the Hartmann flow of a conducting fluid between two horizontal plates, with a layer of non-conducting fluid separating the top wall from the conducting fluid. Lohrasbi and Sahai \cite{Lohrasbi88} derived analytical solutions for the velocity and temperature profiles in two-phase laminar steady MHD flow between two horizontal plates with one conducting phase. A similar problem in an inclined channel was addressed by \cite{Malashetty97}. A subsequent study by \cite{Umavathi10} presented a solution for MHD Poiseuille-Couette flow and heat transfer of two immiscible fluids between inclined parallel plates. Shah et al. \cite{Shah22} derived approximate analytical solutions for the velocity and temperature fields of unsteady MHD generalized Couette flows of two immiscible and electrically conducting fluids flowing between two horizontal plates subjected to an inclined magnetic field and an axial electric field. In all of these studies, the problem solved was for a predefined flow configuration (i.e., for given interface location and pressure gradient). Recently, Parfenov et al. \cite{Parfenov24} presented a solution for holdup and pressure gradient in horizontal and inclined channels for specified input flow rates of a conductive liquid and gas. The solution enabled exploring the potential for pressure gradient reducing and pumping power savings by the gas flow.

None of the above studies on two-phase MHD flows have considered the effect of the channel walls' electrical conductivity on the induced magnetic field and its influence on velocity profiles and two-phase flow characteristics. The presence of a non-conductive layer (e.g., gas) breaks the symmetry of the boundary conditions assumed in single-phase flow studies, impacting not only the velocity profile of the conductive fluid but also the induced magnetic field. As a result, the effects of the induced magnetic field on two-phase flow characteristics remain unclear, even in a simple two-plate geometry. Furthermore, to the best of our knowledge, studies on two-phase MHD flow in a rectangular duct have not yet been explored in the literature. Consequently, it remains uncertain to what extent the presence of sidewalls and the electrical conductivity of the duct walls influence the induced magnetic field and the resulting flow characteristics.

In this study, we refer to two-phase stratified flow with one phase being conductive subject to external magnetic field. Analytical solutions for a two-phase flow between parallel plates are derived for different wall conductivities. The results are comparable to wide rectangular channels, for which numerical solutions are presented. The verified numerical solution can be used to explore two-phase flow characteristics in realistic experimental set-ups.

\section{Problem formulation} \label{Sec: Formulation}

\begin{figure}[h!]
	\centering
	\def\svgwidth{0.8\textwidth}
	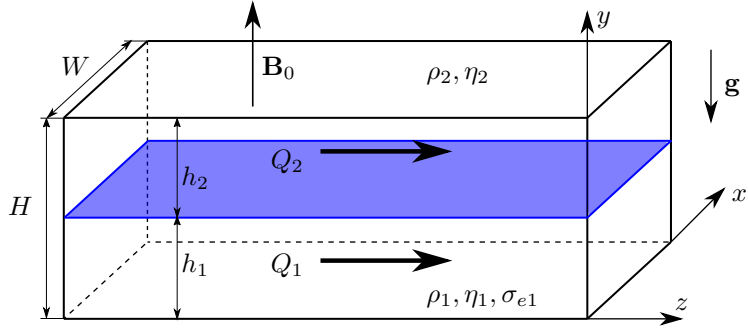
	\caption{Stratified two-phase flow in a rectangular duct under a constant vertical magnetic field.}
	\label{Fig: MHD_geometry}
\end{figure}

We address the magnetohydrodynamics of two-phase stratified flow in a horizontal rectangular duct of height $H$ and width $W$. $AR = W/H$ is the duct aspect ratio (Fig.\ \ref{Fig: MHD_geometry}). The flow is subject to a constant vertical magnetic field $\vec{B_0}=\bigl(0,B_0,0\bigr)$. The volumetric flow rates of the heavy and light fluids are $Q_1$ and $Q_2$, respectively. The interface between two fluids is assumed to be plane, and the height of the lower (heavy) phase layer is denoted as $h_1$. Then the holdup is defined as $\displaystyle h=h_1/H$. The heavy phase is electrically conductive (e.g., liquid metal), whereas phase 2 is a lighter fluid assumed to be electrically non-conductive (e.g., gas). The physical properties of the conducting liquid are $\eta_1$ - dynamic viscosity; $\rho_1$ - density; $\sigma_{e1}$ - electric conductivity. 

For clarity, we present a brief derivation of the relevant MHD equations boundary conditions for the heavy conductive phase. The electromagnetic problem is defined by the Maxwell equations, while assuming a time-independent electromagnetic field:
\begin{subequations}
	\label{Eq: Maxwell's equations}
	\begin{align}
		\nabla\times\vec{E} &= -\frac{\partial \vec{B}}{\partial t} = 0,
		\\
		\nabla\cdot\vec{B} &= 0,
		\\
		\nabla\times\vec{B} &= \mu_0 \bigl(\vec{J} + \epsilon_0 \frac{\partial \vec{E}}{\partial t}\bigr) = \mu_0 \vec{J},
	\end{align}
\end{subequations}
where $\epsilon_0$ and $\mu_0$ are the vacuum permittivity and permeability, respectively. The electric current density in the conductive liquid obeys Ohm's law:
\begin{equation}
	\label{Eq: Ohm's law}
	\vec{J}=\sigma_{e1} \bigl(\vec{E} + \vec{\hat{U}_1}\times\vec{B}\bigr),
\end{equation}
where $\vec{\hat{U}_1} = \bigl(\hat{U}_{1x},\hat{U}_{1y},\hat{U}_{1z}\bigr)$ is the (dimensional) velocity field of the conductive liquid.  In the following, we assume steady, fully developed unidirectional laminar flow in the $z$-direction, so that $\vec{\hat{U}_1} = \bigl(0,0,\hat{U}_{1z}(\hat{x},\hat{y})\bigr)$ and all the flow and electromagnetic variables are independent of $z$. To distinguish between the dimensional and dimensionless velocities and coordinates, for the dimensional values the "hat" notation is used (e.g., $\hat{U}_1$). However, only dimensional values of $\vec{E}$, $\vec{B}$, and $\vec{J}$ are used herein.

The induction equation is derived by applying the curl operator to Eq.\ \ref{Eq: Maxwell's equations}c and accounting for Eq.\ \ref{Eq: Maxwell's equations}b:
\begin{subequations}
	\begin{align}
		\nabla\times\bigl(\nabla\times\vec{B}\bigr) &= \nabla\bigl(\nabla\cdot\vec{B}\bigr) - \Delta\vec{B} = \mu_0 \nabla\times\vec{J}
		\\
		\label{Eq: Laplacian_B_1}
		\Delta\vec{B} &= -\mu_0 \nabla\times\vec{J}
	\end{align}
\end{subequations}

Substituting Ohm's law, Eq.\ \ref{Eq: Ohm's law}, into Eq.\ \ref{Eq: Laplacian_B_1} and taking into account Eq.\ \ref{Eq: Maxwell's equations}a, one gets:
\begin{equation} \label{Eq: Laplacian_B_2}
	\Delta\vec{B} 
	= -\mu_0 \nabla\times\bigl[\sigma_{e1} \bigl(\vec{E} + \vec{\hat{U}_1}\times\vec{B}\bigr)\bigr]
	=  -\mu_0 \sigma_{e1}\nabla\times\bigl(\vec{\hat{U}_1}\times\vec{B}\bigr)
\end{equation}

The r.h.s. of Eq.\ \ref{Eq: Laplacian_B_2} can be evaluated as:
\begin{subequations}
	\begin{align}
		\vec{\hat{U}_1}\times\vec{B} 
		&= - \hat{U}_{1z} B_y \vec{e_x} + \hat{U}_{1z} B_x \vec{e_y}
		\\
		\nabla\times\bigl(\vec{\hat{U}_1}\times\vec{B}\bigr)
		&= \Bigl(B_x \frac{\partial \hat{U}_{1z}}{\partial \hat{x}}
		+ B_y \frac{\partial \hat{U}_{1z}}{\partial \hat{y}}\Bigr)
		\vec{e_z}
	\end{align}
\end{subequations}

Therefore, the induction equation in vector form reads:
\begin{equation} \label{Eq: vector_induction}
	\Delta \vec{B} + \mu_0 \sigma_{e1} 
	\Bigl(B_x \frac{\partial \hat{U}_{1z}}{\partial \hat{x}}
	+ B_y \frac{\partial \hat{U}_{1z}}{\partial \hat{y}}\Bigr)
	\vec{e_z}
	=0
\end{equation}

Eq.\ \ref{Eq: vector_induction} can be written as three scalar equations in Cartesian coordinates:
\begin{subequations} \label{Eq: scalar_induction}
	\begin{align} 
		&\frac{\partial^2 B_x}{\partial \hat{x}^2} 
		+ \frac{\partial^2 B_x}{\partial \hat{y}^2}
		= 0 
		\\
		&\frac{\partial^2 B_y}{\partial \hat{x}^2} 
		+ \frac{\partial^2 B_y}{\partial \hat{y}^2}
		= 0 
		\\		
		&\frac{\partial^2 B_z}{\partial \hat{x}^2} 
		+ \frac{\partial^2 B_z}{\partial \hat{y}^2}
		+ \mu_0 \sigma_{e1} 
		\Bigl(B_x \frac{\partial \hat{U}_{1z}}{\partial \hat{x}}
		+ B_y \frac{\partial \hat{U}_{1z}}{\partial \hat{y}}\Bigr)
		= 0
	\end{align}
\end{subequations}

The magnetic field can be represented as a superposition of the externally applied magnetic field $B_0$ and that induced by the motion (flow) of the conductive liquid, $\displaystyle \vec{B} = B_0 \vec{e_y} + \vec{\hat{B}_{ind}}$. In fact, equations\ \ref{Eq: scalar_induction} with homogeneous boundary conditions for $\vec{\hat{B}_{ind}}$ indicate that the induced magnetic field acts only in the flow direction, i.e., $\displaystyle\vec{\hat{B}_{ind}}=\bigl(0,0,\hat{B}_{ind}(x,y)\bigr)$. The total (dimensional) magnetic field is then expressed as $\displaystyle\vec{B}=\bigl(0,B_0,\hat{B}_{ind}(x,y)\bigr)$.

The dimensionless form of the only non-zero (axial) component of the induction equation reads:
\begin{equation} \label{Eq: induction_dimless}
	\frac{\partial^2 B_{ind}}{\partial x^2} 
	+ \frac{\partial^2 B_{ind}}{\partial y^2}
	+ \Rey_m 
	\frac{\partial U_{1z}}{\partial y}
	= 0,
\end{equation}
where the induced magnetic field is scaled by $B_0$, i.e. $\displaystyle B_{ind}=\hat{B}_{ind}/B_0$, the velocity is scaled by the superficial velocity of the conductive phase, $\displaystyle U_{1S} = Q_1/(H W)$ with $\displaystyle Q_1$ being the corresponding flow rate. The length are scaled by the duct height $H$. Thus, the only dimensionless parameter in Eq.\ \ref{Eq: induction_dimless} is the magnetic Reynolds number, $\displaystyle \Rey_m=\mu_0\sigma_{e1} U_{1S} H$.

It can be noted that upon rescaling $B_{ind}$ by $\Rey_m$, the resulting scaled induced magnetic field is independent of $\Rey_m$. In the book 
of M\"uhler and B\"uhler \cite{Muller01} (equation 3.2 therein), the (dimensional) induced magnetic field ($\hat{B}_{ind}$ in our notation) was scaled by $\displaystyle B_0 \Rey_m/\Ha = \mu_0 U_{1S} / \sqrt{\sigma_{e1} \eta_1}$ and denoted as $b$, where $\displaystyle \Ha = B_0 H \sqrt{\sigma_{e1}/\eta_1}$. In terms of $b=\hat{B}_{ind}/B_0\cdot\Ha/\Rey_m$, Eq.\ \ref{Eq: induction_dimless} can be rewritten as:
\begin{equation} \label{Eq: Induction_b}
	\frac{\partial^2 b}{\partial x^2} 
	+ \frac{\partial^2 b}{\partial y^2}
	+ \Ha
	\frac{\partial U_{1z}}{\partial y}
	= 0.
\end{equation}

The dimensional momentum equations for the fully developed flow of the conductive liquid read:
\begin{subequations} \label{Eq: Momentum_1_dim_1}
	\begin{align} 
		0
		&= -\frac{\partial \hat{P}}{\partial \hat{x}}  + \bigl(\vec{J}\times\vec{B}\bigr)_x, 
		\\
		0
		&= -\frac{\partial \hat{P}}{\partial \hat{y}} + \rho_1 g  + \bigl(\vec{J}\times\vec{B}\bigr)_y,
		\\
		0 
		&= -\frac{\partial \hat{P}}{\partial \hat{z}} + \eta_1 \Delta\hat{U}_{1z} + \bigl(\vec{J}\times\vec{B}\bigr)_z, 
	\end{align}
\end{subequations}

In the following, the (negative) pressure gradient driving the flow is denoted as $\displaystyle\hat{G}=\partial\hat{P}/\partial\hat{z}$($<0$). The last term in Eqs.\ \ref{Eq: Momentum_1_dim_1}, the Lorenz force, can be rewritten in terms of the magnetic field only, using Eq.\ \ref{Eq: Maxwell's equations}c:
\begin{subequations}
	\begin{align}
		&\vec{J}\times\vec{B} = \frac{1}{\mu_0} \bigl(\nabla\times\vec{B}\bigr)\times\vec{B},
		\\
		&\begin{aligned}
			\bigl(\nabla\times\vec{B}\bigr)\times\vec{B} 
			&= \Bigl[\nabla
			\times \bigl(B_0 \vec{e_y} + \vec{\hat{B}_{ind}}\bigr)\Bigr] 
			\times \bigl(B_0 \vec{e_y} + \vec{\hat{B}_{ind}}\bigr)
			\\
			&= \Bigl(\frac{\partial\hat{B}_{ind}}{\partial \hat{y}} \vec{e_x}
			- \frac{\partial\hat{B}_{ind}}{\partial \hat{x}} \vec{e_y}\Bigr)
			\times \bigl(B_0 \vec{e_y} + \vec{\hat{B}_{ind}}\bigr)
			\\
			&= - \hat{B}_{ind} \frac{\partial\hat{B}_{ind}}{\partial \hat{x}} \vec{e_x}
			-  \hat{B}_{ind} \frac{\partial\hat{B}_{ind}}{\partial \hat{y}} \vec{e_y}
			+ B_0 \frac{\partial\hat{B}_{ind}}{\partial \hat{y}} \vec{e_z}.		
		\end{aligned}		
	\end{align}
\end{subequations}
Thereby the momentum equation in the axial $z$-direction, Eq.\ \ref{Eq: Momentum_1_dim_1}c, becomes:
\begin{equation} \label{Eq: Momentum_1_dim}
	\eta_1 \Bigl(\frac{\partial^2 \hat{U}_{1z}}{\partial \hat{x}^2} 
	+ \frac{\partial^2 \hat{U}_{1z}}{\partial \hat{y}^2}\Bigr) 
	+ \frac{1}{\mu_0} B_0 \frac{\partial\hat{B}_{ind}}{\partial \hat{y}} 
	= \hat{G}.
\end{equation}

Due to absence of secondary flows and independence of both the gravitational acceleration, $g$, and the induced magnetic field, $\hat{B}_{ind}$ on the $z$ coordinate, the pressure gradient, $\hat{G}$, is constant across the duct cross-section. This can be proven by differentiating Eqs.\ \ref{Eq: Momentum_1_dim_1}a and \ref{Eq: Momentum_1_dim_1}b with respect to $z$ and changing the order of differentiation, so that the equations reduce to $\displaystyle\partial/\partial x (\partial\hat{P}/\partial\hat{z})=0$ and $\displaystyle\partial/\partial y (\partial\hat{P}/\partial\hat{z})=0$.
The dimensionless form of Eq.\ \ref{Eq: Momentum_1_dim} reads
\begin{equation} \label{Eq: Momentum_b}
	\frac{\partial^2 U_{1z}}{\partial x^2} 
	+ \frac{\partial^2 U_{1z}}{\partial y^2}
	+ \frac{\Ha^2}{\Rey_m} \frac{\partial B_{ind}}{\partial y} 
	=
	\frac{\partial^2 U_{1z}}{\partial x^2} 
	+ \frac{\partial^2 U_{1z}}{\partial y^2}
	+ \Ha \frac{\partial b}{\partial y} 
	= G,
\end{equation}	
where the pressure gradient is scaled by $\displaystyle\eta_1 U_{1S} / H^2$. Similarly to the induction equation \ref{Eq: Induction_b}, the only non-dimensionless group that appears in the momentum equation, Eq.\ \ref{Eq: Momentum_b}, is the Hartmann number.

The dimensionless momentum equation for the non-conductive lighter fluid layer reads:
\begin{equation} \label{Eq: Momentum_2_dim}
	\frac{\partial^2 U_{2z}}{\partial x^2} 
	+ \frac{\partial^2 U_{2z}}{\partial y^2}
	= \eta_{1 2} G,
\end{equation}
where the velocity is scaled by the superficial velocity of conductive liquid $\displaystyle U_{2z}=\hat{U}_{2z} / U_{1S}$ and $\displaystyle\eta_{1 2} = \eta_1/\eta_2$ is the viscosity ratio of fluids.

In order to find the fully developed velocity profile in both phases and the induced magnetic field, the governing equations \ref{Eq: Induction_b}, \ref{Eq: Momentum_b}, and \ref{Eq: Momentum_2_dim} have to be solved with the appropriate boundary conditions. For the velocity, they are no-slip condition on the duct walls and continuity of the velocity and shear stress across the fluid--fluid interface:
\begin{subequations} \label{Eq: BC_U_rectangular}
	\begin{align}
		U_{1z}(x,y=0) &= 0,\qquad U_{1z}(x=0,AR,y) = 0
		\\
		U_{2z}(x,y=1) &= 0,\qquad U_{2z}(x=0,AR,y) = 0 
		\\
		U_{1z}(x,y=h) &= U_{2z}(x,y=h) = U_i
		\\
		\eta_{1 2} \frac{\partial U_1}{\partial y}\Biggr\rvert_{x,y=h} &=
		\frac{\partial U_2}{\partial y}\Biggr\rvert_{x,y=h}
	\end{align}
\end{subequations}

The boundary conditions for the induced magnetic field depends on the conductivity of the duct walls and fluid--fluid interface (see Appendix\ \ref{Sec: BC_induced_B}):
\begin{subequations} \label{Eq: BC_induced_B_rectangular}
	\begin{align}
		&-\frac{\partial b}{\partial y}\Biggr\rvert_{x,y=0}
		+ \frac{1}{c_{bw}} b(x,y=0) 
		= 0 
		\\
		&-\frac{\partial b}{\partial x}\Biggr\rvert_{x=0,AR,y}
		+ \frac{1}{c_{sw}} b(x=0,AR,y) 
		= 0 
		\\
		&-\frac{\partial b}{\partial y}\Biggr\rvert_{x,y=h}
		+ \frac{1}{c_i} b(x,y=0) 
		= 0 
	\end{align}
\end{subequations}
where $c_{bw}$, $c_{sw}$, and $c_i$ are the dimensionless electric conductivities of the bottom and side walls and the interface, respectively. The dimensionless conductivity is obtained by normalizing the dimensional one with $\displaystyle\sigma_{e1} H / t$, where $t$ is the thickness of the corresponding wall. Since the upper phase is non-conductive, the conductivity of the interface is assumed to be zero $\displaystyle \bigl(c_i=0\bigr)$.

In this study, we consider only two limiting cases of the walls being either fully insulating $\displaystyle \bigl(c=0\bigr)$ or perfectly conducting $\displaystyle \bigl(c\to\infty\bigr)$ surfaces. In the former case, $b=0$, in the latter -- the normal derivative of $b$ becomes zero at the wall.

\section{The Two-Plate model analytical solutions}\label{Sec: Analytical}

In the limit of flow between two infinite parallel plates (i.e., Two-Plate model), the velocity and induced magnetic field are independent of the spanwise coordinate $x$ (i.e., $\displaystyle U_z = U_{z}(y)$ and $b=b(y)$). The dimensionless governing equations for the rectangular duct (Eqs. \ref{Eq: Induction_b}, \ref{Eq: Momentum_b}, and \ref{Eq: Momentum_2_dim}) are reduced to:
\begin{subequations} \label{Eq: TP_Governing}
	\begin{align}
		&\frac{\partial^2 b}{\partial y^2}
		+ \Ha
		\frac{\partial U_{1}}{\partial y}
		= 0,
		\\
		&\frac{\partial^2 U_{1}}{\partial y^2}
		+ \Ha \frac{\partial b}{\partial y} 
		= G,
		\\
		&\frac{\partial^2 U_{2}}{\partial y^2}
		= \eta_{1 2} G.
	\end{align}
\end{subequations}

Note that, for convenience, we drop the subscript $z$ in the following derivations (e.g., $\displaystyle U_{1z}\rightarrow U_{1}$). 
The conductivity of the two plates is assumed to be the same, $c_{bw}$, and the superficial velocity is calculated simply through the channel height, $\displaystyle U_{1S,2S} = Q_{1,2} / H$, where the definition of flow rate, $Q_{1,2}$, is volumetric per unit channel width, i.e. in $m^2/s$.

By differentiating Eqs.\ \ref{Eq: TP_Governing}a with respect to $y$ and substituting into \ref{Eq: TP_Governing}b, one can derive an equation for $b$ which is uncoupled from $U_{1}$:
\begin{equation} \label{Eq: TP_b_equation}
	-\frac{1}{\Ha}\frac{\partial^3 b}{\partial y^3}
	+ \Ha \frac{\partial b}{\partial y} 
	= G
\end{equation}

Integration of Eq.\ \ref{Eq: TP_Governing}a yields:
\begin{equation} \label{Eq: b_derivative}
	\frac{\partial b}{\partial y} = 
	-\Ha U_1 + C_b,
\end{equation}
where $C_b$ is a constant. Owing to the no-slip B.C. on the bottom wall $\displaystyle \bigl(U_1(y=0)=0\bigr)$, this constant can be found from the derivative of $b$:
\begin{equation} \label{Eq: BC_Cb}
	\frac{\partial b}{\partial y} \Biggr\rvert_{y=0} = C_b.
\end{equation}

Eq.\ \ref{Eq: TP_Governing}b can be then rewritten:
\begin{equation} \label{Eq: TP_U_equation}
	\frac{\partial^2 U_{1}}{\partial y^2}
	-\Ha^2 U_1 + \Ha C_b
	= G
\end{equation}

When the bottom wall is a perfect conductor, $\displaystyle C_b = \partial b/\partial y (y=0)=0$, so that Eq.\ \ref{Eq: TP_U_equation} becomes independent of $b$:
\begin{equation} \label{Eq: TP_U_equation_perf_cond}
	\frac{\partial^2 U_{1}}{\partial y^2}
	-\Ha^2 U_1
	= G
\end{equation}

Hence, whenever Eq.\ \ref{Eq: TP_U_equation_perf_cond} is considered to represent the momentum equation for the conductive fluid, as in most studies on two-phase MHD flows (e.g., \cite{Shail73,Shah22,Parfenov24}), the solution obtained is actually valid for perfectly conducting walls. In the case of insulating walls (Eq.\ \ref{Eq: TP_U_equation}), there is an additional term on the left-hand side of the momentum equation that counteracts the Lorentz force due to the external magnetic field. The general solution of the coupled equations\ \ref{Eq: TP_Governing}a,c and \ref{Eq: TP_b_equation} is:
\begin{subequations} \label{Eq: General_solution}
	\begin{align}
		b(y) &= -C_1 \cosh(\Ha y) 
		- C_2 \sinh (\Ha y) 
		+ C_3
		+ \frac{G}{\Ha} y
		\\
		U_1(y) &= C_1 \sinh (\Ha y) 
		+ C_2 \cosh (\Ha y) 
		+ C_4
		\\
		U_2(y) &= \frac{\eta_{1 2} G}{2} y^2
		+ D_1 y
		+ D_2
	\end{align}
\end{subequations}
Constants in Eq.\ \ref{Eq: General_solution} can be found from the boundary conditions. Applying the no-slip condition at the bottom and top walls and the continuity of velocity at the interface (i.e., $U_1(y=h) = U_2(y=h) = U_i$, where $U_i$ denotes the interfacial velocity) yields: 
\begin{subequations}
	\begin{align}
		&C_2 + C_4 = 0
		\\
		&C_1 \sinh(\Ha h) + C_2 \cosh (\Ha h) + C_4 = U_i
		\\
		&D_2 = - \dfrac{\eta_{1 2} G}{2} - D_1
		\\
		&\dfrac{\eta_{1 2} G}{2} \biggl(h^2 - 1\biggr) + D_1 (h- 1) = U_i
	\end{align}
\end{subequations}
The tangential shear stress is continuous across the interface:
\begin{equation}
	\eta_{1 2} G h + D_1 
	= \eta_{1 2} \Ha \bigl[C_1 \cosh (\Ha y) 
	+ C_2 \sinh (\Ha y) \bigr]
\end{equation}
and the magnetic field $b$ satisfies Eqs.\ \ref{Eq: BC_induced_B_rectangular}a and \ref{Eq: BC_induced_B_rectangular}c at the bottom wall and non-conductive interface, respectively:
\begin{subequations} \label{Eq: C2_derivation}
	\begin{align}
		&\Ha C_2 - \frac{G}{\Ha} + \frac{1}{c_{bw}}(-C_1 + C_3) = 0
		\\
		&-C_1 \cosh(\Ha h) 
		- C_2 \sinh (\Ha h) 
		+ C_3
		+ \frac{G}{\Ha} h
		= 0
	\end{align}
\end{subequations}
In the following, we present the final expressions for the velocity and induced magnetic field profiles for the two limiting cases for the conductivity of the bottom wall: perfectly insulating wall, i.e., $c_{bw} = 0$, and perfectly conductive one, where $c_{bw} \to \infty$.

\subsection{Insulating bottom wall, $c_{bw} = 0$}
\vspace*{-6mm}
\begin{subequations} \label{Eq: U_TP_profiles_ins}
	\begin{align}
		&\begin{aligned}
			&U_1(y) = 
			\dfrac{1}{\sinh(\Ha h)} \bigl[U_i - C_2 \bigl(\cosh(\Ha h) - 1\bigr)\bigr] \sinh(\Ha y)
			\\
			&+ C_2 \bigl(\cosh(\Ha y) - 1\bigr)
		\end{aligned}
		\\
		&\begin{aligned}
			&U_2(y) =
			\eta_{1 2} \biggl\{
			\dfrac{G}{2} (y + 1)
			+ 		
			\Ha \coth(\Ha h) \bigl[
			U_i - C_2 \bigl(\cosh(\Ha h) - 1\bigr)\bigr]
			\\
			&+ C_2 \Ha \sinh(\Ha h)  
			- G h
			\biggr\}
			(y - 1)
		\end{aligned}
		\\
		&\begin{aligned}
			&b(y) =
			C_2 \biggl\{
			\dfrac{1}{\sinh(\Ha h)}\bigl(\cosh(\Ha h) - 1\bigr)
			\bigl[\cosh(\Ha y)
			- \cosh(\Ha h)\bigr]
			\\
			&+ \bigl[\sinh(\Ha h)
			- \sinh(\Ha y)\bigr]
			\biggr\}
			+ \dfrac{U_i}{\sinh(\Ha h)} \bigl[\cosh(\Ha y)
			- \cosh(\Ha h)\bigr]
			\\
			&- \frac{G}{\Ha} (h-y)
		\end{aligned}
	\end{align}
\end{subequations}
where $C_2$ and $U_i$ are given by:
\begin{subequations}
	\begin{align}
		&C_2 = \frac{G}{\Ha} 
		\dfrac{\eta_{1 2} (1-h) \Ha \bigl[
			(1+h) \cosh(\Ha h) - (1-h)\bigr] + 2 h \sinh(\Ha h)}{4 \bigl[\cosh(\Ha h) - 1\bigr] + 2 \eta_{1 2} (h-1) \Ha \sinh(\Ha h)}
		\\
		&\begin{aligned}
			&U_i = -G \eta_{1 2}
			\\
			&\times
			\dfrac{(1-h)\biggl[
				\dfrac{1}{2}(1+h) + \dfrac{C_2}{G} \bigl\{
				-\Ha \coth(\Ha h)\bigl[
				\cosh(\Ha h) - 1\bigr]
				+ \Ha \sinh(\Ha h)\bigr\}
				- h\biggr]}{1 + \eta_{1 2} (1-h) \Ha \coth(\Ha h)}
		\end{aligned}
	\end{align}
\end{subequations}
It is important to observe that $C_2$ and $U_i$ are proportional to $G$, thereby the velocity and the induced magnetic field are proportional to the dimensionless driving pressure gradient, $G$.
Integration of the velocity profiles over their corresponding cross sections yields the flow rates of the two phases, whereby their ratio is found as:
\begin{align} \label{Eq: Q21_perf_cond}
	\begin{aligned}
		&Q_{2 1} = \frac{Q_2}{Q_1} = \frac{U_{2S}}{U_{1S}}
		=
		\\
		&- \dfrac{\eta_{1 2}(1-h)^2(h+2) G \Ha
		}{
			6 \dfrac{\cosh(\Ha h) - 1}{\sinh(\Ha h)} \bigl\{
			U_i - C_2 \bigl[\cosh(\Ha h) - 1\bigr] 
			\bigr\} 
			+ 6 C_2 \bigl[
			\sinh(\Ha h) - \Ha h\bigr]
		}
		\\
		&+ \dfrac{\eta_{1 2}
			\bigl\{
			\Ha \coth(\Ha h) \bigl[
			U_i - C_2 \bigl(\cosh(\Ha h) - 1\bigr)\bigr]
			+ C_2 \Ha \sinh(\Ha h)  
			- G h
			\bigr\}
			(1-h)^2
			\Ha
		}{
			2 \dfrac{\cosh(\Ha h) - 1}{\sinh(\Ha h)} \bigl\{
			U_i - C_2 \bigl[\cosh(\Ha h) - 1\bigr] 
			\bigr\} 
			+ 2 C_2 \bigl[
			\sinh(\Ha h) - \Ha h\bigr]
		}
	\end{aligned}
\end{align}

Since $C_2$ and $U_i$ are proportional to $G$, the flow rate ratio, $Q_{2 1}$, is independent of the dimensionless pressure gradient. In fact, Eq.\ \ref{Eq: Q21_perf_cond} can be used to find holdup $h$ for specified $Q_{2 1}$, $\Ha$, and $\eta_{1 2}$. 

Once the holdup is obtained, $G$ is given by:
\begin{align} \label{Eq: Press_Gradient_ins}
	\begin{aligned}
		&G (c_{bw}=0) =
		- \dfrac{2 Q_{2 1}}{\eta_{1 2} (1-h)^2} 
		\\
		& \times
		\dfrac{1}{
			\dfrac{1}{3}(h+2) + \dfrac{1}{G}
			\bigl\{
			\Ha \coth(\Ha h) \bigl[
			U_i - C_2 \bigl(\cosh(\Ha h) - 1\bigr)\bigr]
			+ C_2 \Ha \sinh(\Ha h)  
			- G h
			\bigr\}}
	\end{aligned}
\end{align}
Note that, upon substitution of $C_2$ and $U_i$, the right-hand-side of Eq.\ \ref{Eq: Press_Gradient_ins} is, in fact, independent of $G$. 

For given $Q_1$ and $\Ha$, the pressure gradient factor that quantifies the effect of presence of the gas phase (lubrication factor) is defined as the ratio of $G$ by the frictional pressure gradient of the single-phase flow of the conductive phase $G_1$ (see Appendix\ \ref{Sec: Single_phase_TP}): 
\begin{align}
	&\begin{aligned} \label{Eq: P_1f_ins}
		&P^1_f (c_{bw}=0) = \frac{G}{G_1}\Biggr\rvert_{c_{bw}=0}
		= \frac{\Bigl[1-\dfrac{2}{\Ha}\tanh\biggl(\dfrac{\Ha}{2}\biggr)\Bigr]}{
			\eta_{1 2} (1-h)^2  \Ha \tanh \Bigl(\dfrac{\Ha}{2}\Bigr)
		}  
		\\
		&\times
		\dfrac{Q_{2 1} }{\dfrac{1}{3}(h+2) 
			+ \dfrac{1}{G}
			\bigl\{
			\Ha \coth(\Ha h) \bigl[
			U_i - C_2 \bigl(\cosh(\Ha h) - 1\bigr)\bigr]
			+ C_2 \Ha \sinh(\Ha h)  
			- G h
			\bigr\}}
	\end{aligned}
\end{align}

The presence of gas flowing with the flow rate $Q_2$ requires additional pumping power. Hence, the pumping power factor of the two-phase system (compared to that required for the single-phase flow of the conductive fluid) is:
\begin{equation}
	\Po = P^1_f \frac{Q_1 + Q_2}{Q_1} = P^1_f (1 + Q_{2 1})
\end{equation}

The lubrication effect can be also analyzed based on the integral balance of forces acting on the two-phase MHD flow in the cross section. This balance can be found by integration of Eqs.\ \ref{Eq: TP_Governing}b and \ref{Eq: TP_Governing}c across each fluid layer, $y\in[0,h]$ and $y\in[h,1]$, respectively, and then summation of the forces. Taking into account that the induced magnetic field is equal to zero at the interface, the balance reads:
\begin{equation} \label{Eq: Integral_balance}
	- \tau_{bw} - \tau_{tw} - \Ha b(y=0) = G,
\end{equation}
where the dimensionless bottom and top wall shear stresses are defined as:
\begin{subequations} \label{Eq: Wall_stresses}
	\begin{align}
		\tau_{bw} &= \frac{\partial U_1}{\partial y} \Biggr\rvert_{y=0}
		\\		
		\tau_{tw} &= - \frac{1}{\eta_{1 2}}
		\frac{\partial U_2}{\partial y} \Biggr\rvert_{y=1}		
	\end{align}
\end{subequations}
Note that in this work the wall shear stresses acting opposite to the flow direction are defined as positive (i.e., $\displaystyle\tau_{bw},\tau_{tw}>0$), while the negative pressure gradient drives the flow (i.e., $G<0$).

In the case of insulating bottom wall, the induced magnetic field at the wall is equal to zero ($b(y=0)=0$), so that the pressure gradient force (pressure gradient multiplied by the channel height) is balanced by the wall shear stresses:
\begin{equation} \label{Eq: Integral_balance_ins}
	G (c_{bw}=0) = -\tau_{bw} - \tau_{tw}.
\end{equation}

The dimensionless bottom wall shear stress can be found by calculating derivative of Eq.\ \ref{Eq: U_TP_profiles_ins}a:
\begin{equation} \label{Eq: Shear_stress_bottom}
	\tau_{bw} = \dfrac{\Ha}{\sinh(\Ha h)} \bigl[U_i - C_2 \bigl(\cosh(\Ha h) - 1\bigr)\bigr]
\end{equation}

The shear stress factor that quantifies the effect of both external magnetic field and presence of the gas phase is then defined as the ratio of $\tau_{bw}$ by the wall shear stress in the single-phase flow of the conductive phase with the same flow rate ($\displaystyle\tau_{bw\rvert1}(\Ha=0)=6$):
\begin{equation} \label{Eq: Shear_factor_bottom}
	\tau_{bw}^{1,0} = \frac{\tau_{bw}}{\tau_{bw\rvert1}(\Ha=0)} = \dfrac{\Ha}{6\sinh(\Ha h)} \bigl[U_i - C_2 \bigl(\cosh(\Ha h) - 1\bigr)\bigr]
\end{equation}

In order to single out the effect of the gas phase on the bottom wall shear stress, $\tau_{bw}^{1,0}$, should be divided by $\tau_{w\rvert1}^0$ (Eq.\ \ref{Eq: Shear_stress_factor_SP} in Appendix\ \ref{Sec: Single_phase_TP}). This factor is denoted as $\tau_{bw}^{1}$.

Similarly, the top wall shear stress can be found by taking derivative of Eq.\ \ref{Eq: U_TP_profiles_ins}b:
\begin{equation} \label{Eq: Shear_stress_top}
	\tau_{tw} = - \Bigl\{G\bigl(1 - h\bigr) + \Ha\coth(\Ha h) \bigl[U_i - C_2 \bigl(\cosh(\Ha h) - 1\bigr)\bigr] + C_2 \Ha \sinh(\Ha h) \Bigr\}
\end{equation}

The shear stress factors $\tau_{tw}^{1,0}$ and $\tau_{tw}^{1}$ are then defined  by dividing $\tau_{tw}$ by $\tau_{bw\rvert1}(\Ha=0)$ and $\tau_{w\rvert1}^0$ in the same manner as $\tau_{bw}^{1,0}$ (Eq.\ \ref{Eq: Shear_factor_bottom}) and $\tau_{bw}^{1}$, respectively.

\subsection{Perfectly conducting bottom wall, $c_{bw} \to \infty$}
\vspace*{-6mm}
\begin{subequations}
	\begin{align}
		&\begin{aligned}
			&U_1(y) = 
			\dfrac{1}{\sinh(\Ha h)} \bigl[U_i - \dfrac{G}{\Ha^2} \bigl(\cosh(\Ha h) - 1\bigr)\bigr] \sinh(\Ha y)
			\\
			&+ \dfrac{G}{\Ha^2} \bigl(\cosh(\Ha y) - 1\bigr)
		\end{aligned}
		\\       
		&\begin{aligned}
			&U_2(y) =
			\eta_{1 2} \biggl\{
			\dfrac{G}{2} (y + 1)
			+ 
			\Ha \coth(\Ha h)
			\bigl[
			U_i - \dfrac{G}{\Ha^2} \bigl(\cosh(\Ha h) - 1\bigr)\bigr] 
			\\           
			&+ G \bigl(
			\sinh(\Ha h)   
			-  h
			\bigr)
			\biggr\}
			(y - 1)
		\end{aligned}
		\\
		&\begin{aligned}
			&b(y) =
			- \dfrac{1}{\sinh(\Ha h)} \bigl[
			U_i - \dfrac{G}{\Ha^2}\bigl(\cosh(\Ha h) - 1\bigr)
			\bigr]
			\bigl[\cosh(\Ha y)
			- \cosh(\Ha h)\bigr]
			\\
			&- \dfrac{G}{\Ha^2}\bigl[
			\bigl(\sinh(\Ha h)
			- \sinh(\Ha y)
			\bigr)
			+ (h-y)
			\bigr]
		\end{aligned}
	\end{align}
\end{subequations}
where $U_i$ is defined as:
\begin{align} \label{Eq: U_i_cond}
	\begin{aligned}
		&U_i = - G \eta_{1 2}
		\\
		&\times 
		\dfrac{(1-h)\biggl[
			\dfrac{1}{2}(1-h) + \dfrac{1}{\Ha} \bigl\{
			-\coth(\Ha h)\bigl[
			\cosh(\Ha h) - 1\bigr]
			+ \sinh(\Ha h)\bigr\}
			\biggr]}{1 + \eta_{1 2} (1-h) \Ha \coth(\Ha h)} 
	\end{aligned}
\end{align}

The flow rate ratio is then found by integration of the velocity profiles across their respective cross section:
\begin{align}
	\begin{aligned}
		&Q_{2 1} = - \Ha^3 (1-h)^2 
		\\
		&\times
		\Biggl[
		\frac{
			\dfrac{\eta_{1 2}}{6}(1-h)^2(h+2) 
		}{
			\sinh(\Ha h) - \Ha h 
			+ \Ha^2 \dfrac{G}{\sinh(\Ha h)}\biggl[
			U_i - \dfrac{G}{\Ha^2} \bigl(\cosh(\Ha h) - 1\bigr)\biggr]
			\bigl[\cosh(\Ha h) - 1\bigr]
			}
		\\
		&- 
		\frac{
			\dfrac{\eta_{1 2}}{2 G} 
			\bigl\{
			\Ha \coth(\Ha h) \bigl[
			U_i - \dfrac{G}{\Ha^2} \bigl(\cosh(\Ha h) - 1\bigr)\bigr]
			+ \dfrac{G}{\Ha^2} \Ha \sinh(\Ha h)  
			- G h
			\bigr\}
			(1-h)^2
		}{
			\sinh(\Ha h) - \Ha h
			+ \Ha^2 \dfrac{G}{\sinh(\Ha h)}\biggl[
			U_i - \dfrac{G}{\Ha^2} \bigl(\cosh(\Ha h) - 1\bigr)\biggr]
			\bigl[\cosh(\Ha h) - 1\bigr]
			}
		\Biggr]
	\end{aligned}
\end{align}

The corresponding pressure gradient is:
\begin{align} \label{Eq: Press_Gradient_perf_cond}
	&\begin{aligned}
		&G (c_{bw} \to \infty) 
		= - \frac{1}{\eta_{1 2} (1-h)^2}
		\\
		&\times
		\dfrac{2 Q_{2 1}}{
			\dfrac{1}{3}(h+2) + 
			\Ha \coth(\Ha h) 
			\biggl[
			\dfrac{U_i}{G} - \dfrac{1}{\Ha^2} \bigl(\cosh(\Ha h) - 1\bigr)
			\biggr]            
			+ \sinh(\Ha h)   
			-  h
		}
	\end{aligned}
\end{align}
Note that here too, as in Eq.\ \ref{Eq: Press_Gradient_ins}, the right-hand-side of Eq.\ \ref{Eq: Press_Gradient_perf_cond} is independent of $G$.

The pressure gradient factor that quantifies the lubrication effect for the case of a perfectly conducting wall reads: 
\begin{align} \label{Eq: P_1f_perf_cond}
	&\begin{aligned}
		&P^1_f (c_{bw} \to \infty) = \frac{G}{G_1}\Biggr\rvert_{c_{bw}\to\infty}
		= \frac{1-\dfrac{2}{\Ha}\tanh\biggl(\dfrac{\Ha}{2}\biggr)}{
			\eta_{1 2} (1-h)^2  \Ha^2
		}  
		\\
		&\times
		\dfrac{2 Q_{2 1} }{\dfrac{1}{3}(h+2) + 
			\Ha \coth(\Ha h) 
			\biggl[
			\dfrac{U_i}{G} - \dfrac{1}{\Ha^2} \bigl(\cosh(\Ha h) - 1\bigr)
			\biggr]            
			+ \sinh(\Ha h)   
			-  h
		}
	\end{aligned}
\end{align}

In the case of perfectly conducting bottom wall, $b(y=0)\ne 0$, so that the balance for the pressure gradient is obtained from Eq.\ \ref{Eq: Integral_balance} as following:
\begin{equation} \label{Eq: Integral_balance_cond}
	G (c_{bw}\to\infty) = -\tau_{bw} - \tau_{tw} - \Ha b(y=0).
\end{equation}

The expressions for the bottom and top wall shear stresses, Eq.\ \ref{Eq: Shear_stress_bottom} and Eq.\ \ref{Eq: Shear_stress_top}, are the same as for the case of insulating bottom wall, with $U_i$ is given by Eq.\ \ref{Eq: U_i_cond} and $C_2 = G/\Ha^2$ (see Eq.\ \ref{Eq: C2_derivation}a with $c_{bw}\to\infty$). 

\section{Numerical method}\label{Sec: Numerics}

For a rectangular duct of a finite aspect ratio, we solve numerically the coupled dimensionless governing equations (Eqs. \ref{Eq: Induction_b}, \ref{Eq: Momentum_b}, and \ref{Eq: Momentum_2_dim}) for the velocity in each phase $U_1(x,y)$ and $U_2(x,y)$ and the scaled magnetic field $b(x,y)$, which is non-zero only in the lower conductive phase. $U_{1,2}(x,y)$ and $b(x,y)$ are subject to the boundary conditions Eq.\ \ref{Eq: BC_U_rectangular} and Eq.\ \ref{Eq: BC_induced_B_rectangular}, respectively. 

The finite--volume method is adopted in this work with the channel cross-section divided into quadrilateral cells and the unknowns variables defined in the cell centers $\displaystyle[x_i,y_j] \in [0...AR, 0...1]$, where $i=0,1,..., N_x$ and $j=0,1,..., N_y$. The number of cells in each sublayer (phase) is defined proportional to its relative height, i.e. $\displaystyle N_{lower}/N_y=h$ for the lower phase. The heavy-phase holdup $h$ is found for specified input flow rates and Hartmann number. In both $x$- and $y$-directions, the grid is stretched using the sine function near the walls and the interface (for $y$). Thereby the cell center coordinates in the vertical direction are obtained from originally uniformly distributed grid cells between 0 and 1 for each sublayer, $\displaystyle y \rightarrow y - a_y \sin(2 \pi y)$ ($a_x=a_y = 0.06$). 

After finite--volume discretization of the governing equations and boundary conditions, the system of linear algebraic equations is obtained in a form of sparse matrix of size $2 N_x N_y$. The problem is solved by a direct solver for sparse linear equation systems, MUMPS package in FORTRAN (see, e.g., \cite{Gelfgat07}). For specified flow rate ratio, the equations are solved with a unit value pressure gradient by varying the holdup (and as a consequence changing the computational grid) until the desired $Q_{2 1}$ is reached. Once the correct value of the holdup is found he dimensionless pressure gradient is rescaled to bring the sum of the dimensionless superficial velocity of the conductive phase to its correct value of 1 (since the velocity is scaled by $U_{1S}$).

Table\ \ref{Tab: Grid convergence} and Table\ \ref{Tab: Grid convergence_cond} demonstrate that with an increase in the number of computational cells, the numerical solution for a rectangular duct of high aspect ratio (e.g., $AR=100$) converges to the analytical Two-Plate (TP) model solution presented in Sec.\ \ref{Sec: Analytical}. Except for high flow rate ratios and Hartmann numbers (i.e., $\displaystyle Q_{2 1} > 10, \Ha > 100$), a computational grid with $\displaystyle N_x=N_y=200$ is found to be sufficient to predict holdup and pressure gradient up to three decimal digits compared to the analytical solution (resulting in less than $0.1\%$ deviation in most cases). For high $Q_{2 1}$ and $\Ha$, a finer computational grid with $N_x=N_y=400$ is adopted.

It is important to mention that the analytical solution contains hyperbolic functions that should be evaluated with high floating-point precision (i.e., 128 or even 256 decimal points), so that the numerical solution can be advantageous for investigation of a wide range of the parameter values. In addition, it accounts for side-wall effects, which are neglected in the TP model, but can affect the two-phase MHD flow for high, but finite aspect ratios (see the discussion below).
{\renewcommand{\arraystretch}{1.5}
	\begin{table}[h!]
		\caption{\label{Tab: Grid convergence} Grid convergence. Mercury--air stratified flow in a rectangular duct ($AR=100$) with insulating walls.}
		\begin{tabular*}{\textwidth}{l l c c c c c c c c}
			\hline%
			\hline
			\multirow{3}{*}{$N_x$} & 
			\multirow{3}{*}{$N_y$} &
			\multicolumn{4}{c}{$Q_{2 1} = 0.1$} &
			\multicolumn{4}{c}{$Q_{2 1} = 10$} 
			\\
			\cmidrule{3-10}
			& & 
			\multicolumn{2}{c}{$\Ha=5.181$} &
			\multicolumn{2}{c}{$\Ha=103.62$} &
			\multicolumn{2}{c}{$\Ha=5.181$} &
			\multicolumn{2}{c}{$\Ha=103.62$} 
			\\
			\cmidrule{3-10}
			& & $h$ & $P^1_f$ & $h$ & $P^1_f$ & $h$ & $P^1_f$ & $h$ & $P^1_f$
			\\
			\hline
			200  & 200 & 0.9107 & 0.4673 & 0.9540 & 0.5297 
			& 0.5626	& 0.9833 & 0.7863 & 0.6917
			\\
			400  & 400 & 0.9108 & 0.4680 & 0.9541 & 0.5316 
			& 0.5629 & 0.9858 & 0.7869 & 0.6986
			\\
			& 600 & 0.9108 & 0.4682 & 0.9541 & 0.5322 
			& 0.5630 & 0.9865 & 0.7871 & 0.7009		
			\\
			& 800 & 0.9108 & 0.4683 & 0.9541 & 0.5325 & 
			0.5630 & 0.9868 & 0.7872 & 0.7020		
			\\
			& 1000 & 0.9108 & 0.4683 & 0.9540 & 0.5326 
			& 0.5630 & 0.9870 & 0.7873 & 0.7027						
			\\
			\hline
			\multicolumn{2}{@{}c@{}}{\textbf{Analytical}} &
			\textbf{0.9105}	& \textbf{0.4656} & \textbf{0.9539}	& \textbf{0.5328} & \textbf{0.5618} & \textbf{0.9860}	& \textbf{0.7869}	& \textbf{0.7054}
			\\
			\hline%
			\hline
		\end{tabular*}
	\end{table}
}
	
{
	\renewcommand{\arraystretch}{1.5}
	\begin{table}[h!]
		\caption{\label{Tab: Grid convergence_cond} Grid convergence for a rectangular duct with perfectly conducting walls. Same two-phase system as in Table\ \ref{Tab: Grid convergence}.}
		\begin{tabular*}{\textwidth}{l l c c c c c c c c}
			\hline%
			\hline
			\multirow{3}{*}{$N_x$} & 
			\multirow{3}{*}{$N_y$} &
			\multicolumn{4}{c}{$Q_{2 1} = 0.1$} &
			\multicolumn{4}{c}{$Q_{2 1} = 10$} 
			\\
			\cmidrule{3-10}
			& & 
			\multicolumn{2}{c}{$\Ha=5.181$} &
			\multicolumn{2}{c}{$\Ha=103.62$} &
			\multicolumn{2}{c}{$\Ha=5.181$} &
			\multicolumn{2}{c}{$\Ha=103.62$} 
			\\
			\cmidrule{3-10}
			& & $h$ & $P^1_f$ & $h$ & $P^1_f$ & $h$ & $P^1_f$ & $h$ & $P^1_f$
			\\
			\hline
			200  & 200 & 0.9385	& 0.8046 & 0.9893 &	0.9915	
			& 0.6794 & 0.9694 & 0.9493 & 1.0041
			\\
			400  & 400 & 0.9385	& 0.8060 & 0.9893 & 0.9937
			& 0.6797 & 0.9726 & 0.9494 & 1.0102
			\\
			& 600 & 0.9386 & 0.8064 & 0.9893 & 0.9944
			& 0.6798 & 0.9735 & 0.9494 & 1.0123	
			\\
			& 800 & 0.9386 & 0.8066 & 0.9893 & 0.9948
			& 0.6798 & 0.9740 & 0.9494 & 1.0133	
			\\
			& 1000 & 0.9386 & 0.8067 & 0.9893 & 0.9950
			& 0.6799 & 0.9743 & 0.9494 & 1.0139					
			\\
			\hline
			\multicolumn{2}{@{}c@{}}{\textbf{Analytical}} &
			\textbf{0.9386}	& \textbf{0.8099} & \textbf{0.9893}	& \textbf{1.0050} & \textbf{0.6799} & \textbf{0.9823}	& \textbf{0.9495}	& \textbf{1.0264}
			\\
			\hline%
			\hline
		\end{tabular*}
	\end{table}
}

\section{Results and discussion}\label{Sec: Results} 

In this work, we investigate the effect of a gas layer, the external magnetic field intensity and the wall electrical conductivity on the two-phase MHD flow characteristics in horizontal rectangular ducts of high width-to-height aspect ratios. To this aim, we consider air and mercury under standard conditions (i.e., at room temperature and atmospheric pressure) as a representative conductive fluid--gas two-phase system. The physical properties are summarized in Table\ \ref{Tab: Properties}. An additional dimensionless parameter that characterizes conductive fluids in MHD flows is the magnetic Prandtl number, defined as the ratio of momentum diffusivity (kinematic viscosity) to the magnetic diffusivity, or as the ratio of the magnetic to bulk Reynolds numbers ($\displaystyle\Rey=\rho_1 U_{1S} H / \eta_1$): 
\begin{equation}
	\Pran_m = \frac{\Rey_m}{\Rey}=\frac{\mu_0 \sigma_{e1}}{\rho_1/\eta_1} 
\end{equation}
This parameter is small for liquid metals (e.g., for mercury see Table\ \ref{Tab: Properties}). Therefore, $\Rey_m$ is small for laminar flows of moderate $\Rey$. Moreover, as shown above in Sec.\ \ref{Sec: Formulation}, the fully-developed steady flow is independent of $\Rey_m$. In fact, the TP model analytical solution (Sec.\ \ref{Sec: Analytical}) indicates that the dimensionless two-phase flow characteristics (e.g., holdup, velocity profile, pressure gradient and the scaled induced magnetic field) are determined only by the flow rate ratio, viscosity ratio, the Hartmann number, and the conductivity of the bottom wall. In this work, the range of the considered intensities of external magnetic field is $\displaystyle B_0 \in [0.01, 0.2]$Tesla, which represents standard laboratory conditions, whereby the corresponding Hartmann numbers are $\Ha \in [5.181, 103.62]$ in a channel of height $H=0.02$m.
{\renewcommand{\arraystretch}{2}
	\begin{table}[h!]
			\caption{\label{Tab: Properties} Properties of mercury (Hg) and air at standard conditions.}	
			\begin{tabular*}{\textwidth}{@{\extracolsep{\fill}}ccccccc@{\extracolsep{\fill}}}
				\hline%
				\hline				
				& $\rho$ & $\eta$ & $\sigma_e$ & $\Pran_m$ & $\eta_{1 2}=\dfrac{\eta_1}{\eta_2}$
				\\				
				& $\bigg[\dfrac{kg}{m^3}\bigg]$ & $\bigg[\dfrac{kg}{m \cdot s}\bigg]$ & $\bigg[\dfrac{1}{\Omega \cdot m}\bigg]$ & &
				\\
				\hline
				Mercury (1) & 13,534.0 & 0.00149 & 1.0$\times 10^{6}$ &
				1.383$\times 10^{-7}$  &
				\multirow{2}{*}{82.778}					
				\\
				Air (2) & 1.0 & 1.818$\times 10^{-5}$ & 0 & 0 
				\\
				\hline%
				\hline
			\end{tabular*}
	\end{table}
}

\subsection{The Two-Plate model}

\begin{figure}[h!]
	\centering
	\includegraphics[width=0.5\textwidth,clip]{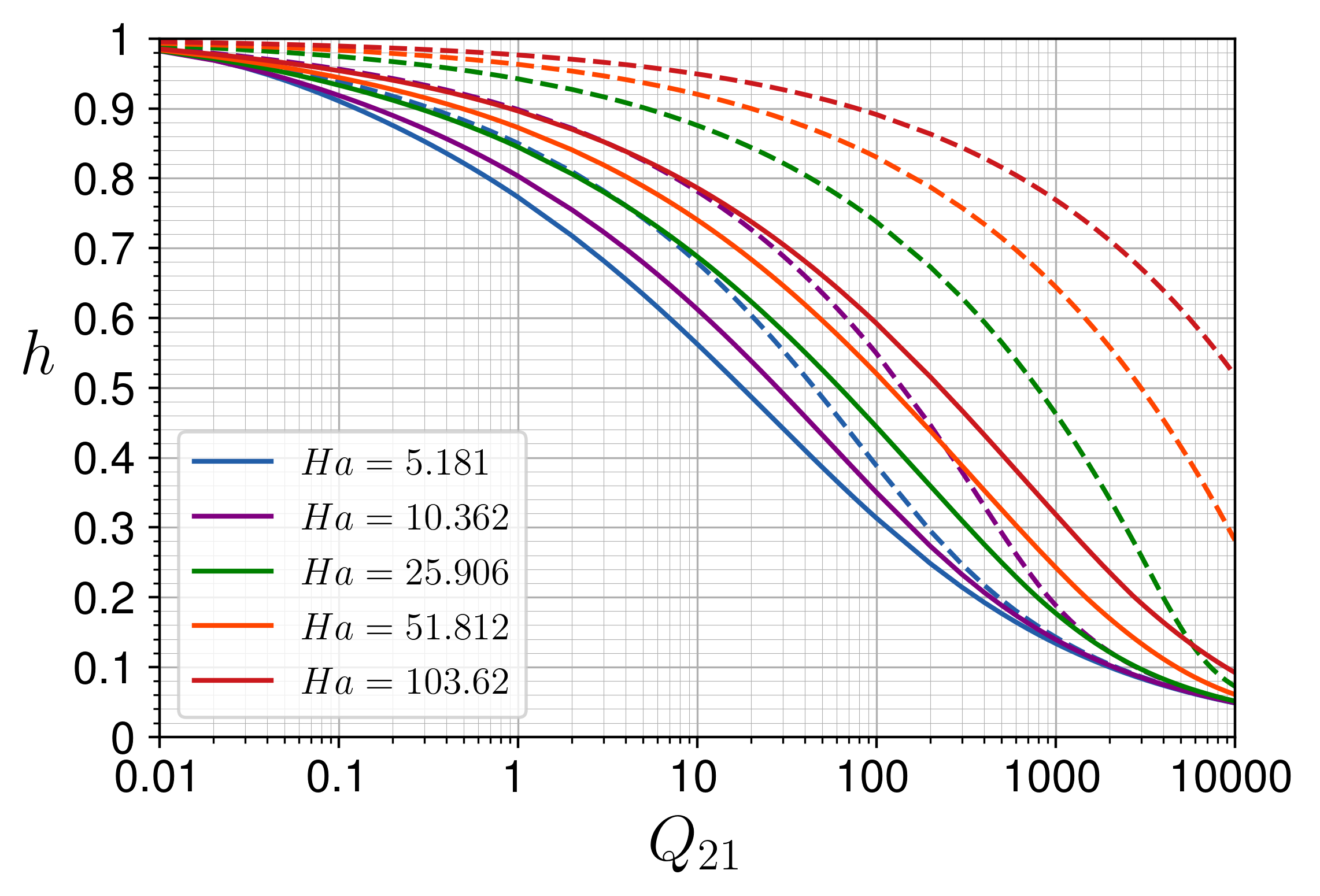}
	\caption{\label{Fig: holdup_TP}Variation of the holdup, $h$, with the air-to-mercury flow rate ratio, $Q_{21}$, predicted by the TP model for different Hartmann numbers and two cases of the bottom wall conductivity. Solid lines -- insulating bottom wall; dashed lines -- perfectly conducting bottom wall.}
\end{figure}

Figure\ \ref{Fig: holdup_TP} presents the holdup as a function of the flow rate ratio of air to mercury predicted by the Two-Plate model, which is approached for very wide rectangular ducts ($AR\to\infty$).The results are shown for various $\Ha$, comparing cases with an insulating and perfectly conducting bottom wall. Since the upper (gas) phase is non-conductive, only conductivity of the bottom wall, which is in contact with the conductive phase, affects the flow characteristics. Figure\ \ref{Fig: holdup_TP} shows that for a given wall conductivity and $\Ha$, the holdup is uniquely determined by the flow rate ratio. The Lorentz force slows down the conductive phase, regardless of whether the bottom wall is conducting or insulating. As a result, a larger flow area is required to maintain a specified flow rate of the conductive phase, leading to an increase in the holdup with $\Ha$. A stronger external magnetic field (higher $\Ha$) reduces the rate at which holdup decreases with increasing gas flow. However, when the bottom wall is insulating, the retarding Lorentz force is smaller, and therefore a smaller flow area is required to maintain the same flow rate of the conductive liquid compared to the case of conducting wall. This results in smaller holdups for the same $\Ha$ and $Q_{2 1}$ (compare solid and dashed lines of the same color in Fig.\ \ref{Fig: holdup_TP}). The effect of wall conductivity on the holdup is noticeable even at low $\Ha$ values and becomes more pronounced as the magnetic field strength increases. In the case of conducting wall and $\Ha>25$, the conductive layer occupies almost the entire flow cross section already when the flow rates of two phases are equal, i.e., $Q_{2 1}=1$.	
\begin{figure}[h!]
	\centering
	\subfloat[Insulating bottom wall; entire cross section]{\includegraphics[width=0.48\textwidth,clip]{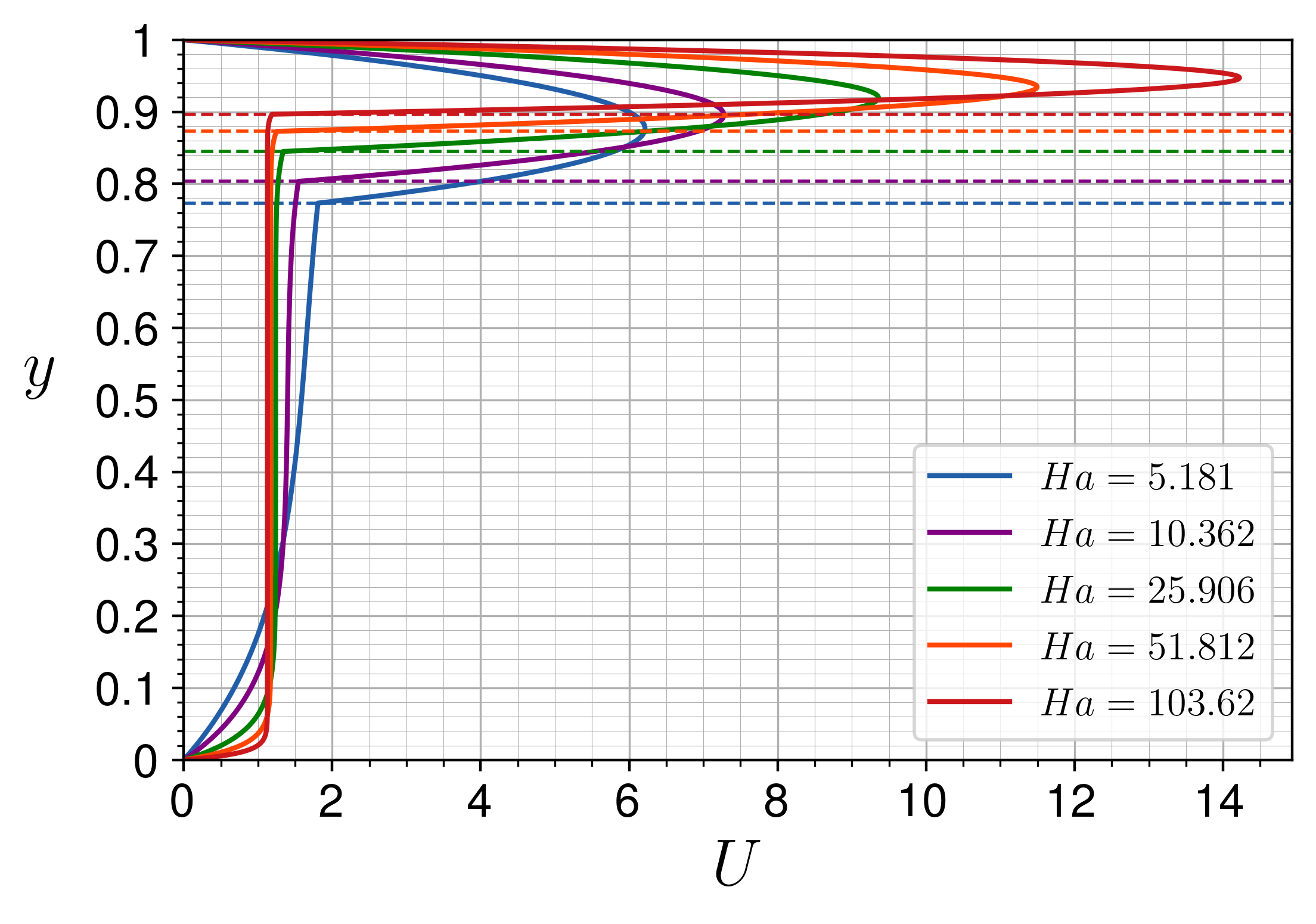}}
	\subfloat[Conductive layer]{\includegraphics[width=0.48\textwidth,clip]{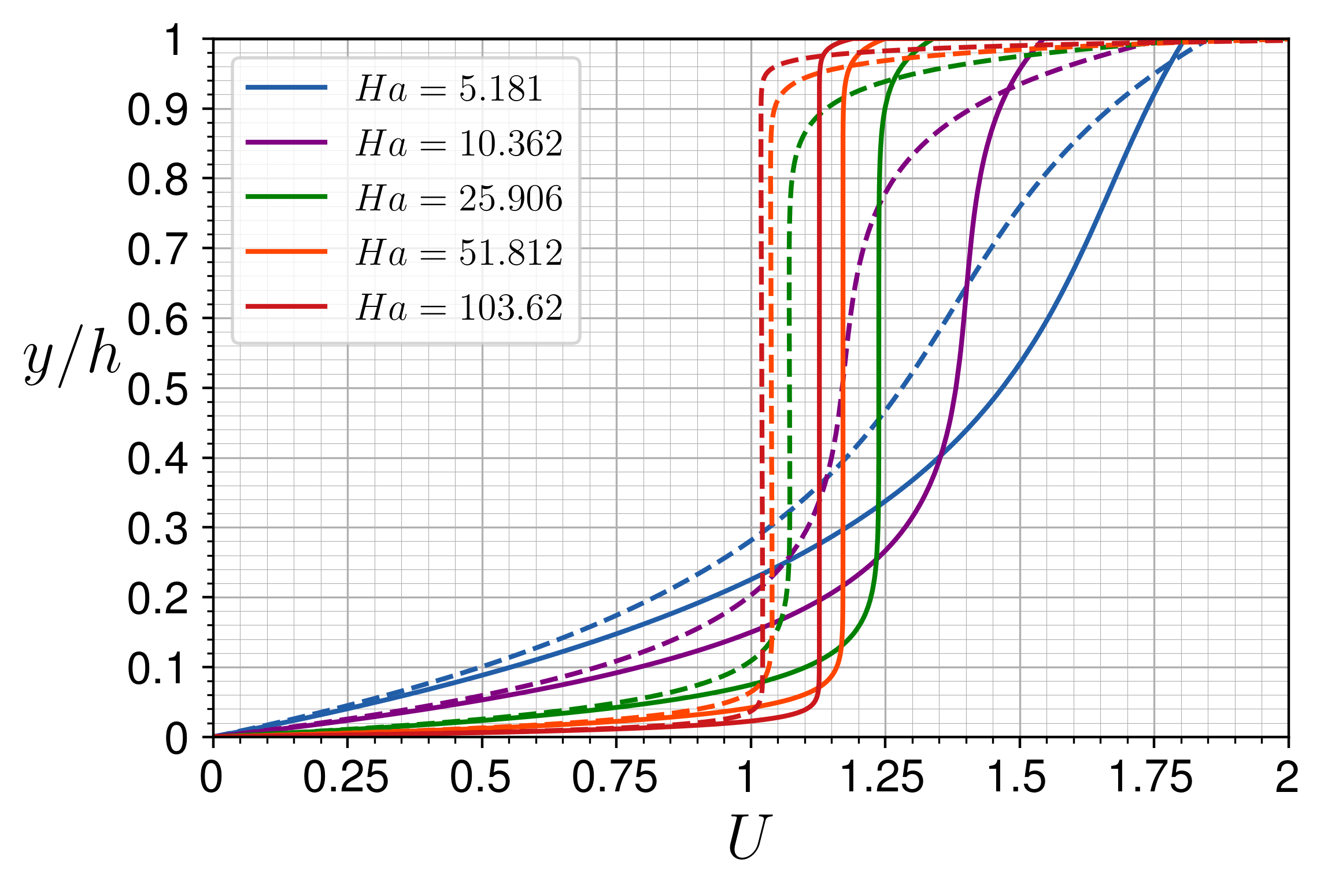}}
	\caption{\label{Fig: U_TP}Dimensionless velocity profiles for different $\Ha$ (air--mercury MHD flow, $Q_{2 1}=1$): (a) across the entire cross section for the case of insulating bottom wall (solid lines); (b) across the lower conductive phase ($y$ scaled by lower phase holdup) for two cases of the bottom wall conductivity.  Solid lines -- insulating bottom wall; dashed lines -- perfectly conducting bottom wall. Horizontal dashed lines denote the location of the interface in (a). }	
\end{figure}

\begin{figure}[h!]
	\centering
	\subfloat[Insulating]{\includegraphics[width=0.4\textwidth,clip]{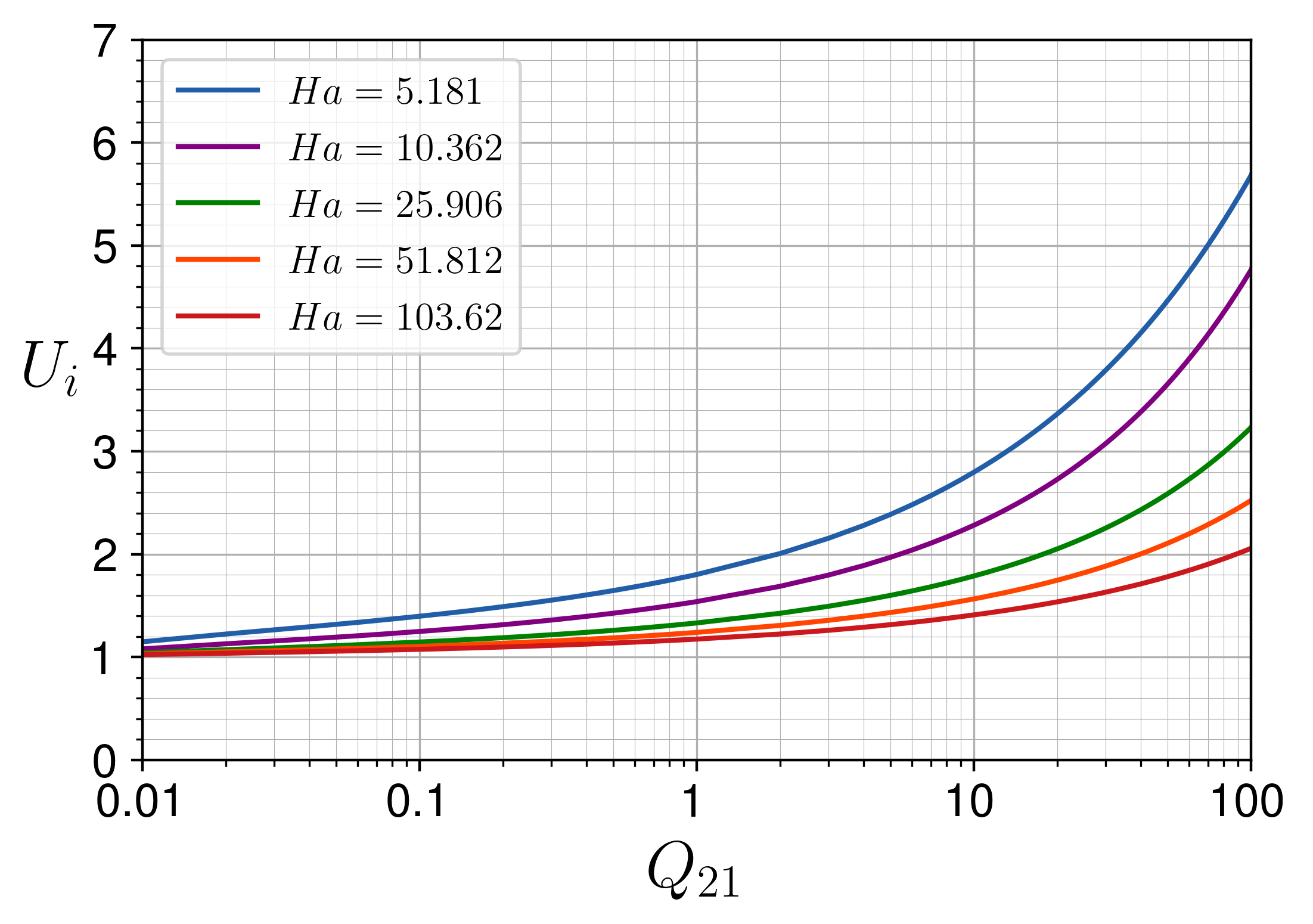}}
	\subfloat[Conducting]{\includegraphics[width=0.4\textwidth,clip]{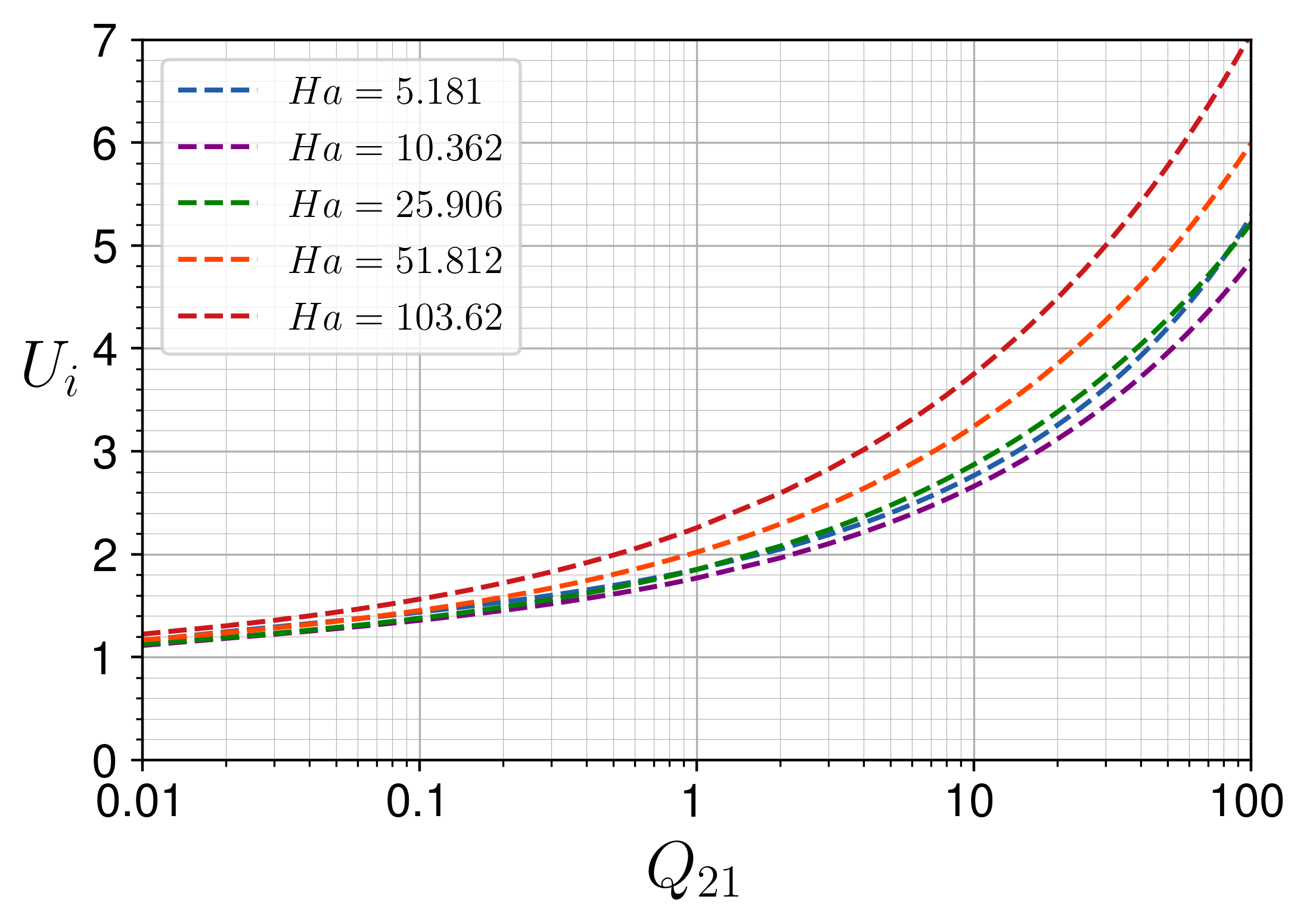}}
	\caption{\label{Fig: U_i}Variation of the interfacial velocity with $Q_{21}$ for different $\Ha$ for (a) insulating and (b) conducting bottom wall.	}
\end{figure}

The effect of wall conductivity is further elaborated by examining the velocity profiles. Figure\ \ref{Fig: U_TP} presents the dimensionless velocity profiles (normalized by the conductive liquid superficial velocity, $U_{1S}$) for different $\Ha$ numbers, while keeping the same flow rate ratio, $Q_{2 1}=1$. Unlike in single-phase flow of a conductive fluid, where the shape of the dimensionless velocity profile (scaled by the average, or maximal, velocity) is independent of the walls conductivity (see Appendix\ \ref{Sec: Single_phase_TP}), in the two-phase flow, the velocity profiles are of different shapes when the bottom wall is insulating (solid lines in Fig.\ \ref{Fig: U_TP}) or conductive (dashed lines). Specifically, the average velocity of the conductive fluid and its core velocity are higher when the bottom wall is insulating (compare solid and dashed lines in Fig.\ \ref{Fig: U_TP}b). This is because the same flow rate is transported through a smaller cross-sectional area (i.e., smaller holdup, see Fig.\ \ref{Fig: holdup_TP}). The reduced holdup also results in a significantly lower air velocity in the upper layer, and thus a lower interfacial velocity for $\Ha>10$ in the case of insulating wall (compare solid and dashed lines in Figs.\ \ref{Fig: U_i}a and \ref{Fig: U_i}b, respectively). In both cases, the air, which is non-conductive and less viscous, moves much faster than the mercury and its velocity profile is practically parabolic. At higher $\Ha$, the increased holdup causes the mercury average velocity and the interfacial velocity to decrease (Fig.\ \ref{Fig: U_TP}), whereas the air velocity increases. Hence, the velocity gradients at the top wall become larger and obviously increase upon further increasing the air flow rate (i.e., at higher $Q_{21}$, see Fig.\ \ref{Fig: Top_shear_TP} below). Figure\ \ref{Fig: U_TP} also shows that with an increase of $\Ha$, the velocity gradient in mercury near the bottom wall and at the interface becomes steeper at a fixed value of $Q_{21}$. This results in a thinner Hartmann boundary layer on the bottom wall. The mercury velocity quickly grows from the bottom wall and attains an almost flat profile, similarly to the single-phase Hartmann flow. However, contrarily to the latter, in the upper part of the conductive layer the velocity does not decay, but steeply grows towards the air velocity at the interface. Hence, the mercury is subject to high wall and interfacial shear stresses. 

\begin{figure}[h!]
	\centering
	\subfloat[Insulating]{\includegraphics[width=0.3\textwidth,clip]{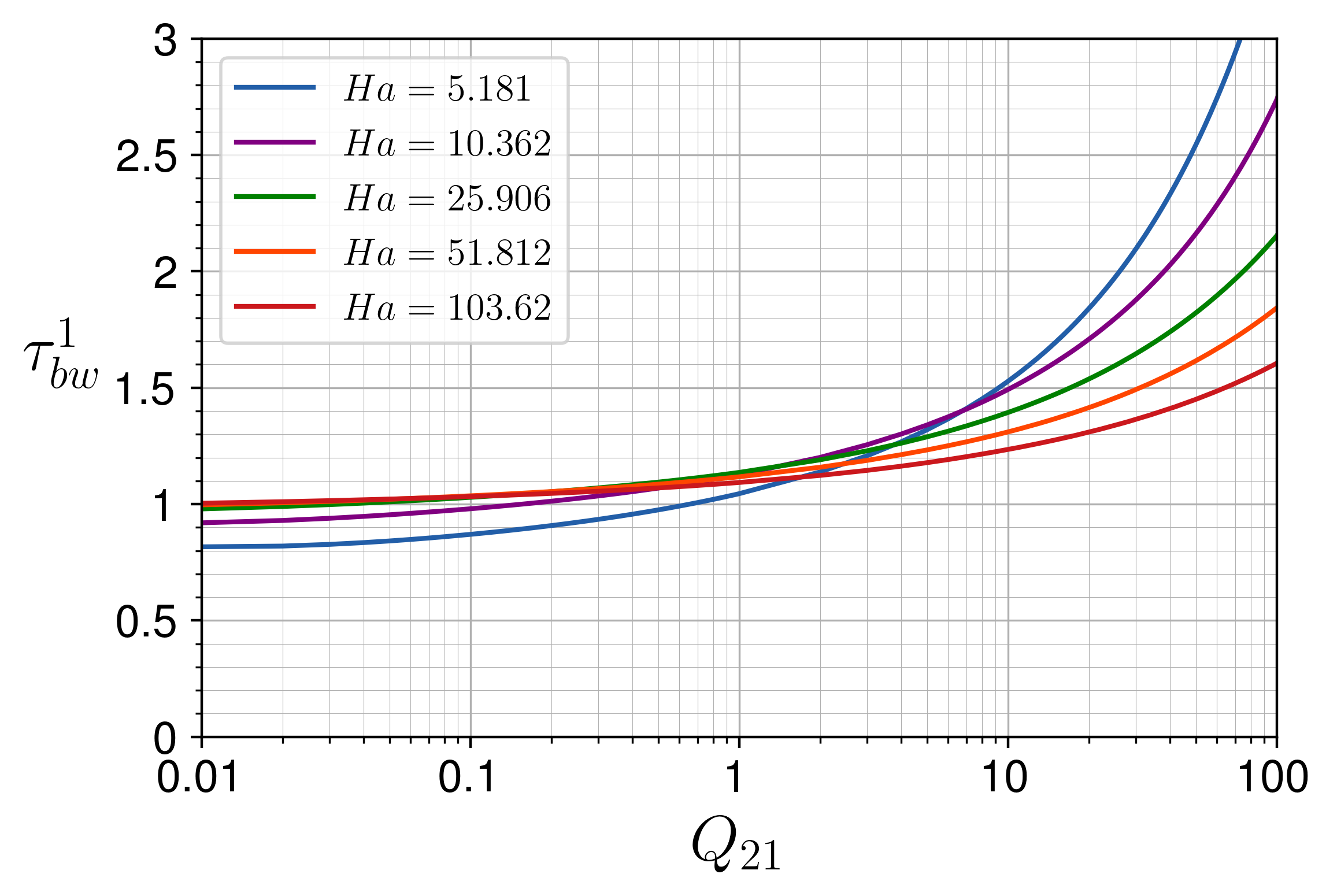}}
	\subfloat[Conducting]{\includegraphics[width=0.3\textwidth,clip]{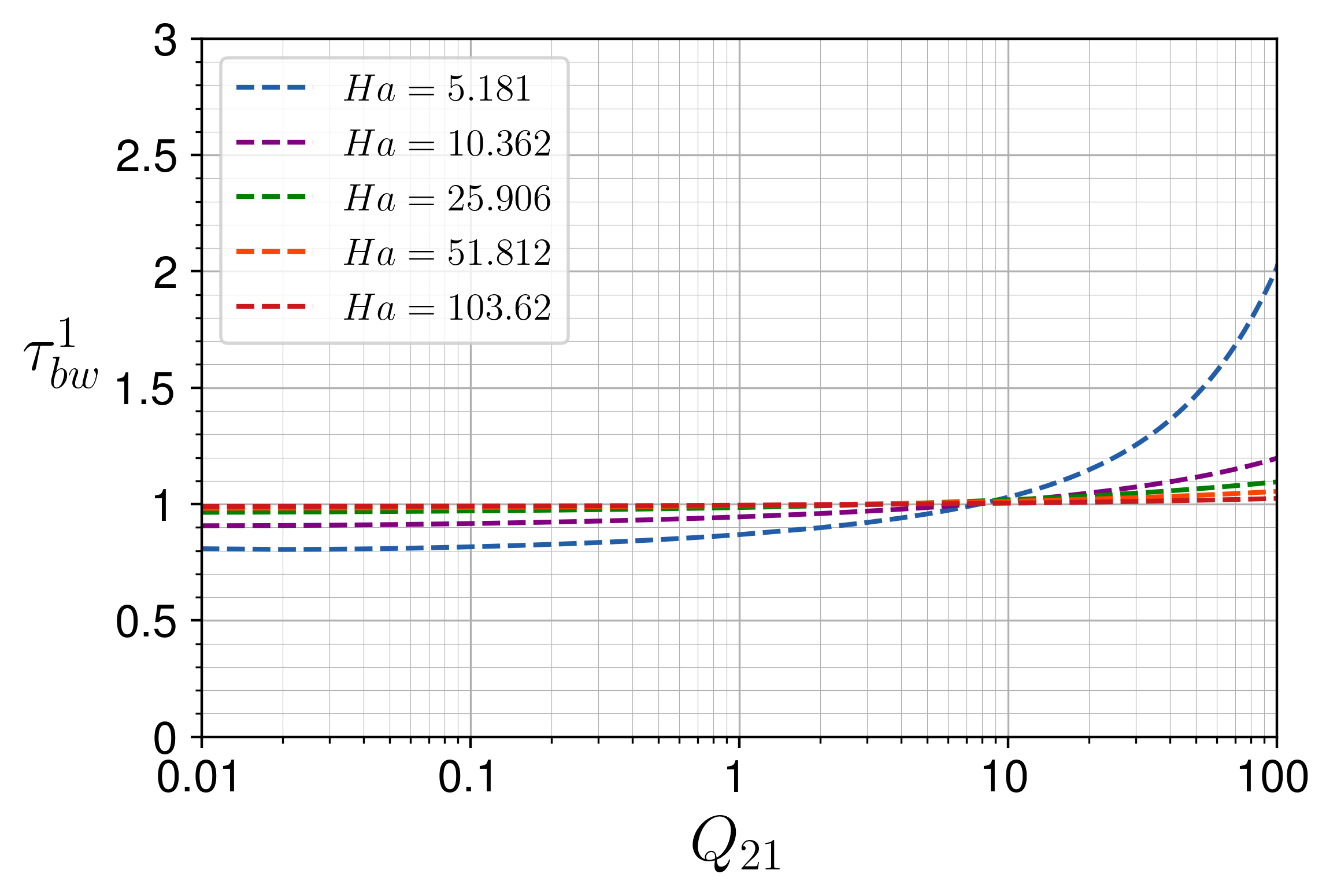}}
	\subfloat[$\tau_{bw}^1 h$ vs. $Q_{21}$]{\includegraphics[width=0.3\textwidth,clip]{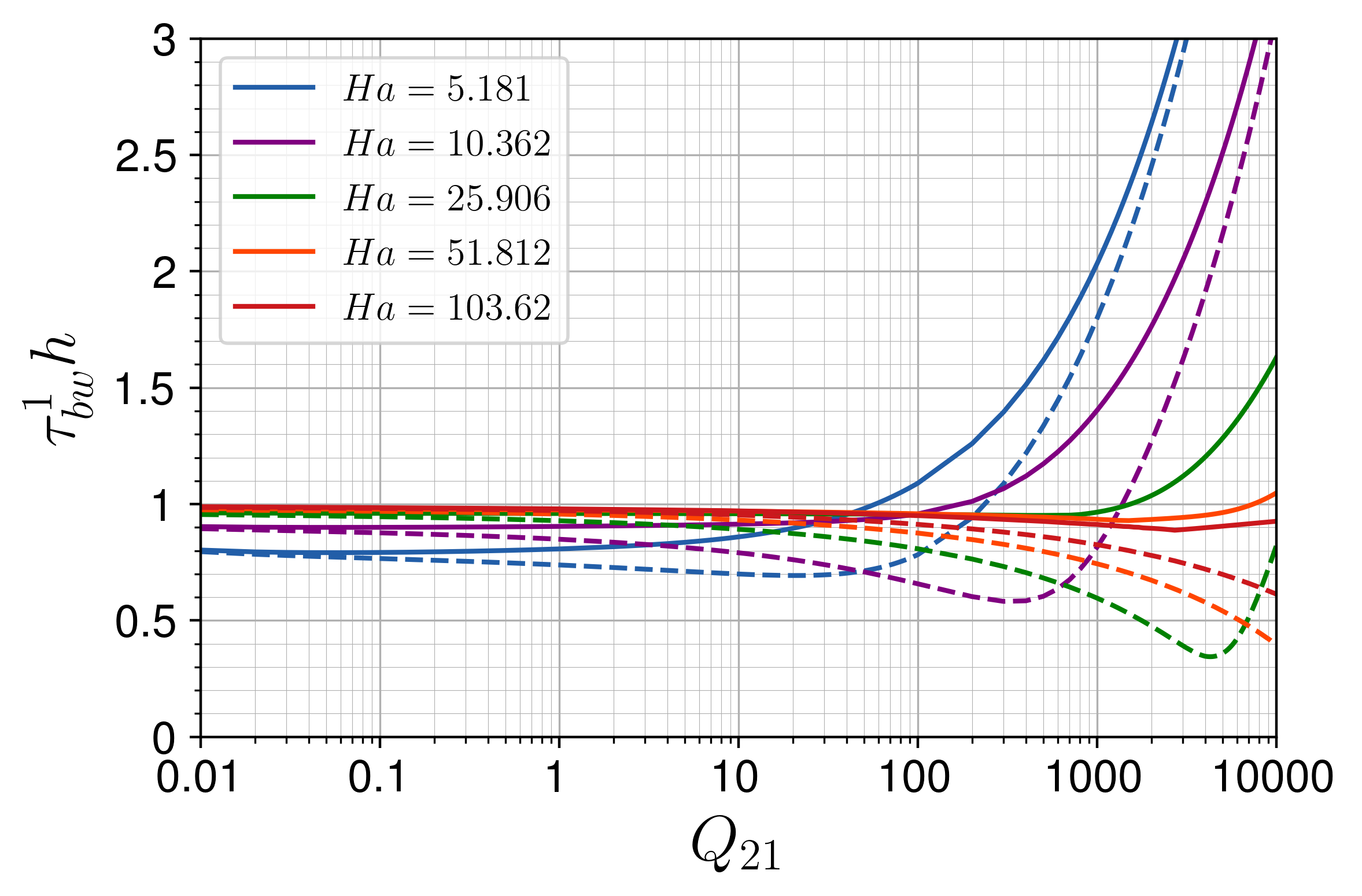}}
	\caption{\label{Fig: Bottom_shear_TP}Variation of the bottom wall shear stress factor, $\tau_{bw}^1$,  with $Q_{21}$ for different $\Ha$ for (a) insulating and (b) conducting bottom wall. (c) Modified factor that takes into account the holdup. Solid lines -- insulating bottom wall; dashed lines -- perfectly conducting bottom wall.}
\end{figure}

As pointed above, in two-phase flows, the shape of the dimensionless velocity profile and, consequently, the wall shear stress are affected by the walls conductivity. Figures\ \ref{Fig: Bottom_shear_TP} and \ref{Fig: Top_shear_TP} show the dimensionless bottom- and top-wall shear stress factors, respectively (see Eqs.\ \ref{Eq: Shear_stress_bottom} and \ref{Eq: Shear_stress_top}). They are defined as the dimensionless wall shear stress normalized by its respective value in single phase flow of the conductive phase. The latter is independent of the wall conductivity (see Eq.\ \ref{Eq: Shear_stress_factor_SP} in Appendix\ \ref{Sec: Single_phase_TP})). As shown in Fig.\ \ref{Fig: Bottom_shear_TP}, the dimensionless bottom-wall shear stress is higher for insulating wall. However, upon rescaling the distance from the bottom wall by the holdup, the bottom-wall shear stress for insulating and conducting walls become the same at high Hartmann numbers (see Fig.\ \ref{Fig: Bottom_shear_TP}c) and both are of the order of 1 over a wide range of $Q_{21}$. Thereby, the bottom-wall shear stress rescaled with the holdup, $\tau_{bw}h$, is independent of the wall conductivity and, similarly to single phase flow, increases proportionally to $\Ha$.  
\begin{figure}[h!]
	\centering
	\subfloat[Insulating]{\includegraphics[width=0.3\textwidth,clip]{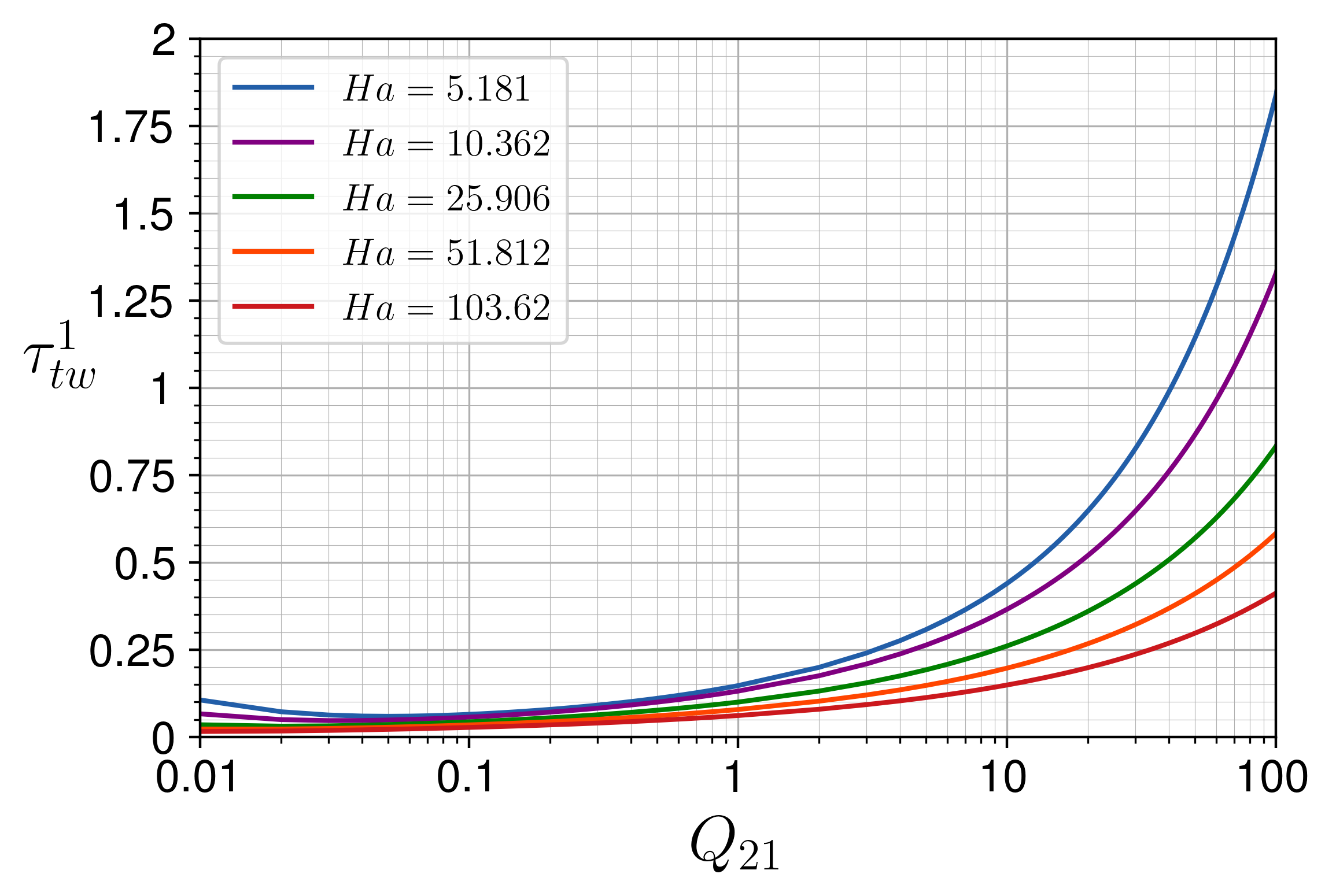}}
	\subfloat[Conducting]{\includegraphics[width=0.3\textwidth,clip]{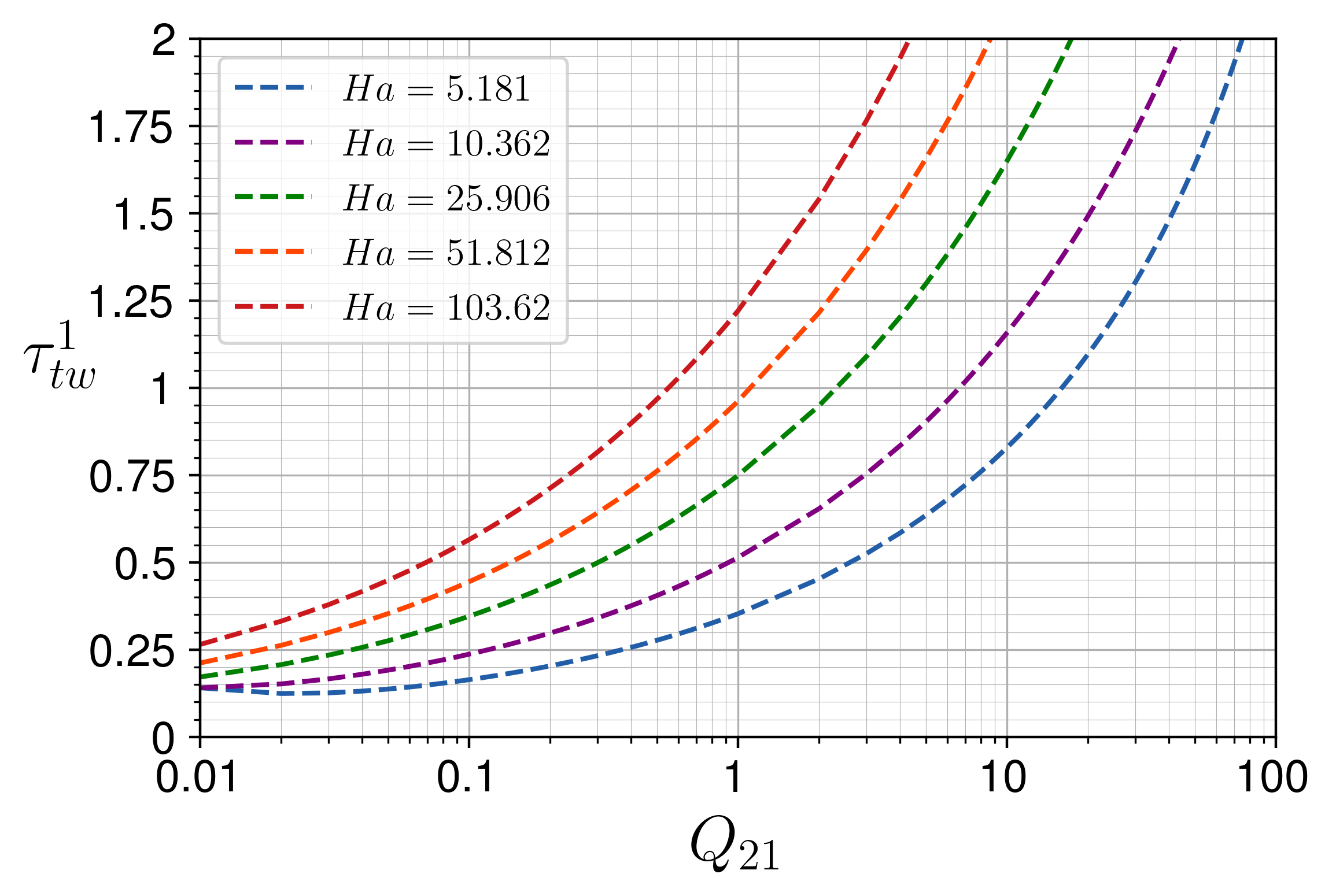}}
	\subfloat[$\tau_{tw}^1 (1 - h)$ vs. $Q_{21}$]{\includegraphics[width=0.3\textwidth,clip]{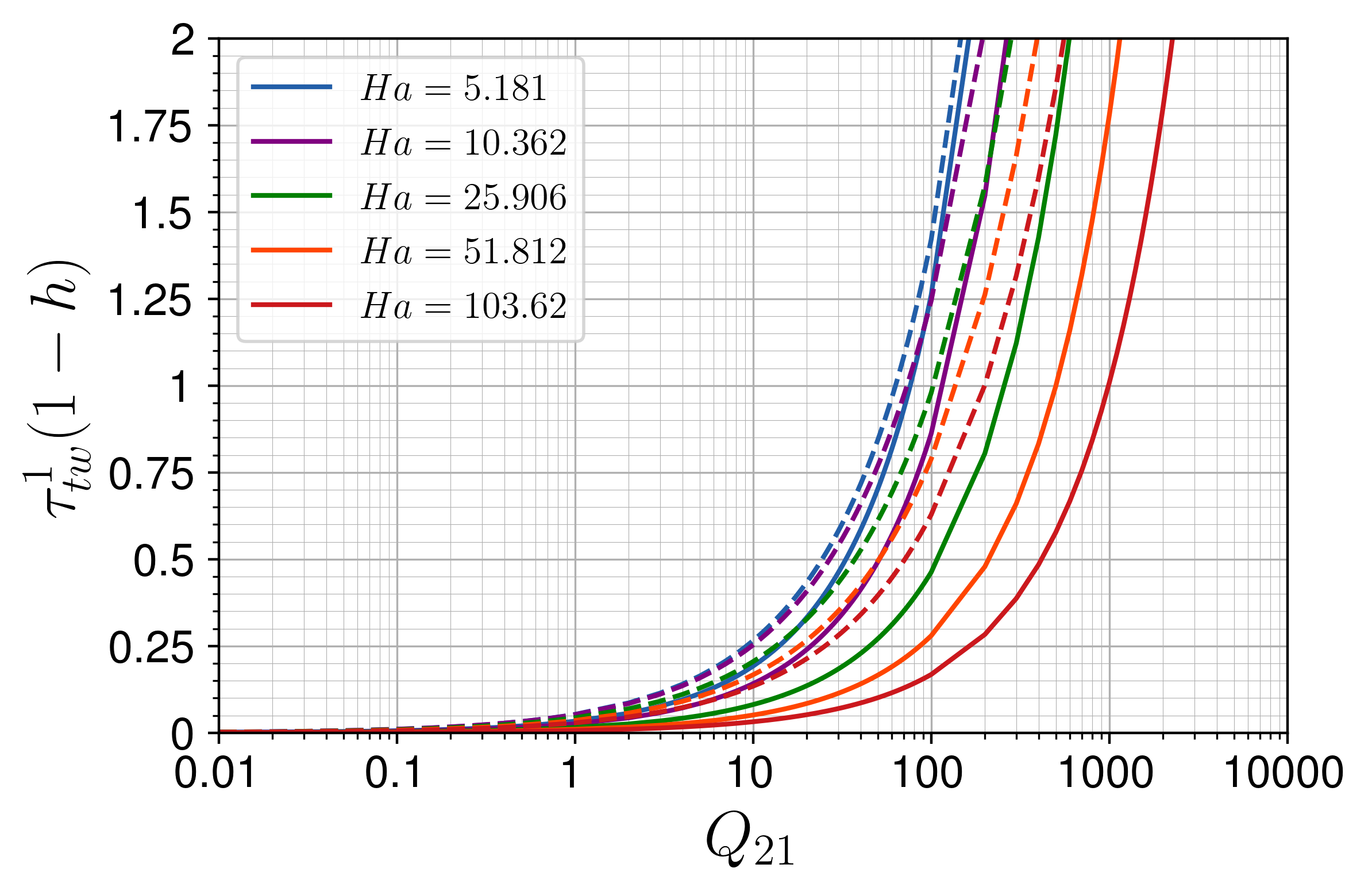}}
	\caption{\label{Fig: Top_shear_TP}Variation of the top wall shear stress factor, $\tau_{tw}^1$, corresponding to Fig.\ \ref{Fig: Bottom_shear_TP}.}
\end{figure}

The top-wall (and interfacial) shear stresses are observed to be lower in the case of insulating bottom wall (Fig.\ \ref{Fig: Top_shear_TP}), due to the larger gas flow cross section obtained in this case. However, for low air flow rates (i.e., $Q_{21} < 1$), the top wall and interfacial shear stresses are much smaller compared to that at the bottom wall (compare Figs.\ \ref{Fig: Bottom_shear_TP} and \ref{Fig: Top_shear_TP}), suggesting that in this region the air layer can lubricate the mercury flow. However, for larger $Q_{21} (> 1)$, a steep growth of the top wall shear stress due to the air flow is observed, so that this shear stress becomes much higher than the bottom-wall shear stress. In the case of perfectly conducting bottom wall the steep growth occurs at lower $Q_{21}$, and the top-wall shear stress is much higher than with the insulating bottom wall. It can be seen in Figs.\ \ref{Fig: Bottom_shear_TP} and \ref{Fig: Top_shear_TP} that even though the contribution of the top- and bottom-wall shear stresses to the pressure gradient in the region of $Q_{21} < 1$ is similar in both cases, when the bottom wall is perfectly conducting there is an additional retarding force due to the non-zero induced magnetic field at the wall, $b(y=0)\ne0$, which the pressure gradient has to balance (see the overall force balance, Eq.\ \ref{Eq: Integral_balance_cond}). This results in increased pressure gradient when compared to the case of an insulating bottom wall (see Fig.\ \ref{Fig: Pressure_TP} below).         
\begin{figure}[h!]
	\centering
	\subfloat[$Q_{2 1}=1$: Insulating]{\includegraphics[width=0.25\textwidth,clip]{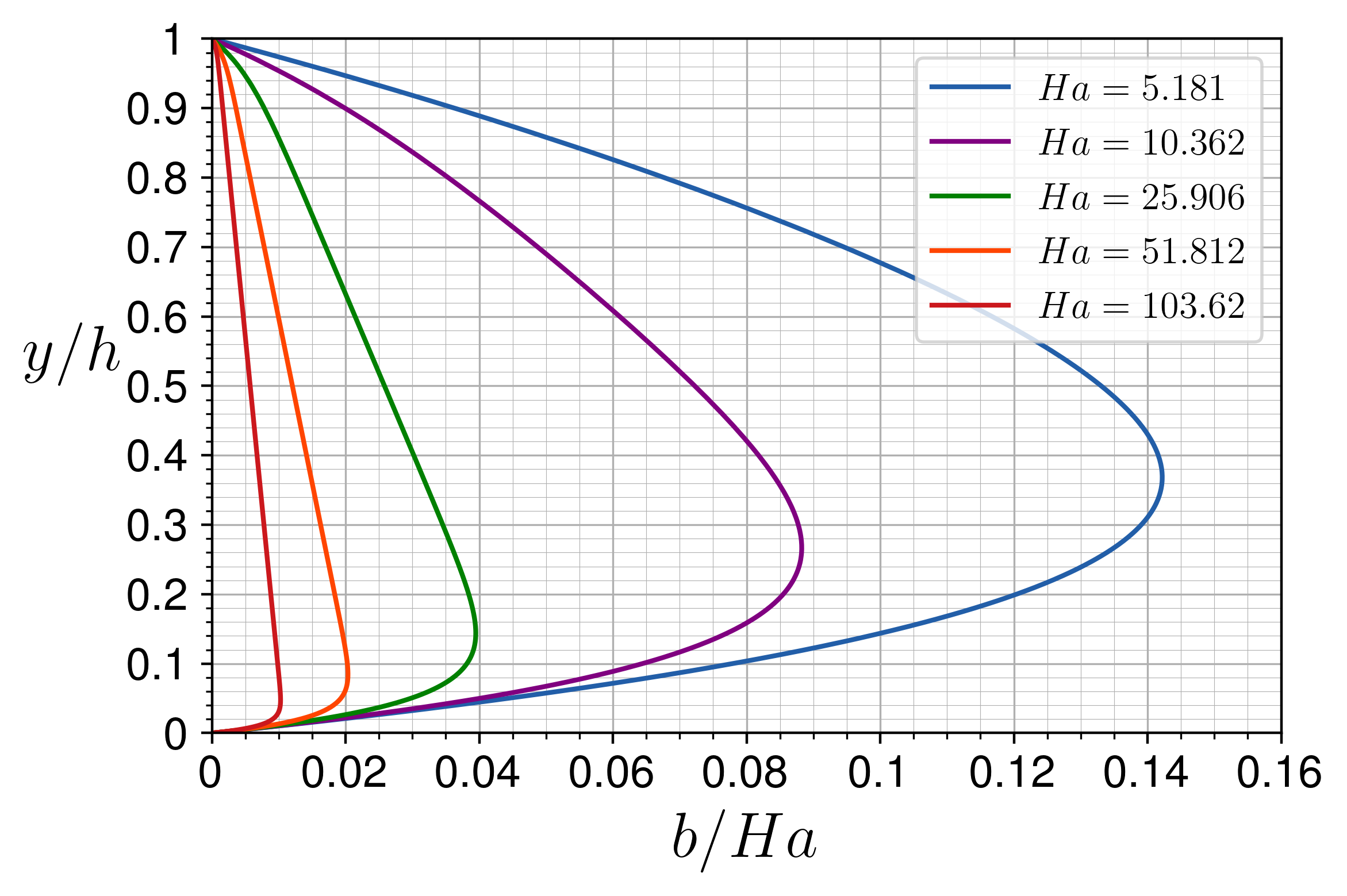}}
	\subfloat[Conducting]{\includegraphics[width=0.25\textwidth,clip]{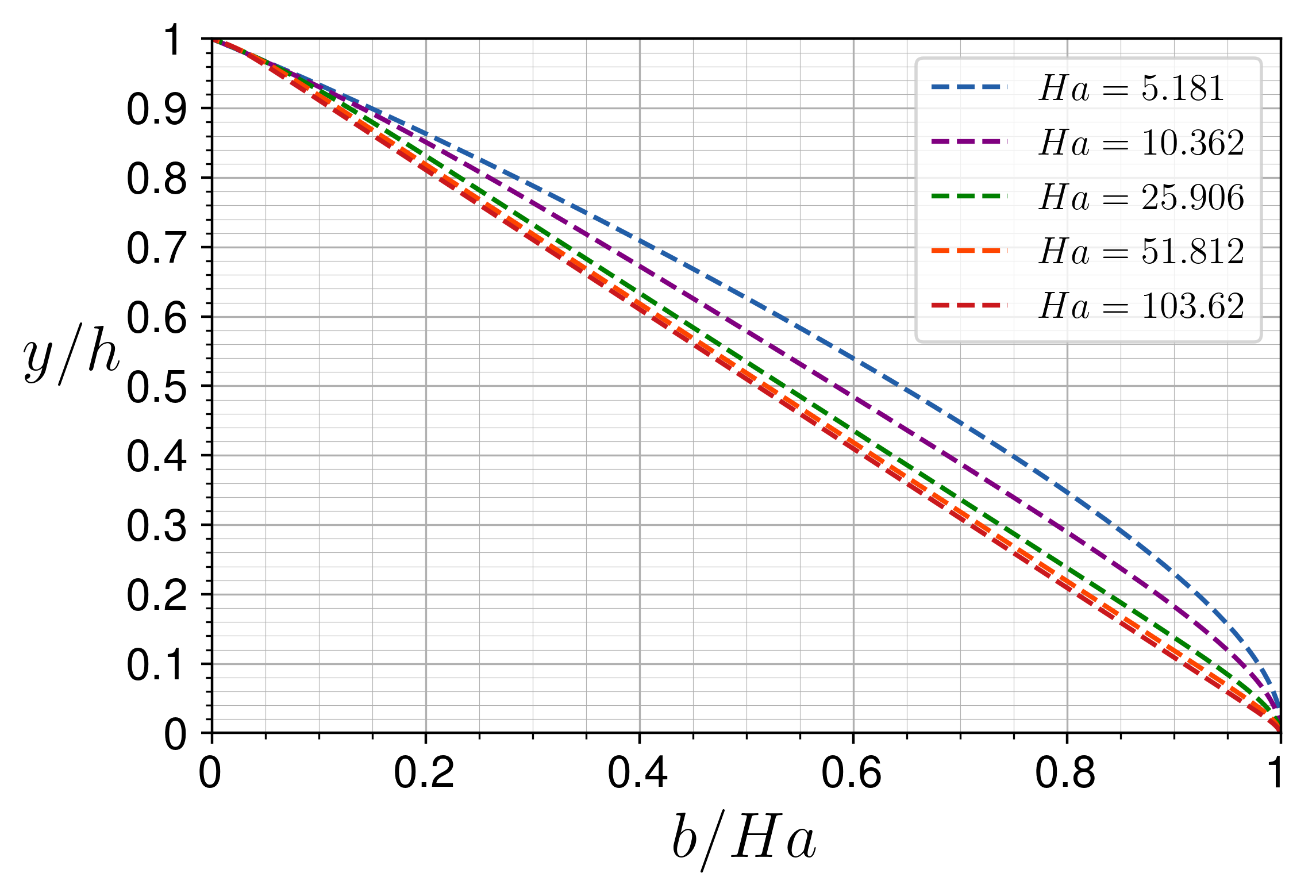}}
	\subfloat[$b_{max}/Ha$ vs. $Q_{2 1}$]{\includegraphics[width=0.25\textwidth,clip]{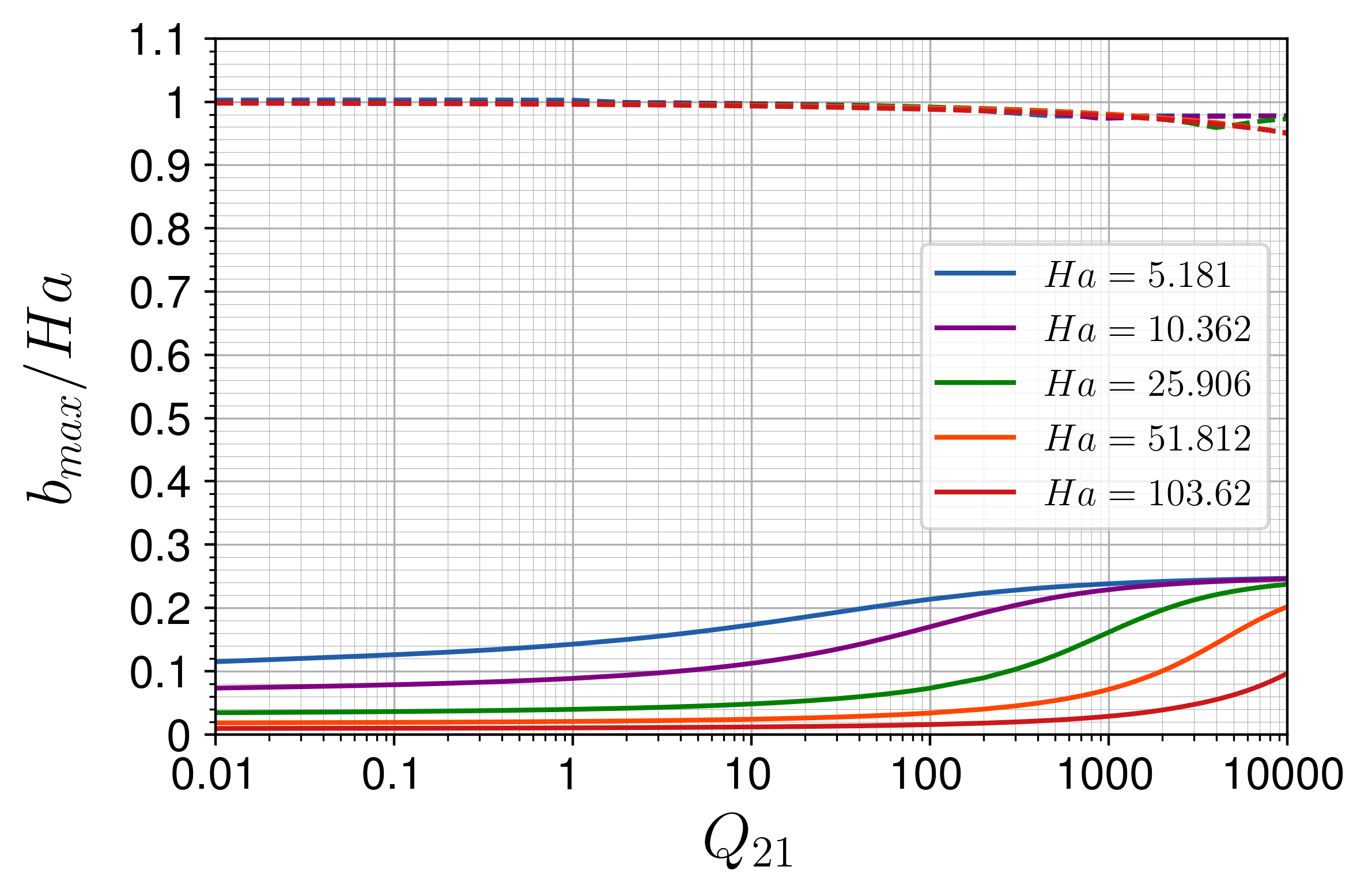}}
	\subfloat[$b_{max}Ha/G$ vs. $Q_{2 1}$]{\includegraphics[width=0.25\textwidth,clip]{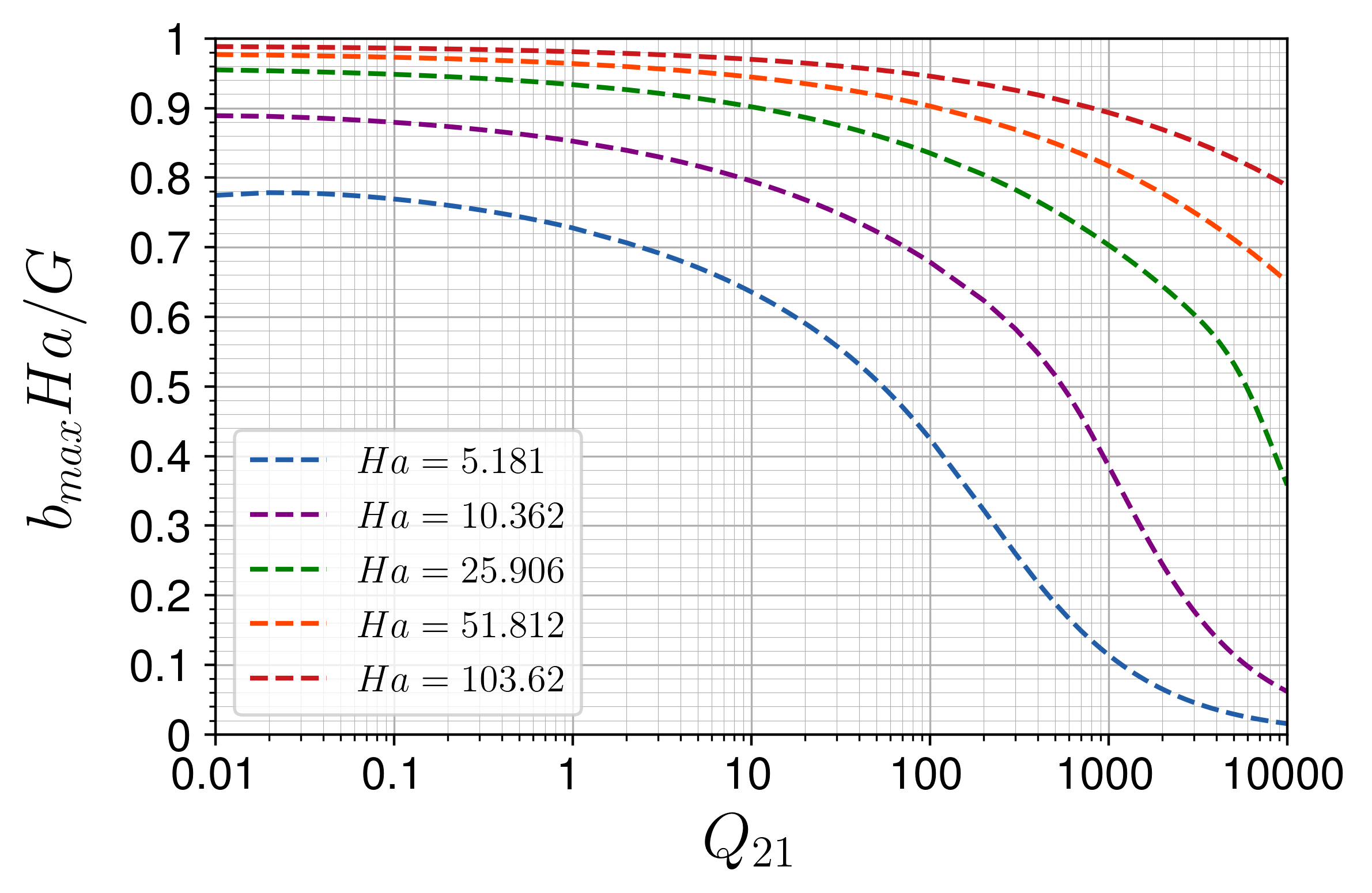}}
	\caption{\label{Fig: B_induced_TP}Induced magnetic field, $b/\Ha$, for different $\Ha$ for (a) insulating and (b) conducting bottom wall. (c) Variation of the maximal value, $b_{max}/\Ha$, with $Q_{21}$. (d) Relative input for the force balance for the case of conducting bottom wall. }
\end{figure}

The variation of the induced magnetic field profiles with $\Ha$ obtained for $Q_{2 1} = 1$ is presented in Fig.\ \ref{Fig: B_induced_TP}. It illustrates the $b/\Ha$ profiles (i.e., dimensionless induced magnetic field, $B_{ind}$, scaled by $\Rey_m$) for both insulating and conducting bottom walls. It is clearly observed that the intensity of the induced magnetic field is significantly lower when the bottom wall is insulating. In that case (Fig.\ \ref{Fig: B_induced_TP}a), the induced magnetic field decreases with increasing $\Ha$. By analyzing the values of $B_{ind}$ (not shown), they are found to be almost proportional to $\Ha$ and become very small at large $\Ha$ (i.e., dimensional induced magnetic intensity, $\hat{B}_{ind}$, becomes negligible compared to $B_0$). However, even for small $\Ha$, $B_{ind}$ is of the order of magnitude smaller than $\Rey_m$. The direction of the Lorentz force is defined by a sign of the gradient of induced magnetic field, $\displaystyle \partial b/\partial y$, and varies across the conductive phase. In a lower part of the layer  $\displaystyle \partial b/\partial y > 0$, since the induced magnetic field grows from $b=0$ at the insulating wall to its maximum, so that locally the Lorentz force acts in the flow direction. This region occupies less than half of the layer height even for the smallest $\Ha$ considered and shrinks with an increase of $\Ha$ until it nearly vanishes near the bottom wall for high $\Ha$. Thereby, in a larger part of the conductive layer (above the location of $b_{max}$), $\displaystyle \partial b/\partial y < 0$ and the Lorentz force acts opposite to the flow direction. It is important to note that the Lorentz force has zero integral contribution into the pressure gradient across the conductive layer (Eq.\ \ref{Eq: Integral_balance_ins}), when the bottom wall is insulating.

When the bottom wall is perfectly conducting (Fig.\ \ref{Fig: B_induced_TP}b), the intensity of the induced magnetic field, $b/\Ha$, is maximal at the bottom wall (i.e., $b_{max}=b(y=0)$), where it is of the order of $1$. It decreases throughout the conductive layer toward the insulating interface, where it attains the zero value. The $b/\Ha$ profile has an almost constant negative slope (except at the bottom wall where $\displaystyle \partial b/\partial y = 0$). $b/\Ha$, as well as its gradient, decreases only slightly as $\Ha$ increases. Hence, the Lorentz force acts opposite to the flow direction over the entire mercury layer. The variation of the maximal value of the induced magnetic field, $b_{max}/\Ha$, with $Q_{21}$ is shown in Fig.\ \ref{Fig: B_induced_TP}c for different $\Ha$. It can be seen that for the case of perfectly conductive bottom wall, $b_{max}/\Ha\approx 1$ over a wide range of $Q_{21}$ and $\Ha$, whereas for the insulating wall, $b_{max}/\Ha$ increases with $Q_{21}$, but stays much lower than $1$. The values of $b(y=0) \Ha/G$ (Fig.\ \ref{Fig: B_induced_TP}d) represent the fraction of the pressure gradient resulting from the Lorentz force in the case of conducting bottom wall (see Eq.\ \ref{Eq: Integral_balance_cond}). For high $\Ha$, this fraction approaches $1$ across a broad range of $Q_{21}$ (up to approximately $10$), indicating that the retarding Lorentz force is the dominant contributor to the pressure gradient. Naturally, its significance diminishes with increasing $Q_{21}$. With an insulating bottom wall, the pressure gradient arises solely from wall shear stresses (see Eq.\ \ref{Eq: Integral_balance_ins}).
\begin{figure}[h!]
	\centering
	\subfloat[$P_f^{1,0}$]{\includegraphics[width=0.3\textwidth,clip]{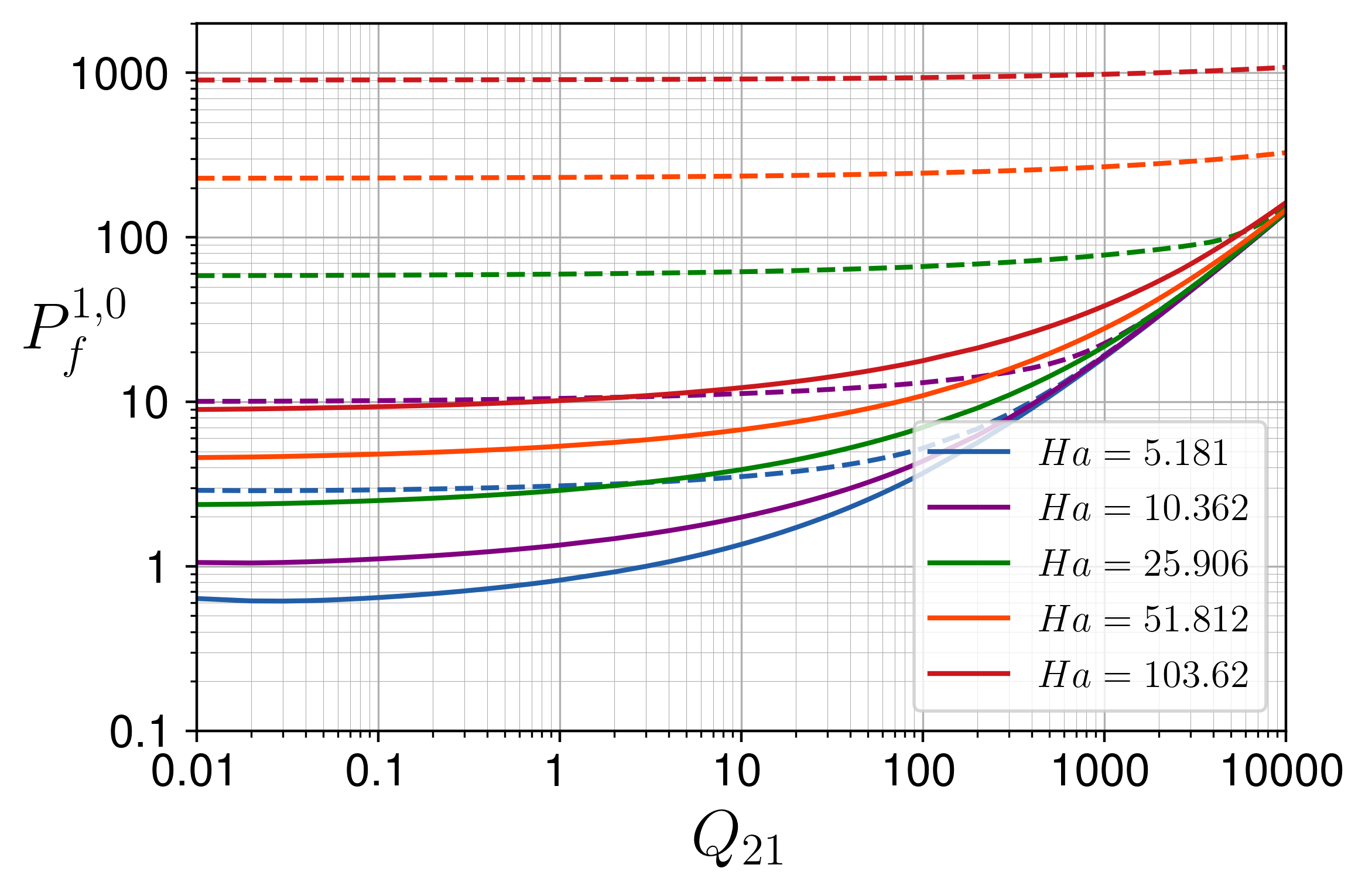}}
	\subfloat[$P_f^1$]{\includegraphics[width=0.3\textwidth,clip]{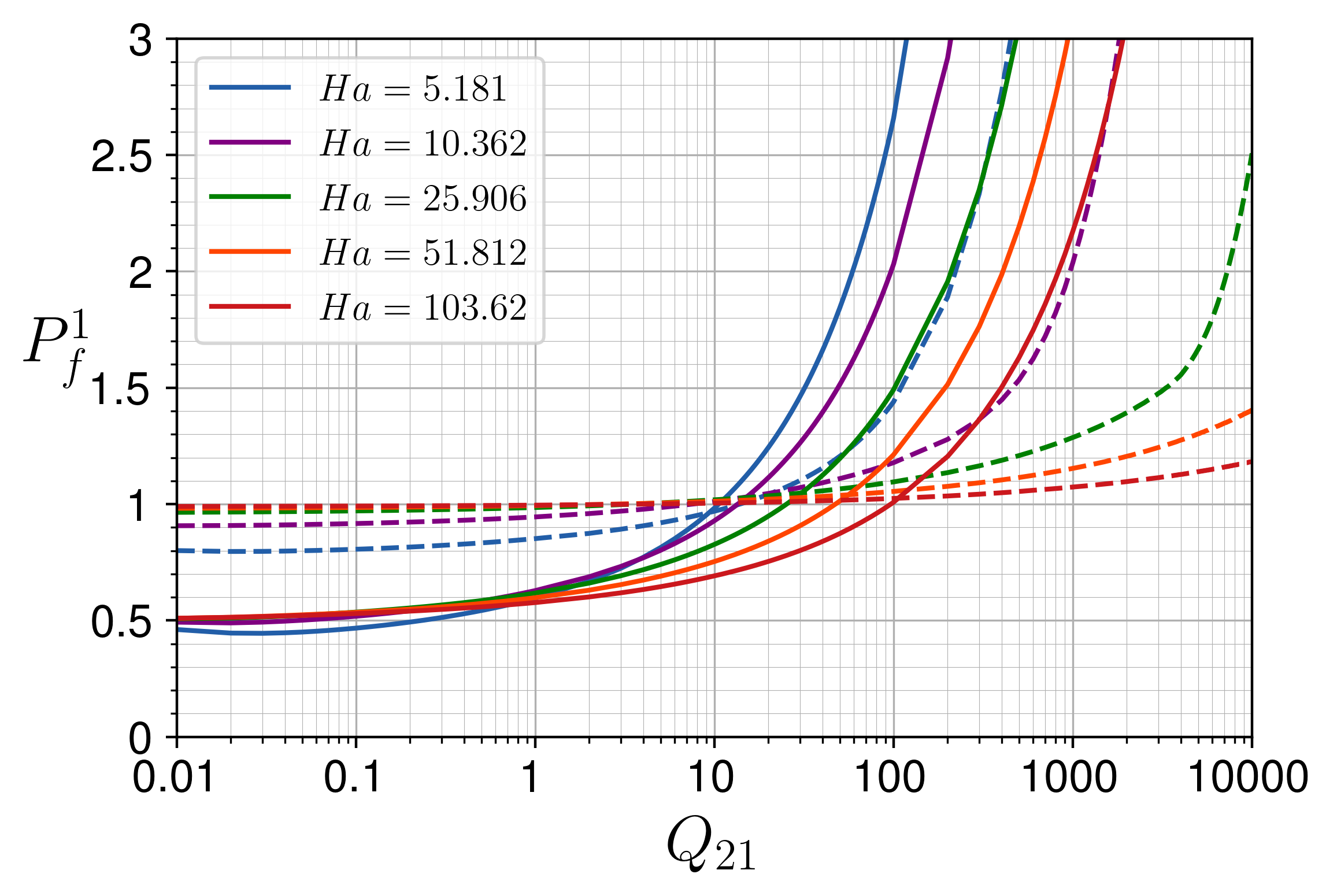}}
	\subfloat[Power factor]{\includegraphics[width=0.3\textwidth,clip]{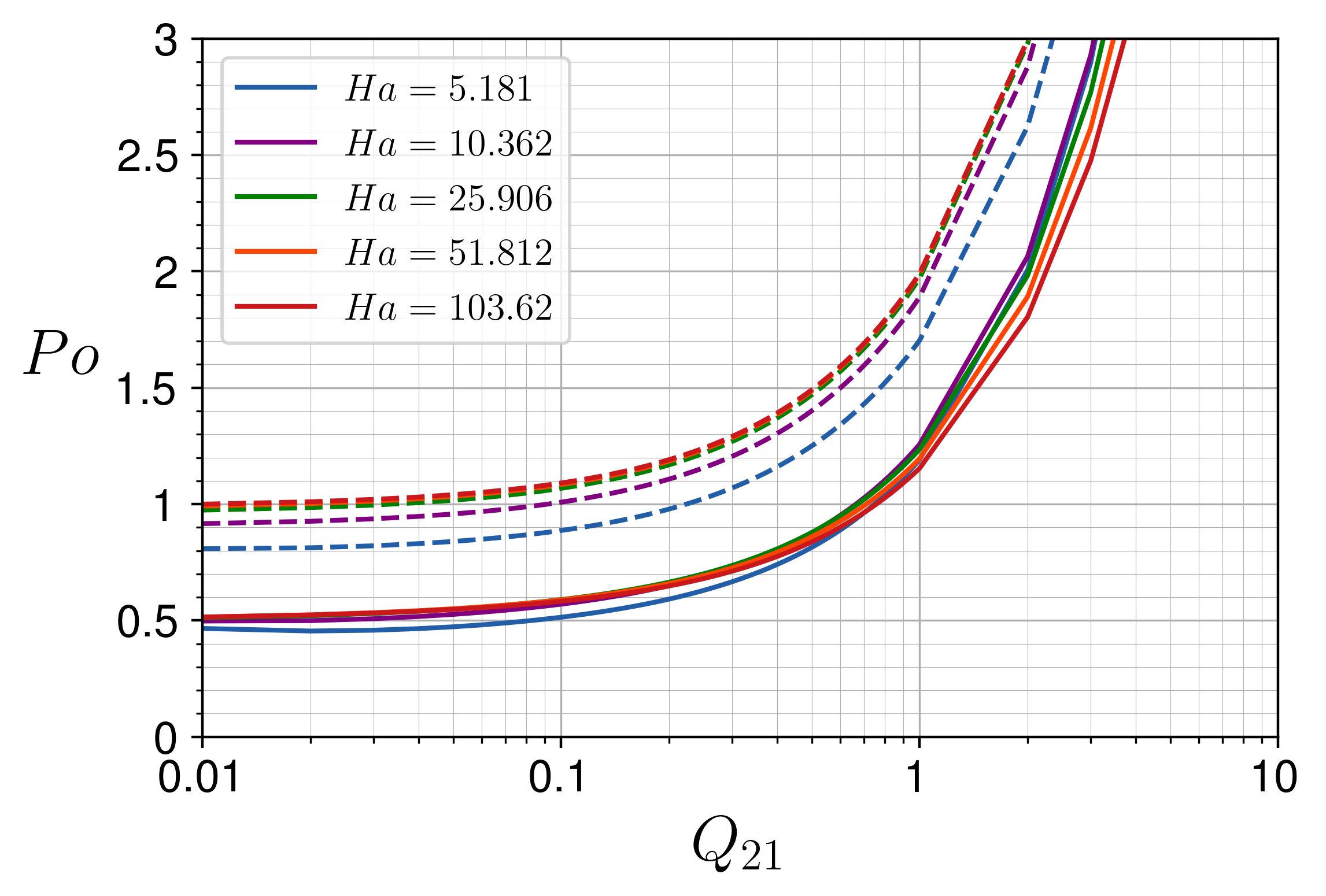}}
	\caption{\label{Fig: Pressure_TP}Variation of (a) the dimensionless pressure gradient, $P_f^{1,0} = -G/12$, (b) the pressure gradient (lubrication) factor, $P_f^{1}$, and (c) power factor, $\Po$,  with $Q_{21}$ for different $\Ha$. Solid lines -- insulating bottom wall; dashed lines -- perfectly conducting bottom wall.}
\end{figure} 

The combined effects of the different Lorenz force, conductive phase holdup, and velocity profiles in the cases of insulating and perfectly conducting bottom walls result in different pressure gradients required to maintain specified flow rates of the fluids. This is illustrated in Fig.\ \ref{Fig: Pressure_TP}a, where the dimensionless pressure gradient, $P_f^{1,0}=-G/12$, is plotted as a function of $Q_{21}$ for different $\Ha$. The $P_f^{1,0}$ values represent the pressure gradient normalized with respect to its value for single-phase flow of the conductive phase without external magnetic field ($\Ha=0$). As shown in the figure, for both conductive and insulating walls, the increased wall and interfacial shear stresses at higher $\Ha$ lead to a larger pressure gradient. However, for the same $\Ha$, the pressure gradient is significantly lower in the case of insulating wall across the entire range of  $Q_{21}$. As the values of the top- and bottom-wall shear stresses are similar in the range of $Q_{21} < 1$ (see Figs.\ \ref{Fig: Bottom_shear_TP} and \ref{Fig: Top_shear_TP}), the higher pressure drop in the case of conducting bottom wall is mainly due to the non-zero Lorentz force at that wall (Eq.\ \ref{Eq: Integral_balance}). For the case of insulating wall and $\Ha=5.181$, $P_f^{1,0} < 1$ is obtained in a range of low $Q_{21}$ up to slightly above $1$. This indicates that, due to the lubrication effect of the air layer on the viscous mercury flow, the pressure gradient can still be lower than that of single-phase mercury flow in absence of the magnetic field ($\Ha=0$). Note, however, that in mercury-air flow with $\Ha=0$, the lubrication effect is even more pronounced and $P_f^{1,0}$ can reach a value as low as approximately $0.35$ ($65\%$ reduction of the pressure gradient with respect to the single-phase flow) for small air flow rates, such that $Q_{21}=0.05$. Moreover, $P_f^{1,0}$ stays less than $1$ up to $Q_{21}\approx 10$. 

The effect of the air flow is better understood by examining the pressure gradient factor  $P_f^1$ (Fig.\ \ref{Fig: Pressure_TP}b). This factor represents the ratio between the axial pressure gradient in two-phase flow and that in single phase flow of the conductive fluid under the same applied magnetic field intensity, $B_0$ (i.e., same $\Ha$). Although not shown in the figure, it was verified that all curves converge to $P_f^1 = 1$ for  $Q_{21}=0$ (i.e., single-phase flow). Therefore, the values of $P_f^1$ below $1$ indicate that introducing air into the channel provides a lubrication effect on the mercury flow. Figure\ \ref{Fig: Pressure_TP}b shows that a significant reduction of the pressure gradient by the gas flow ($P_f^1 \approx 0.5$, or about $50\%$ reduction) can be achieved at low $Q_{21}$ of the order of $10^{-2}$ when the bottom wall is insulating. In this case, for $Q_{21}$ of up to approximately $1$, the extent of pressure drop reduction in the studied range of $\Ha$ is insensitive to its value. However, the lubrication effect is already somewhat lower than that achieved in the absence of magnetic field (compare with $P_f^{1,0}\approx 0.35$, which is discussed above). For higher $Q_{21}$, the lubrication effect is reduced, but the region of $P_f^1 < 1$ extends to higher $Q_{21}$ values upon increasing $\Ha$ (e.g., up to $Q_{21}\approx 100$ for $\Ha=103.62$). Further increase of the gas flow rate (higher $Q_{21}$) beyond the value of $Q_{21}$ corresponding to $P_f^1=1$ results in a rather steep increase of the pressure gradient. The corresponding pumping power factor, $\Po = P_f^1 (1 + Q_{21})$, is shown in Fig.\ \ref{Fig: Pressure_TP}c. In the case of insulating wall (solid lines), the values of Po are insensitive to the Hartmann number, and pumping power saving can be achieved by introducing the air flow up to $Q_{21} \approx 1$.  

In the case of perfectly conducting bottom wall, Figure\ \ref{Fig: Pressure_TP}b (dashed lines) shows that the lubrication effect is significantly weaker, with $P_f^1$ values slightly below $1$ only for $\Ha$ up to $\approx 25$ and $Q_{21}<1$. The narrower range of $Q_{21}$ where $P_f^1 < 1$ is a result of the steep increase of the top-wall shear stress (see Fig.\ \ref{Fig: Top_shear_TP}), which occurs at lower $Q_{21}$ when the wall is perfectly conducting. Consequently, the potential for pumping power saving is also much lower, restricted to a narrower range of $Q_{21}$ up to $\approx 0.1$ and to smaller $\Ha$ compared to the potential power saving achieved in channels with insulating walls. This results clearly demonstrates that even at low magnetic Reynolds number, the conductivity of the wall has a significant impact on flow characteristics and should not be overlooked. 

\subsection{Wide rectangular ducts}

In practice, the results obtained in a two-plate configuration cannot be realized, but only approached when the duct aspect ratio is sufficiently high. Yet, the effects of the side walls and their conductivity on the velocity and induced magnetic field profiles as well as on the integral flow characteristics, such as the holdup and pressure gradient, should be examined. 
\begin{figure}[h!]
	\centering
	\subfloat[$c_{bw}=0$, $Q_{21}=0.1$: $h$]{\includegraphics[width=0.25\textwidth,clip]{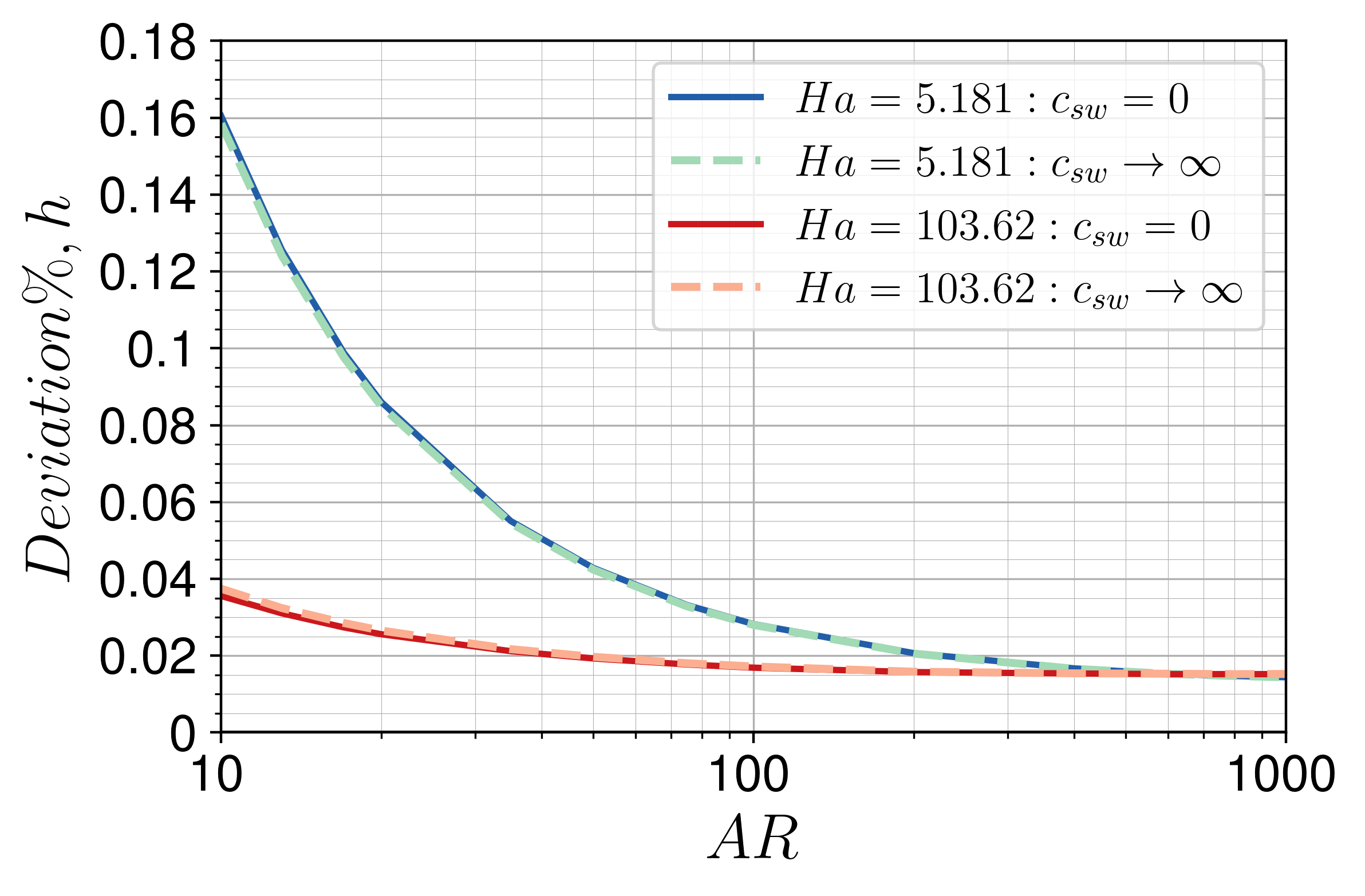}}
	\subfloat[$P_1^{f}$]{\includegraphics[width=0.25\textwidth,clip]{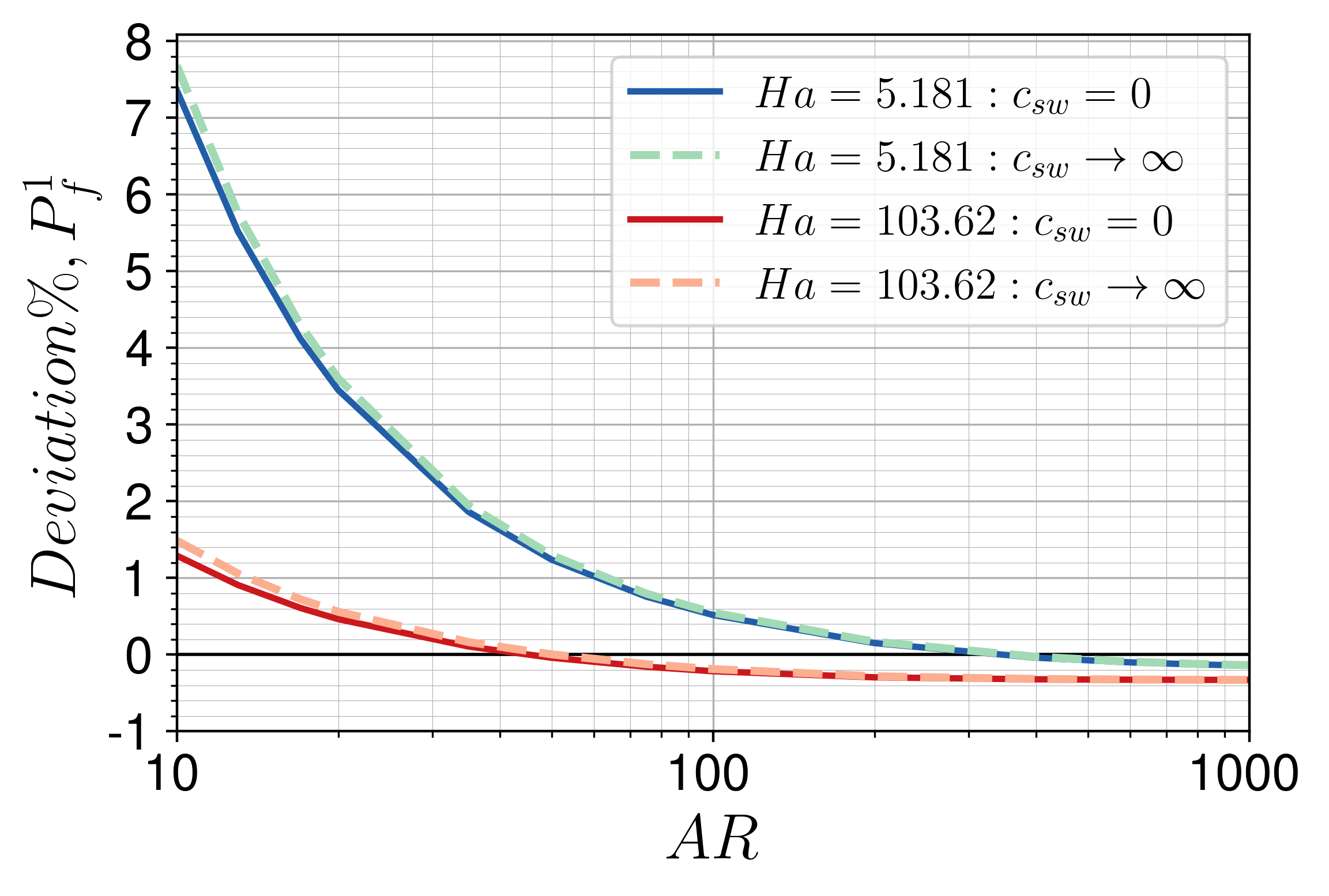}}
	\subfloat[$c_{bw}=0$, $Q_{21}=10$: $h$]{\includegraphics[width=0.25\textwidth,clip]{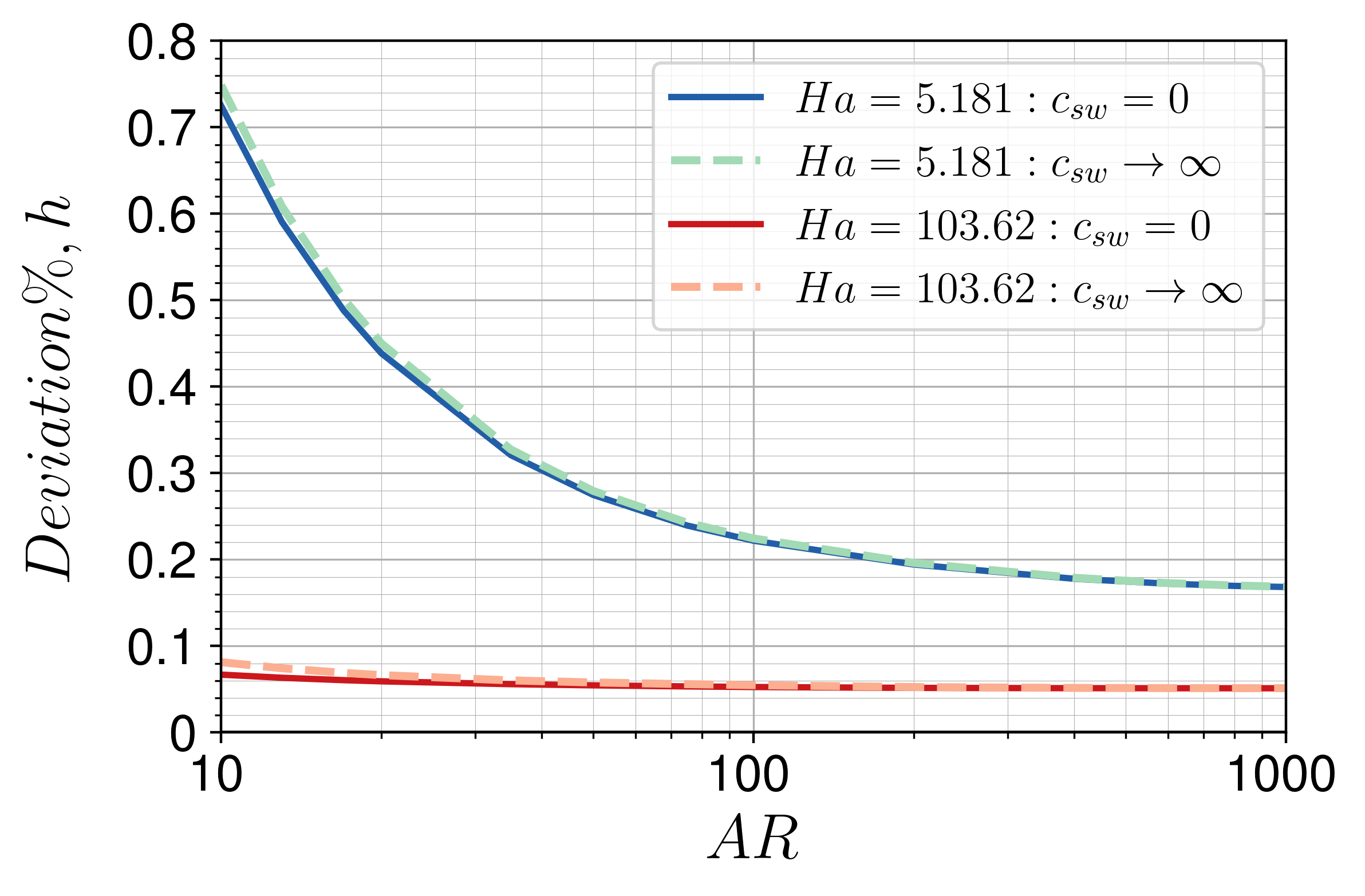}}
	\subfloat[$P_1^{f}$]{\includegraphics[width=0.25\textwidth,clip]{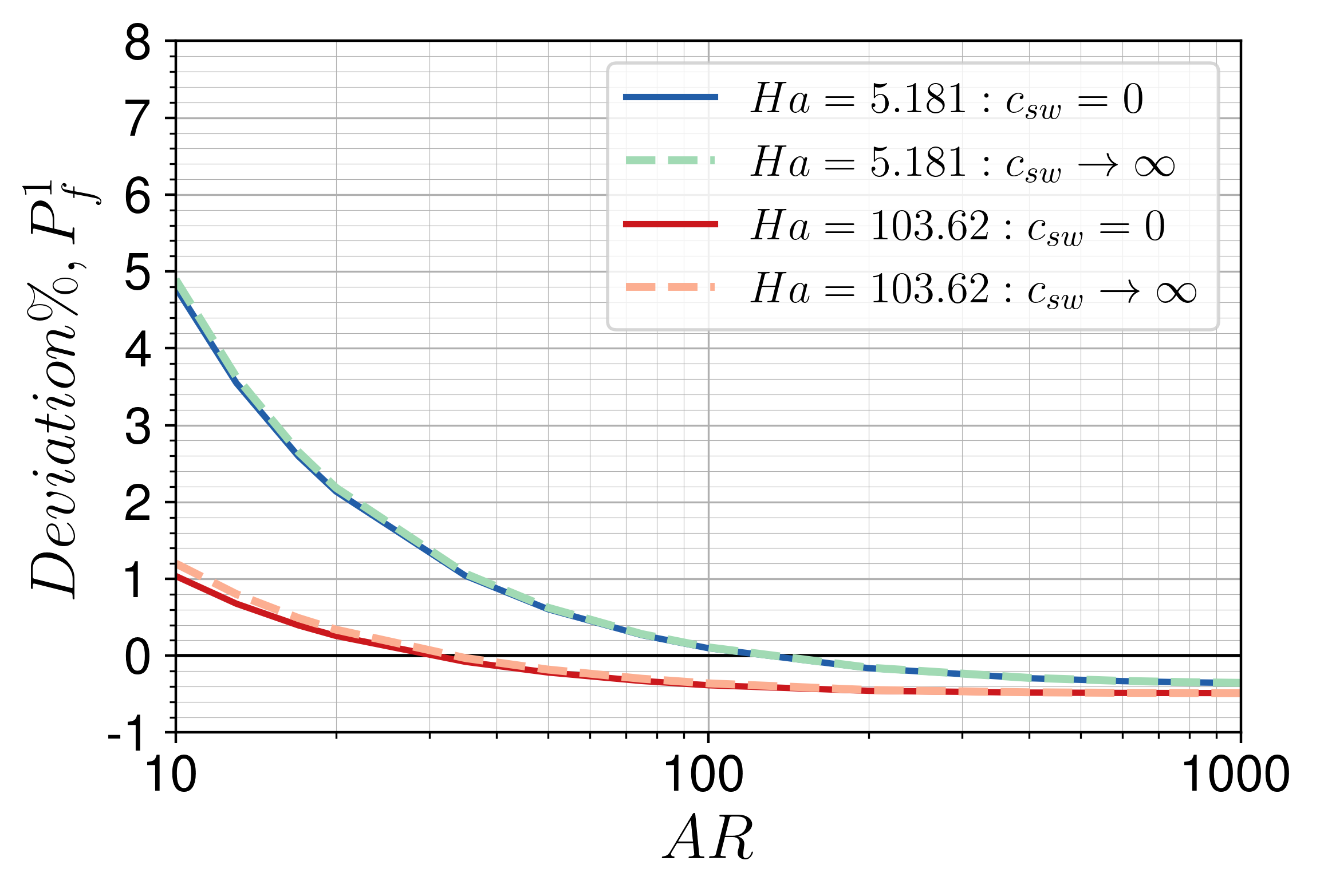}}
	\\
	\subfloat[$c_{bw}\to\infty$, $Q_{21}=0.1$: $h$]{\includegraphics[width=0.25\textwidth,clip]{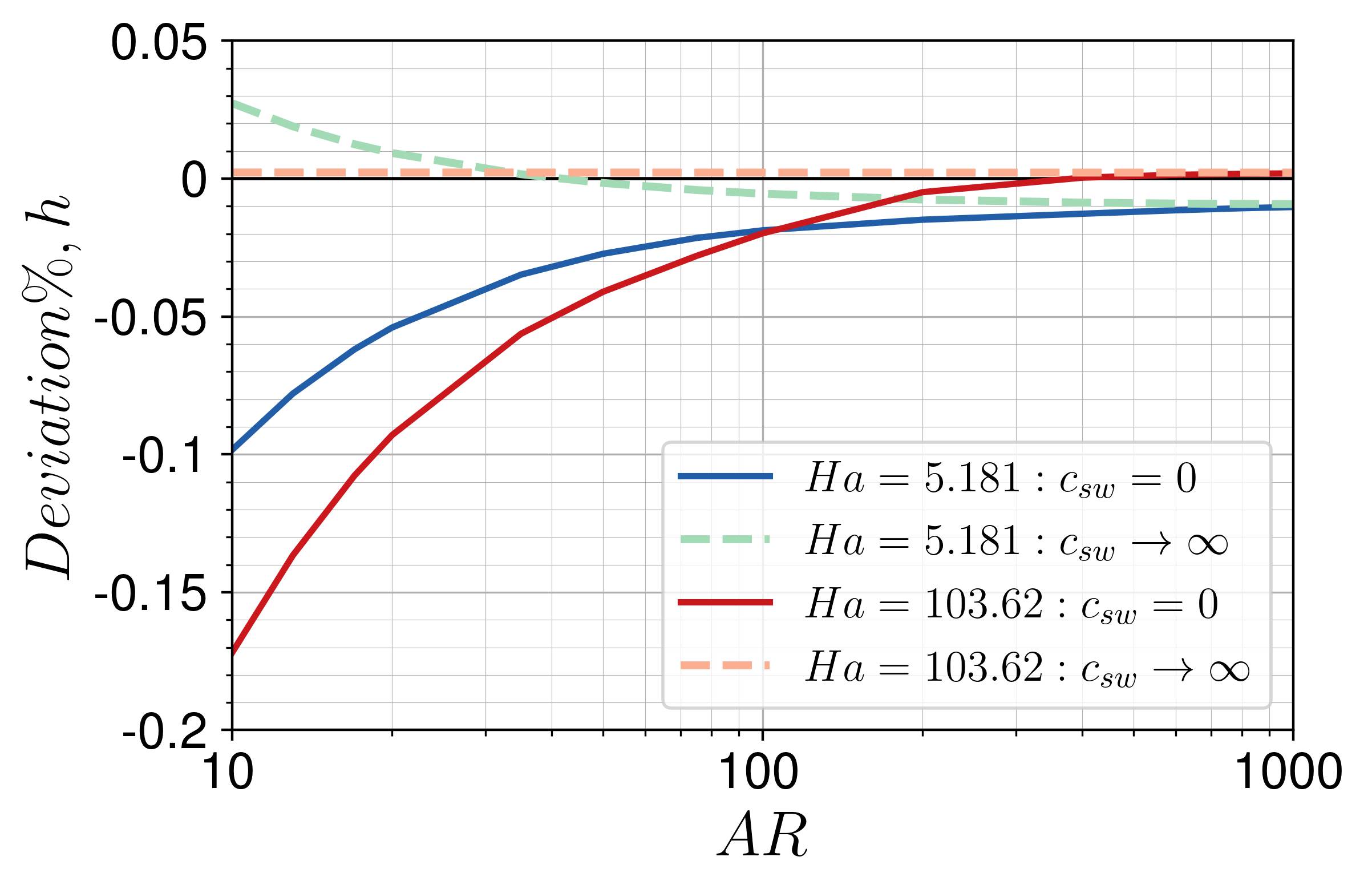}}
	\subfloat[$P_1^{f}$]{\includegraphics[width=0.25\textwidth,clip]{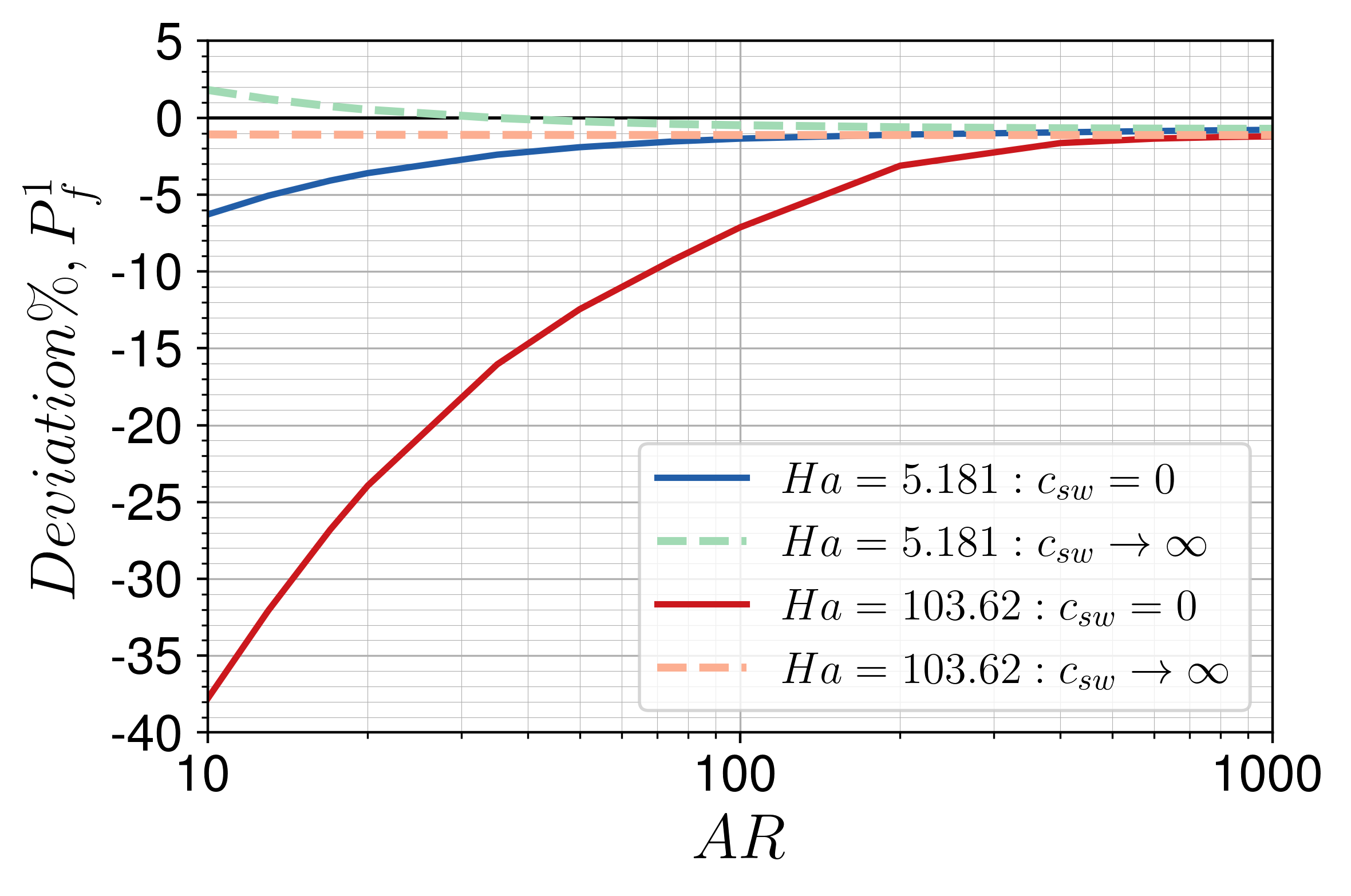}}
	\subfloat[$c_{bw}\to\infty$, $Q_{21}=10$: $h$]{\includegraphics[width=0.25\textwidth,clip]{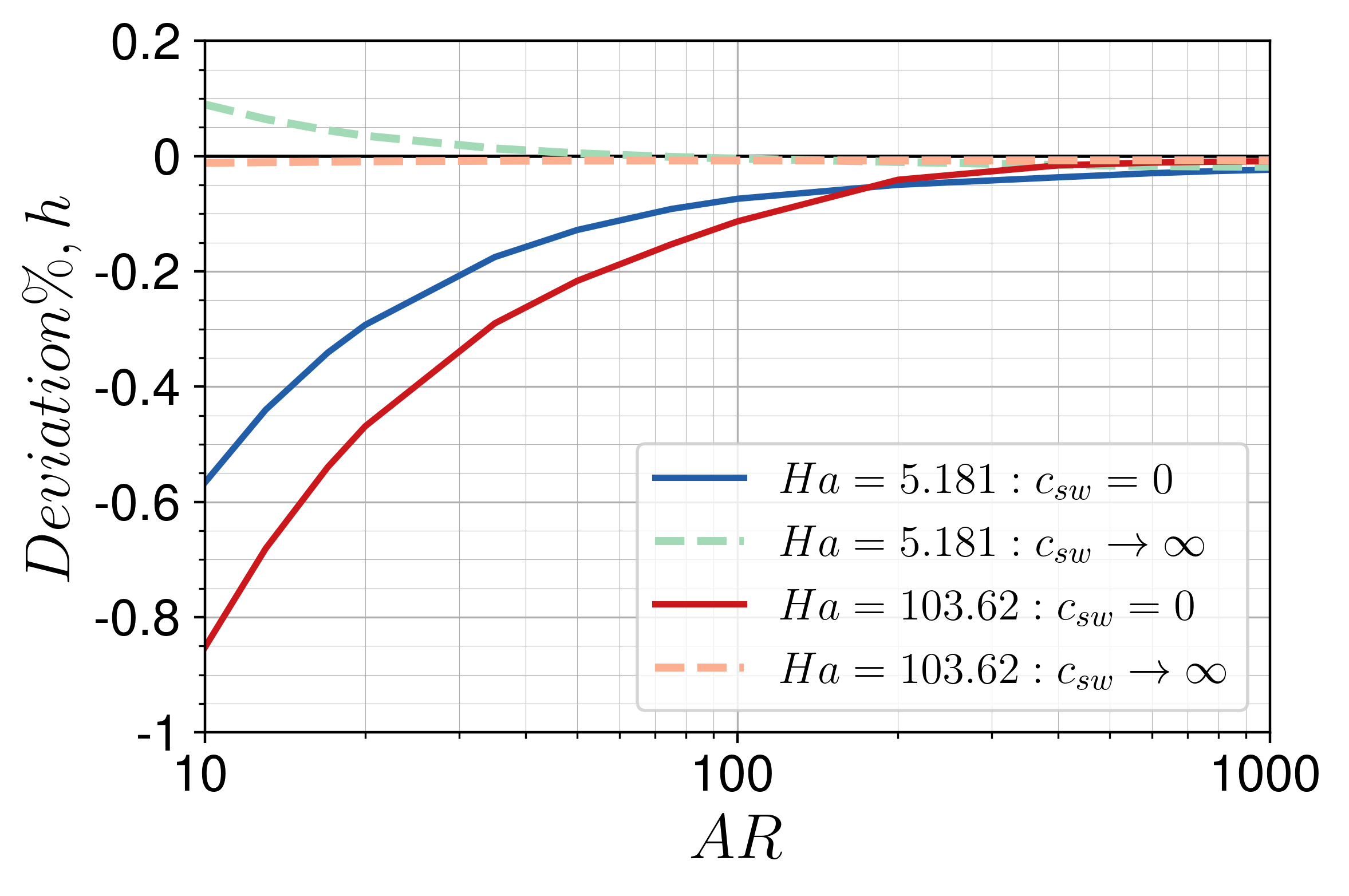}}
	\subfloat[$P_1^{f}$]{\includegraphics[width=0.25\textwidth,clip]{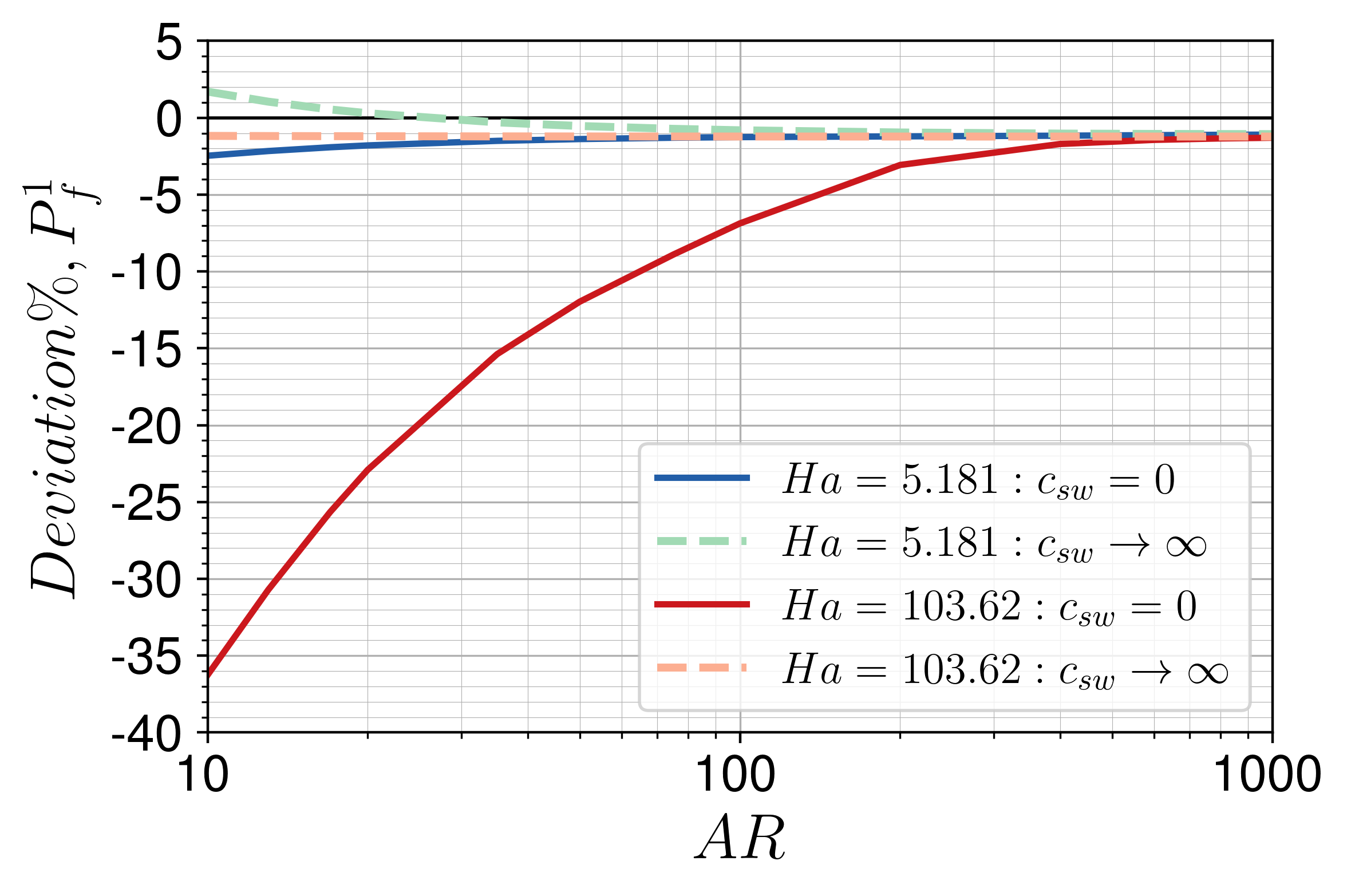}}
	\caption{\label{Fig: AR_convergence}Deviation of the holdup and pressure gradient in wide rectangular ducts from the corresponding TP values. (a)-(d) Insulating bottom wall. (e)-(h) Perfectly conducting bottom wall. Solid lines correspond to the case of insulating side walls, dashed lines -- perfectly conducting side walls.}	
\end{figure}

For this purpose, Figure\ \ref{Fig: AR_convergence} shows a deviation of the holdup and pressure gradient factor, $P_f^1$, for rectangular ducts with high aspect ratios between $10$ and $1000$ from the corresponding TP values (i.e., for the holdup - $(h_{AR} - h_{TP}) / h_{TP} \cdot 100\%$). The top subfigures, Figs. \ref{Fig: AR_convergence}(a)-(d), show the results for the ducts with insulating bottom wall. It is clearly seen that these graphs differ significantly from the bottom subfigures, Figs. \ref{Fig: AR_convergence}(e)-(h), that correspond to the ducts with perfectly conducting bottom wall. When the bottom wall is insulating, the conductivity of side wall has a minor effect on the holdup and pressure gradient for aspect ratios larger than $10$, as can be seen by almost complete overlapping of the solid and dashed lines that correspond to the insulating and conducting side walls, respectively. For higher Hartmann numbers, the TP values are approached faster, i.e., a smaller deviation is obtained for the same $AR$ (compare red and blue lines in Figs. \ref{Fig: AR_convergence}(a)-(d)). In the case of insulating bottom wall, the holdup in the rectangular duct stays always slightly higher than the TP value due to the slowdown of the flow near the side walls. Higher flow rate ratios results in higher deviation in the holdup, though even for $Q_{21}=10$ and AR=10 it is still below $0.8\%$ (compare with a value of $\approx0.16\%$ for $Q_{21}=0.1$). The deviation in the pressure gradient factor is much more significant for low Hartmann number (blue lines) and accounts for several per cents for AR up to about $50$. It decreases quickly for high $AR$ and eventually converges to values that underestimates the TP value by less than $0.5\%$.

The behavior of the holdup and pressure gradient with increasing the duct aspect ratio is very different for the case of perfectly conducting bottom wall (bottom row of Fig. \ref{Fig: AR_convergence}(e)-(h)). In case of the insulating side walls (solid lines), both the holdup and pressure gradient are lower in rectangular ducts than between two plates. The higher Ha (red lines) results in lower holdup, and even more significant, lower pressure gradient for $AR$ below $100$. Decrease in the value of $P_f^1$ with respect to its TP analogue reaches as low as about $-38\%$ for $AR=10$ and $Q_{21}=0.1$. This result will be discussed below (see discussion with reference to Fig.\ \ref{Fig: U1_vs_x_bot_cond}, which shows the very peculiar velocity field near the side walls observed in this case). On the other hand, when the side walls are perfectly conducting, already for rectangular ducts of $AR$ slightly above $10$ (dashed lines in Figs.\ \ref{Fig: AR_convergence}(e)-(h)), the holdup and pressure gradient become similar to the TP model values.  
\begin{figure}[h!]
	\centering
	\subfloat[$AR=10$, $\text{max}(U)=6.432$, $h=0.776$]{\includegraphics[width=0.32\textwidth,clip]{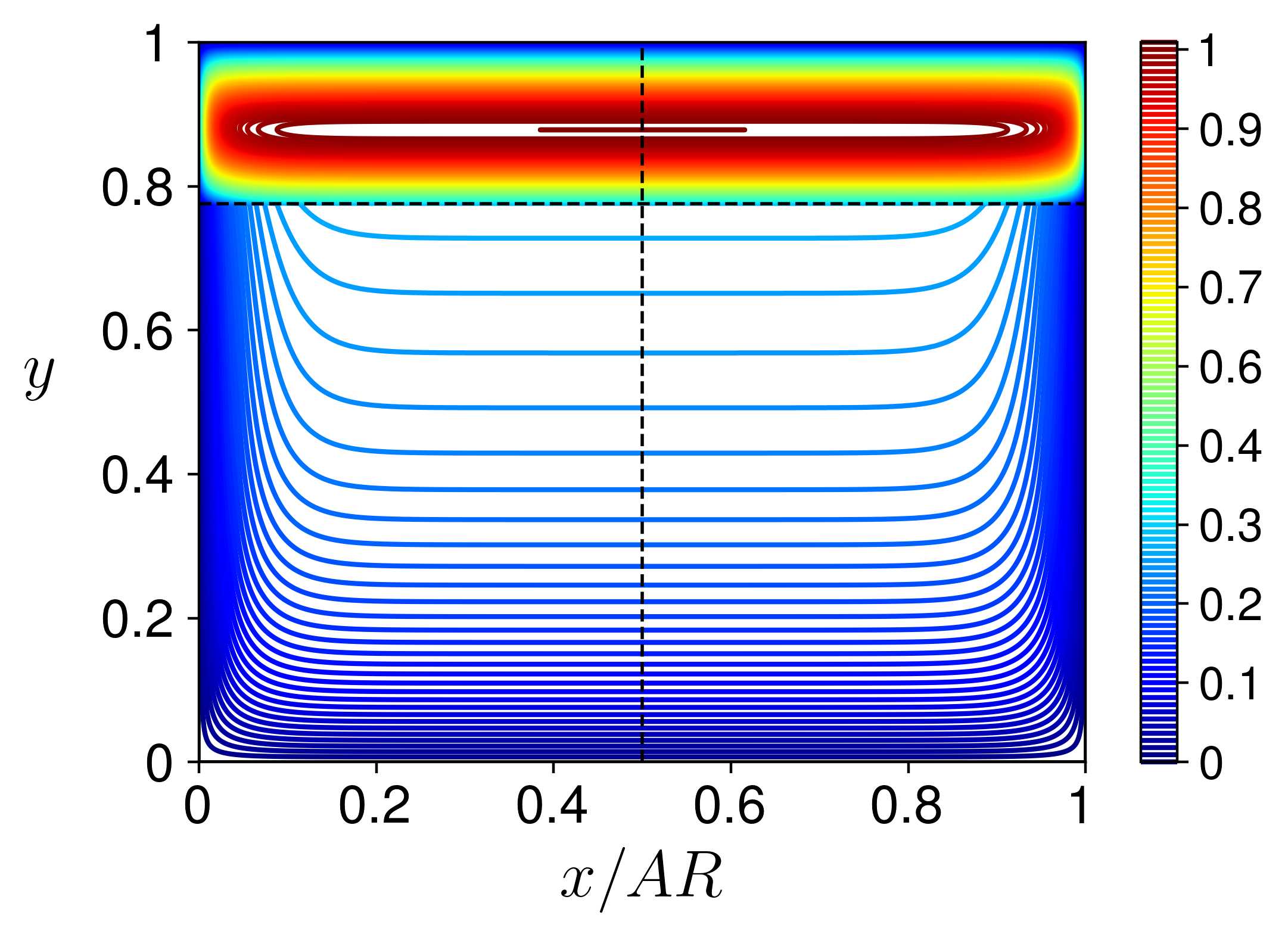}}
	\subfloat[$AR=10$]{\includegraphics[width=0.32\textwidth,clip]{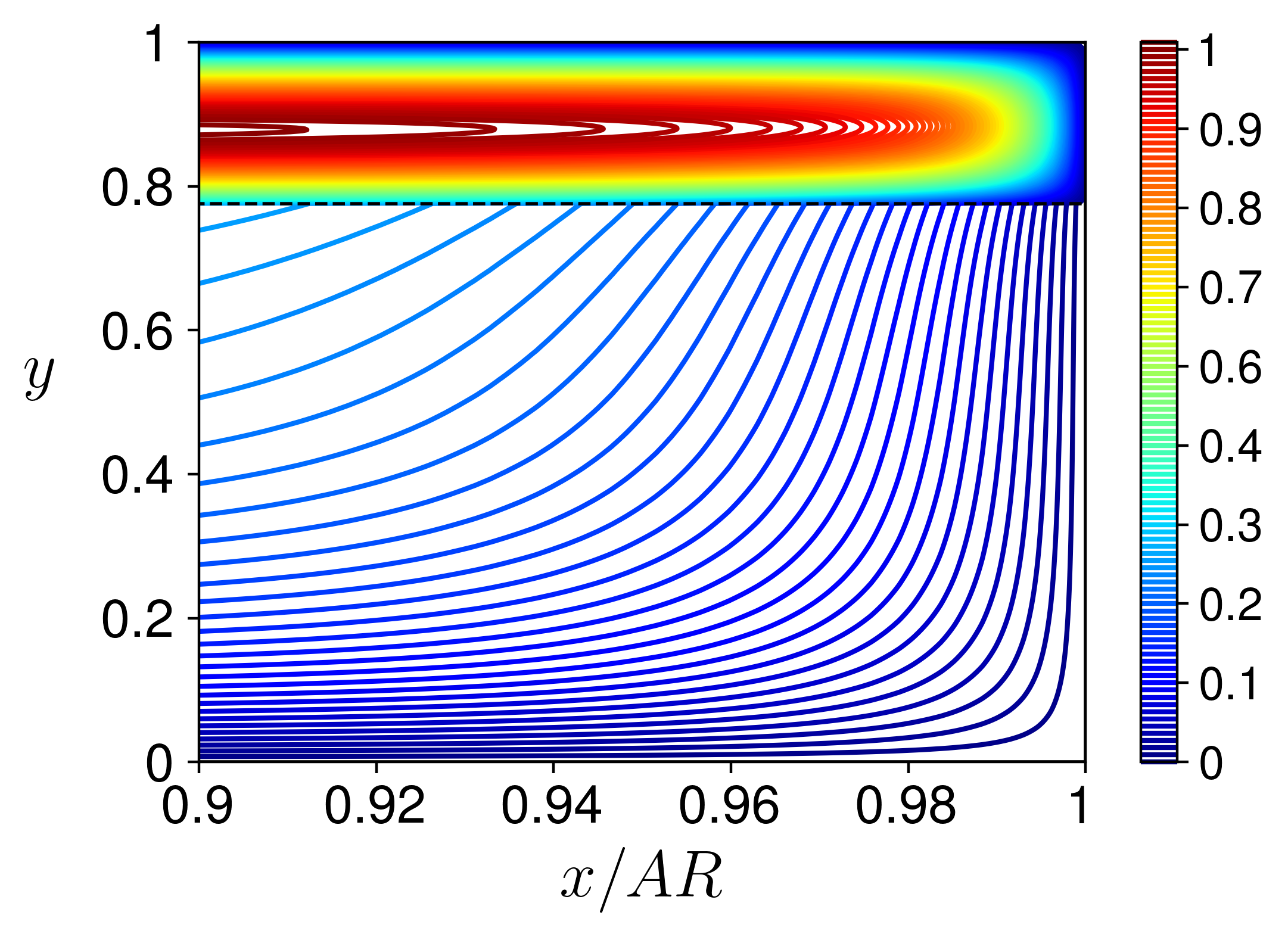}}
	\subfloat[$AR=100$, $\text{max}(U)=6.214$, $h=0.773$]{\includegraphics[width=0.32\textwidth,clip]{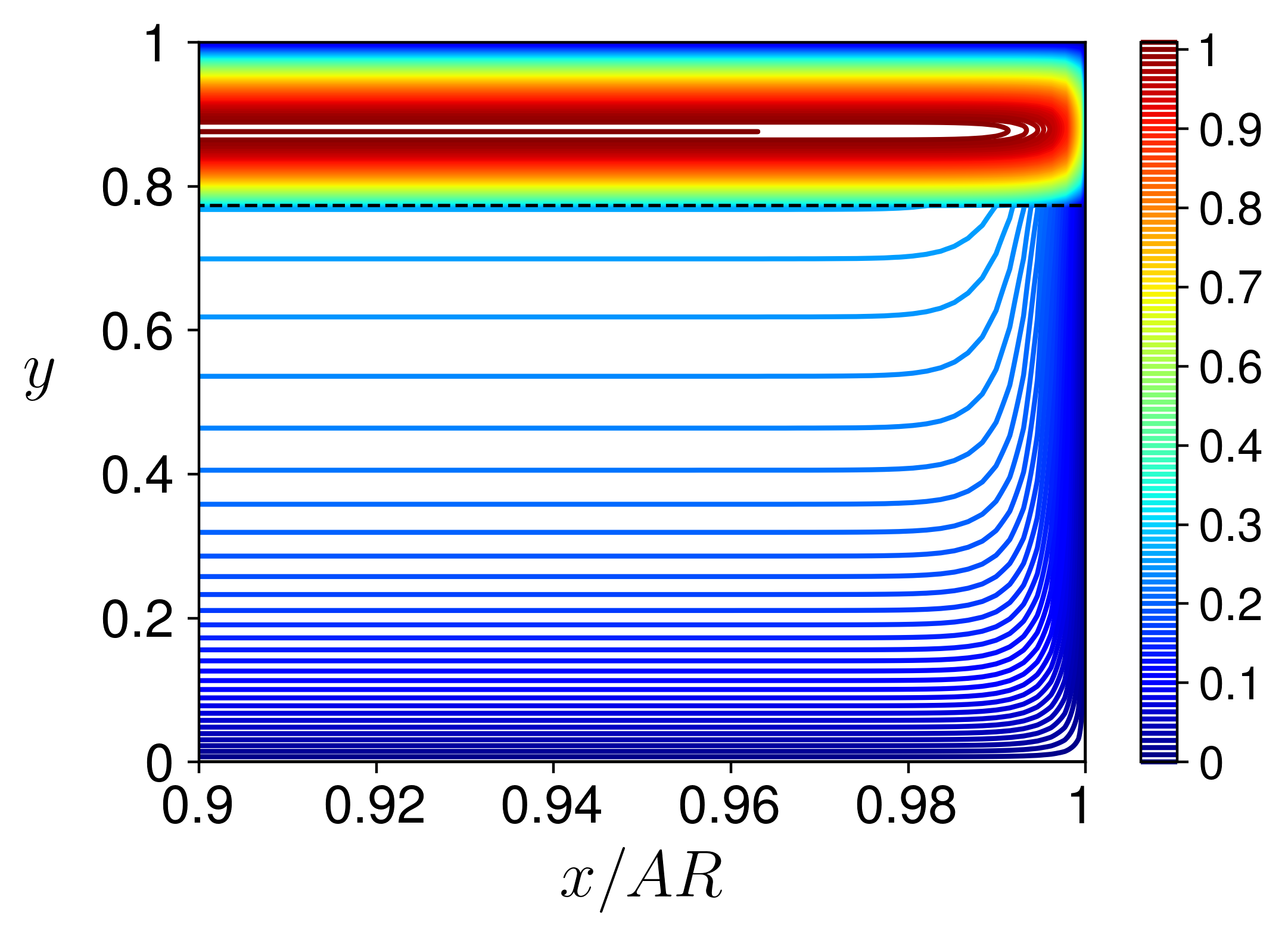}}
	\caption{\label{Fig: U_contours_Ha_5}Contours of the velocity scaled by its maximum value, $U/\text{max}(U)$. Insulating walls, $Q_{2 1}=1$, $\Ha=5.181$. (a) The entire cross section; (b) and (c): enlargement of the near-side-wall region of ducts of $AR=10$ and $100$, respectively. The unperturbed interface is denoted by horizontal dashed black line; the cross-section centerline -- vertical dash-dot black line.}	
\end{figure} 

The velocity contours in wide rectangular ducts ($AR=10$ and $100$) with insulating walls are demonstrated in Fig.\ \ref{Fig: U_contours_Ha_5} for $Q_{21}=1$ and $\Ha=5.181$. As discussed above, for high width-to-height aspect ratios, the integral parameters, such as holdup and pressure gradient converges to the two-plate model values. However, because of the no-slip condition, there is still an effect of the side walls on the velocity field even for high aspect ratios. To illustrate the spanwise variation of the velocity for wide ducts, the spanwise coordinate, x, is scaled by the aspect ratio. Moreover, to allow comparison of contours in the different frames, the velocities are scaled by the maximal velocity in the flow cross section, which is located in less viscous and non-conductive gas phase. The maximal velocity as well as the mercury holdup is already practically the same for the aspects ratios of 10 and 100. Figure\ \ref{Fig: U_contours_Ha_5} shows that across most of the duct cross section the velocity isolines are parallel to horizontal walls, and the velocity profiles in the $y$ direction are similar to that obtained in the TP model (shown in Fig.\ \ref{Fig: U_TP}) becoming indistinguishable from it as $AR$ is increased. The difference is found to be significant only in the near-side-wall region of about $10\%$ of the duct width, when significant velocity gradients (Shercliff boundary layer) are observed. 

\begin{figure}[h!]
	\centering
	\subfloat[Insulating walls]{\includegraphics[width=0.48\textwidth,clip]{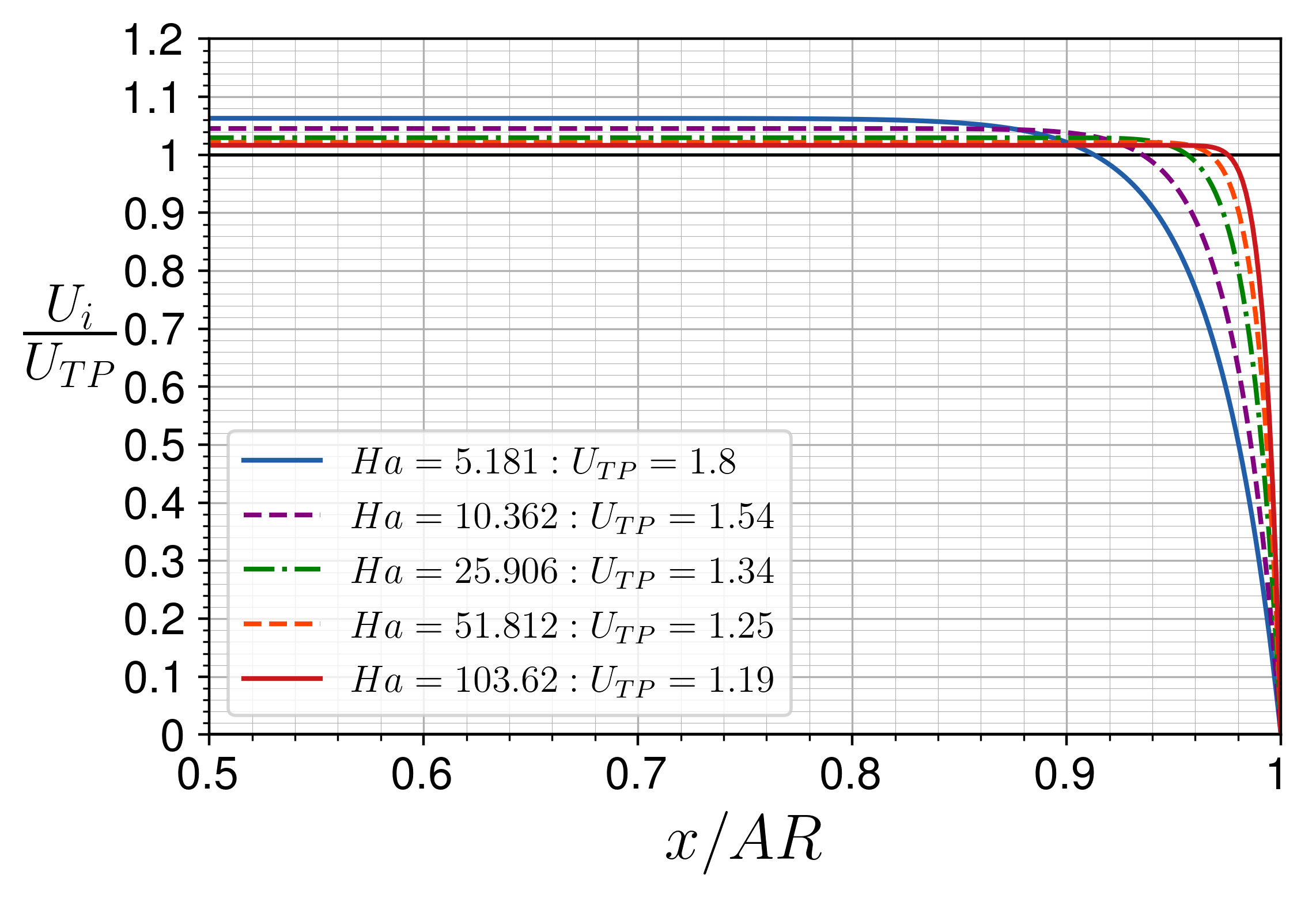}}
	\subfloat[Perfectly conducting walls]{\includegraphics[width=0.48\textwidth,clip]{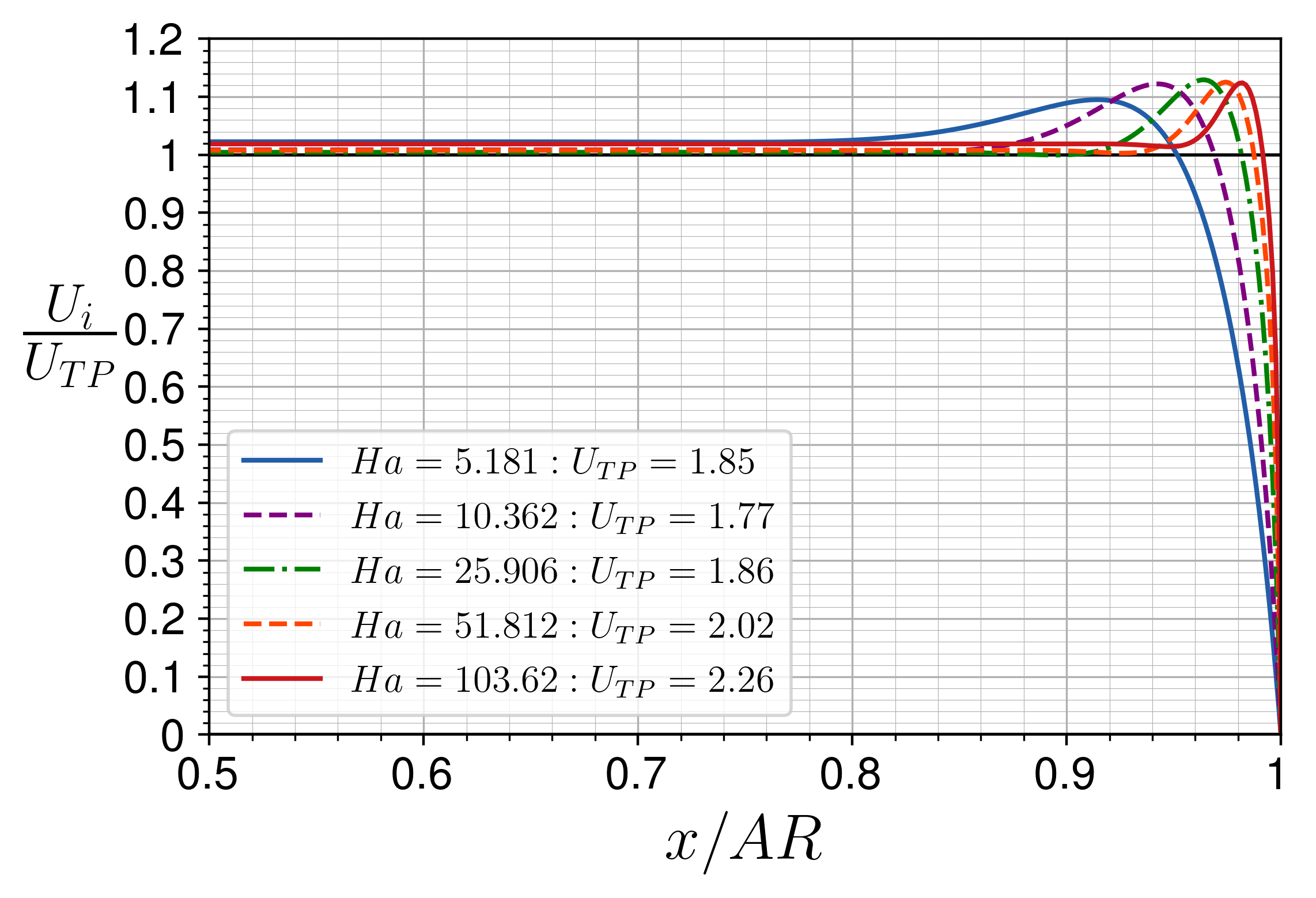}}
	\caption{\label{Fig: U_i_duct}Interfacial velocity for different Hartmann numbers. $Q_{2 1}=1$, $AR = 10$.}	
\end{figure}

\begin{figure}[!htb]
	\centering
	\subfloat[$\text{max}(U_1)=1.201$, $h=0.896$]{\includegraphics[width=0.32\textwidth,clip]{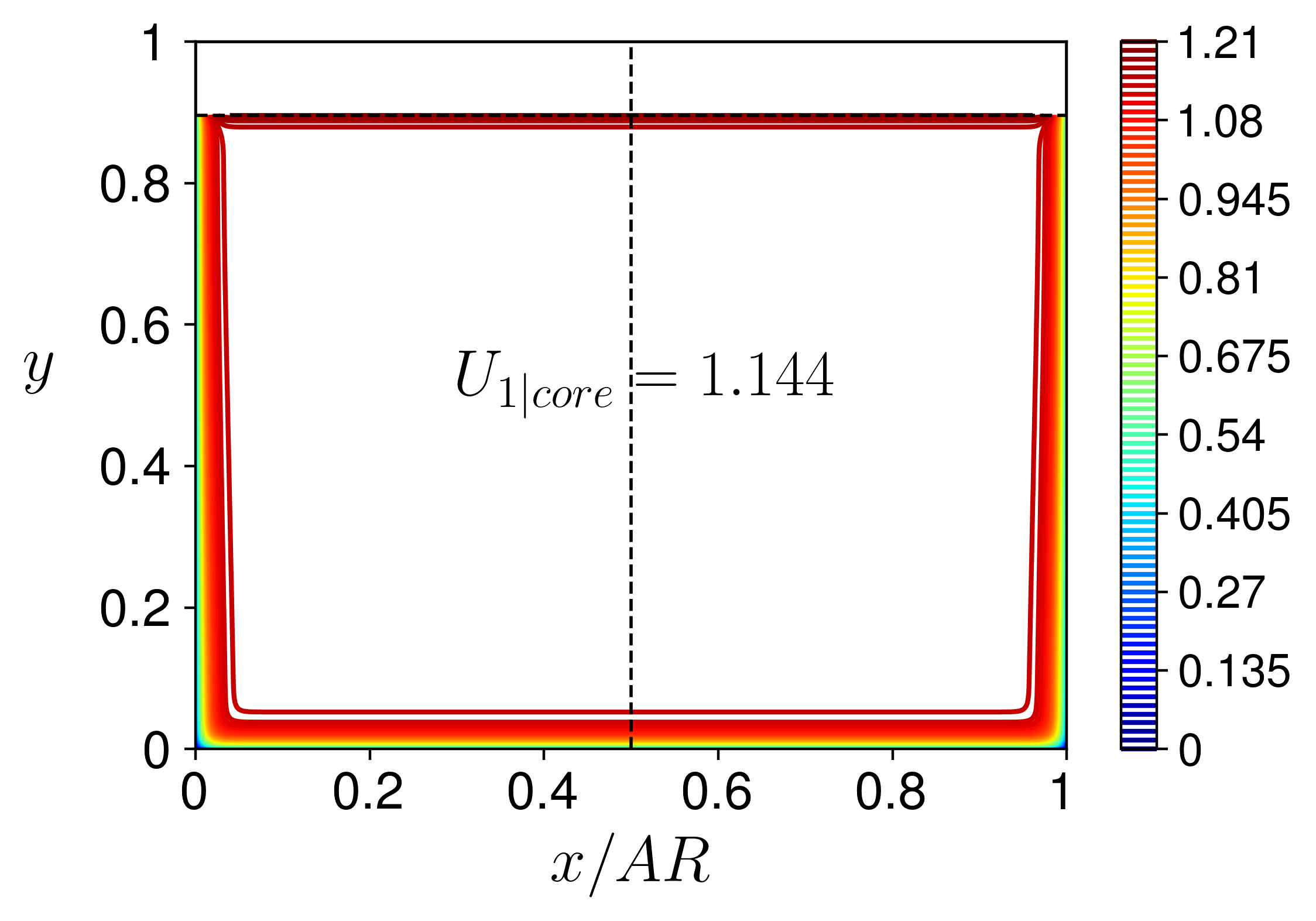}}
	\subfloat[$\text{max}(U_1)=2.398$, $h=0.976$]{\includegraphics[width=0.32\textwidth,clip]{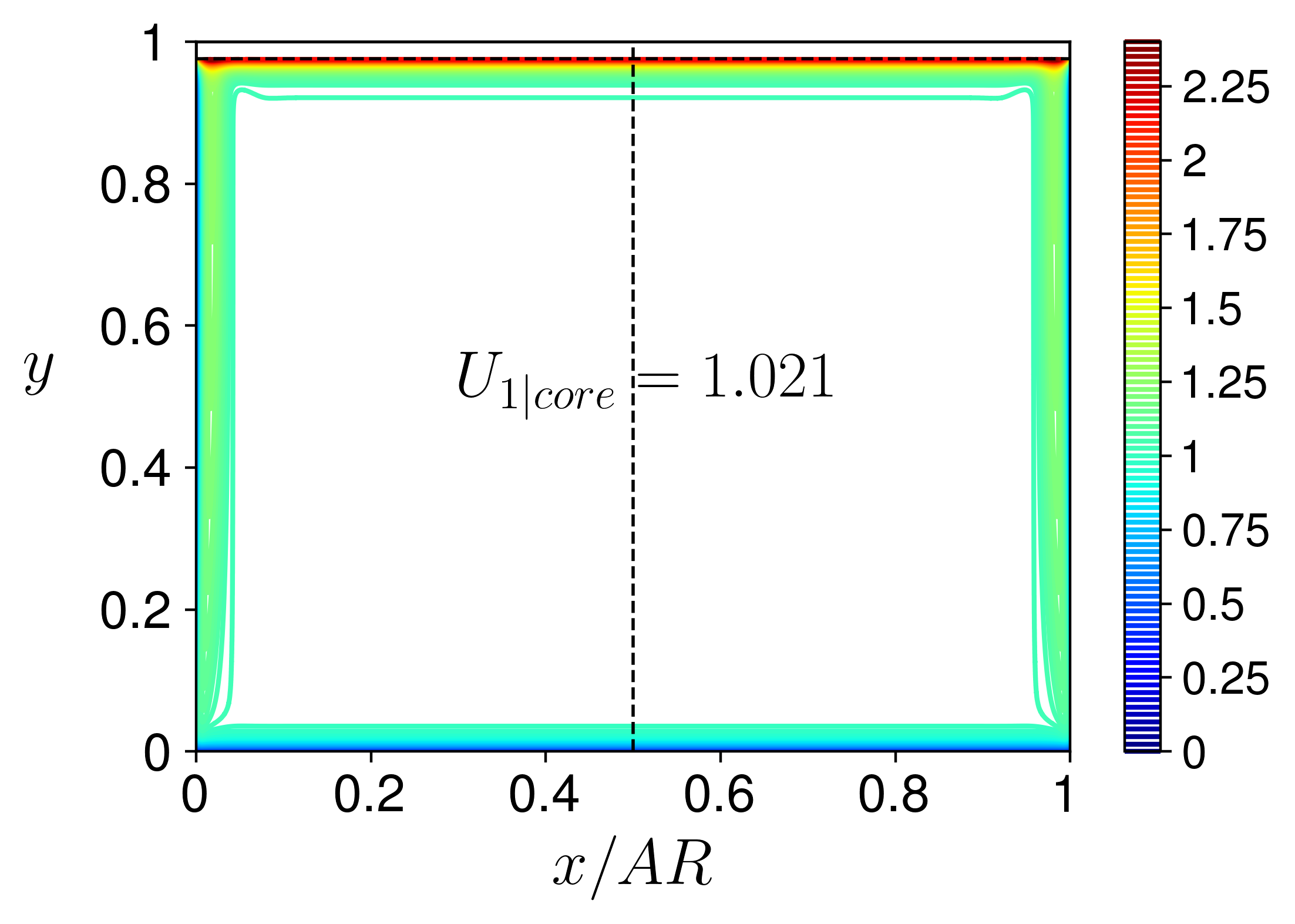}}
	\subfloat[$-0.03\le U_1 \le 18.97$, $h=0.973$]{\includegraphics[width=0.32\textwidth,clip]{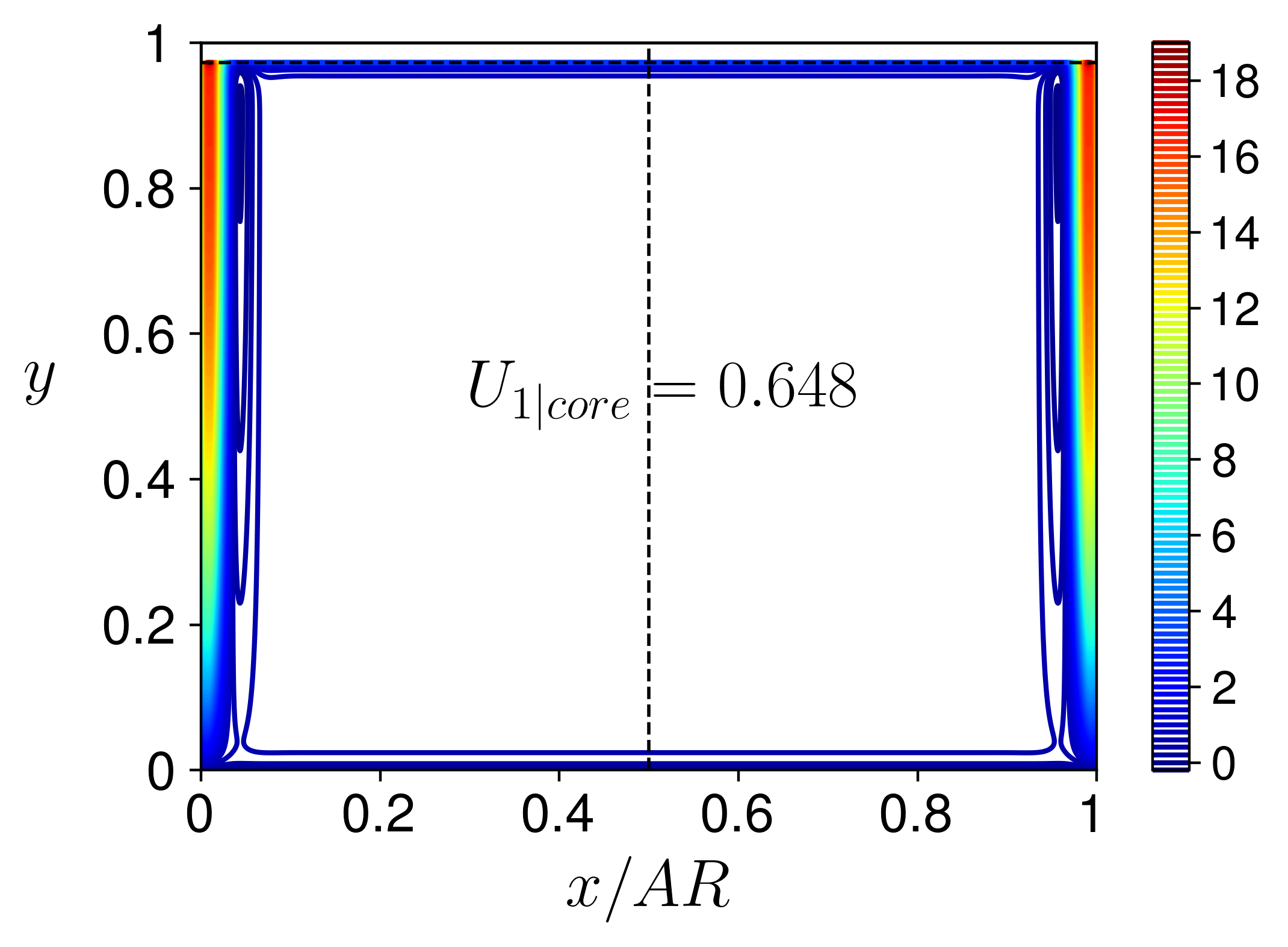}}
	\\
	\subfloat[$\text{max}(U_1)=1.201$, $h=0.896$]{\includegraphics[width=0.32\textwidth,clip]{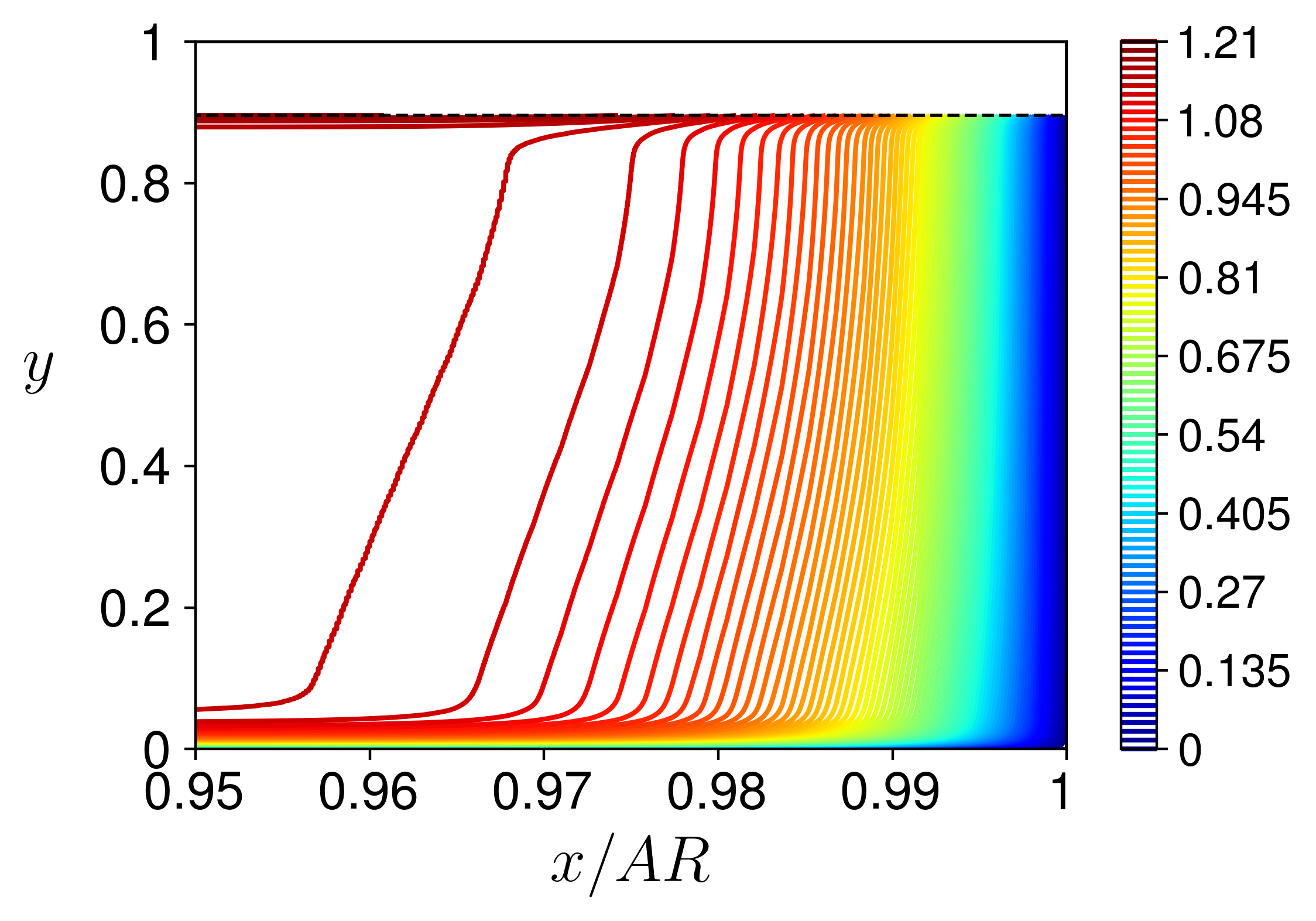}}
	\subfloat[$\text{max}(U_1)=2.398$, $h=0.976$]{\includegraphics[width=0.32\textwidth,clip]{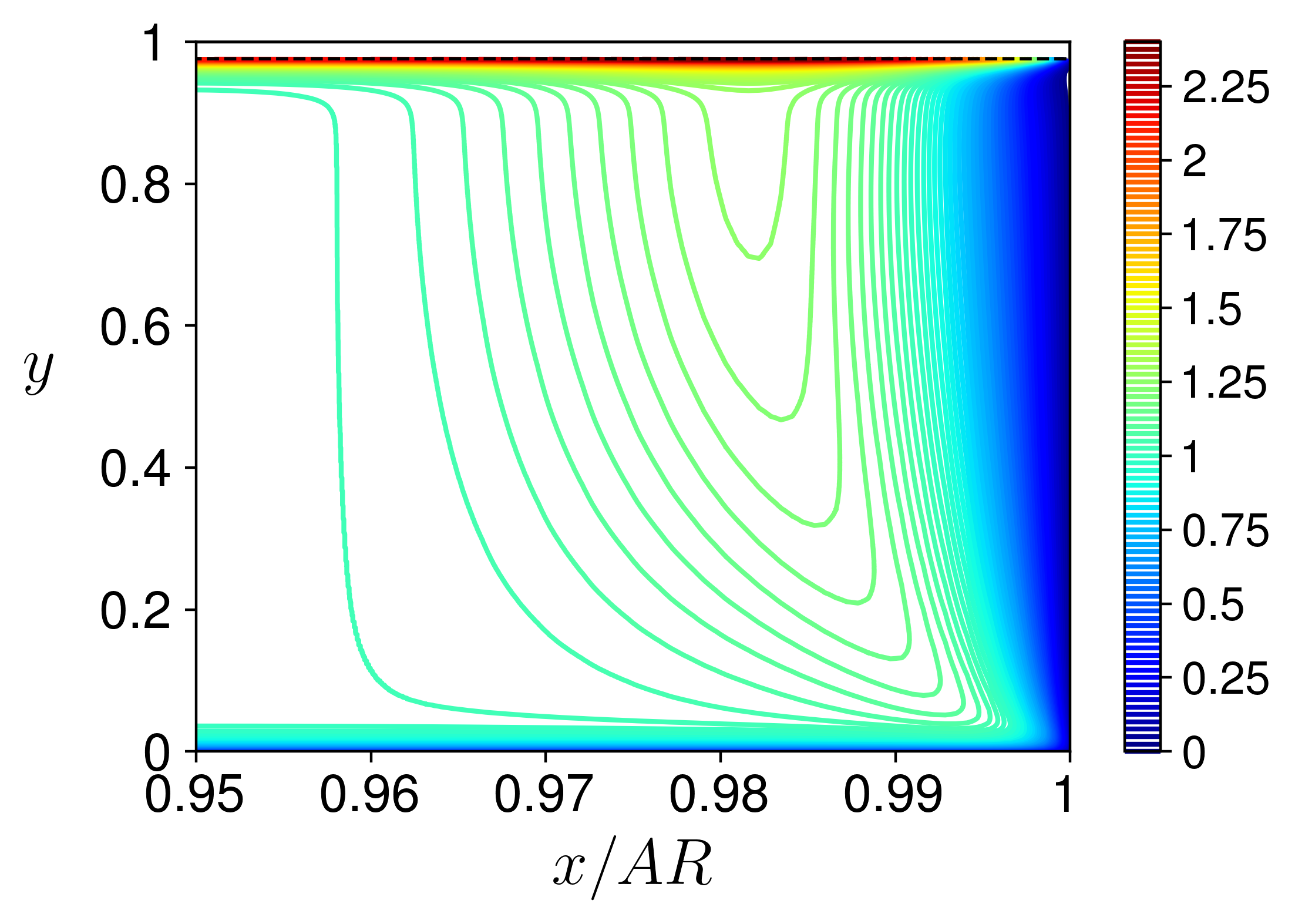}}
	\subfloat[$-0.03\le U_1 \le 18.97$, $h=0.973$]{\includegraphics[width=0.32\textwidth,clip]{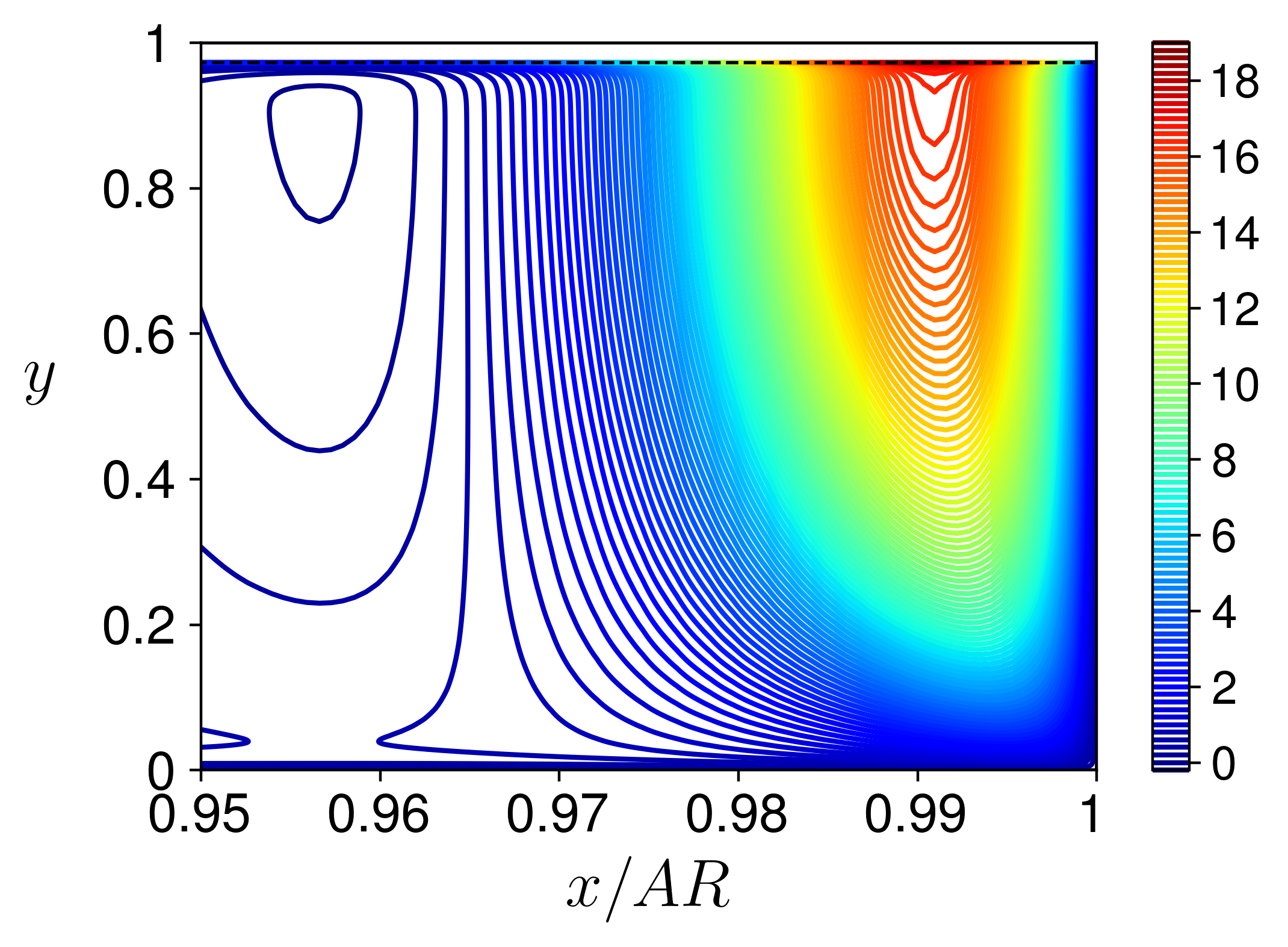}}
	\caption{\label{Fig: U_contours_Ha_103}Velocity contours in the conductive phase, $U_1$. Effect of the wall conductivity in a rectangular duct ($AR=10$), $Q_{2 1}=1$, $\Ha=103.62$. (a) Insulating walls; (b) perfectly conducting walls; (c) bottom wall - conducting, side walls - insulating. (d)-(f) Enlargement of the near-side-wall region of the velocity contours shown in (a)-(c).}	
\end{figure}

Inspection of Fig.\ \ref{Fig: U_contours_Ha_5} shows that the air velocity is much higher than that of the mercury, and its velocity contours closely resemble Poiseuille flow in a much smaller duct of the air-layer height, with an effective aspect ratio of $AR/(1-h)$. The mercury velocity is maximal at the interface and its contours resemble the pattern found in MHD Couette-Poiseuille flow, the difference being the spanwise variation of the interfacial velocity in the considered two-phase MHD flow. The two types of velocity boundary layers -- the Shercliff B. L. over the side walls and the Hartmann B. L. over the bottom wall, can be observed in the figures. As shown in Figs.\ \ref{Fig: U_contours_Ha_5}a,b for $AR=10$ and $Ha=5.181$, the Shercliff B. L. is thicker near the interface and occupies up to $15\%$ of the duct width near each side wall. Obviously, its relative thickness is significantly reduced with increasing $AR$ (e.g., $<2\%$ for $AR=100$, Fig.\ \ref{Fig: U_contours_Ha_5}c). The effect of the side walls is also evident when examining the spanwise variation of the interfacial velocity (Fig.\ \ref{Fig: U_i_duct}). While for $AR=10$ (Fig.\ \ref{Fig: U_i_duct}) the interfacial velocity is slightly higher than its value in the TP model even at the duct centerline, for $AR=100$ they are the same across most of the duct width (not shown). The effect of $\Ha$ on the spanwise variation of the interfacial velocity shown in this figure suggests that similarly to the Hartmann B. L., the thickness of the Shercliff B. L. decreases proportionally to approximately $\Ha^{-1}$ (compare also Figs.\ \ref{Fig: U_contours_Ha_5}a,b with Figs.\ \ref{Fig: U_contours_Ha_103}a,d below).
\begin{figure}[h!]
	\centering
	\subfloat[$AR=10$, $b_\text{max}/\Ha=0.151$, $h=0.776$]{\includegraphics[width=0.32\textwidth,clip]{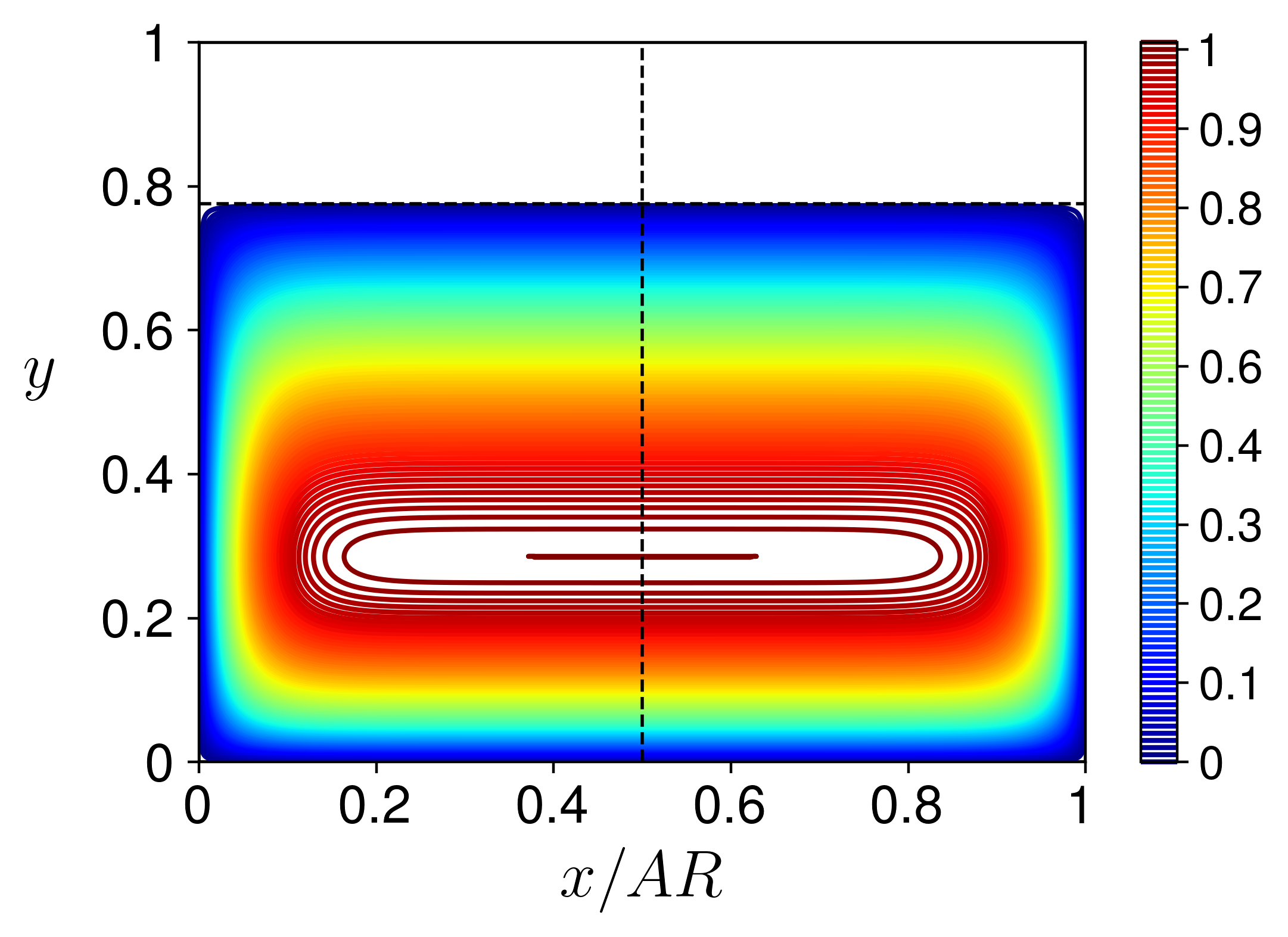}}
	\subfloat[$AR=10$]{\includegraphics[width=0.32\textwidth,clip]{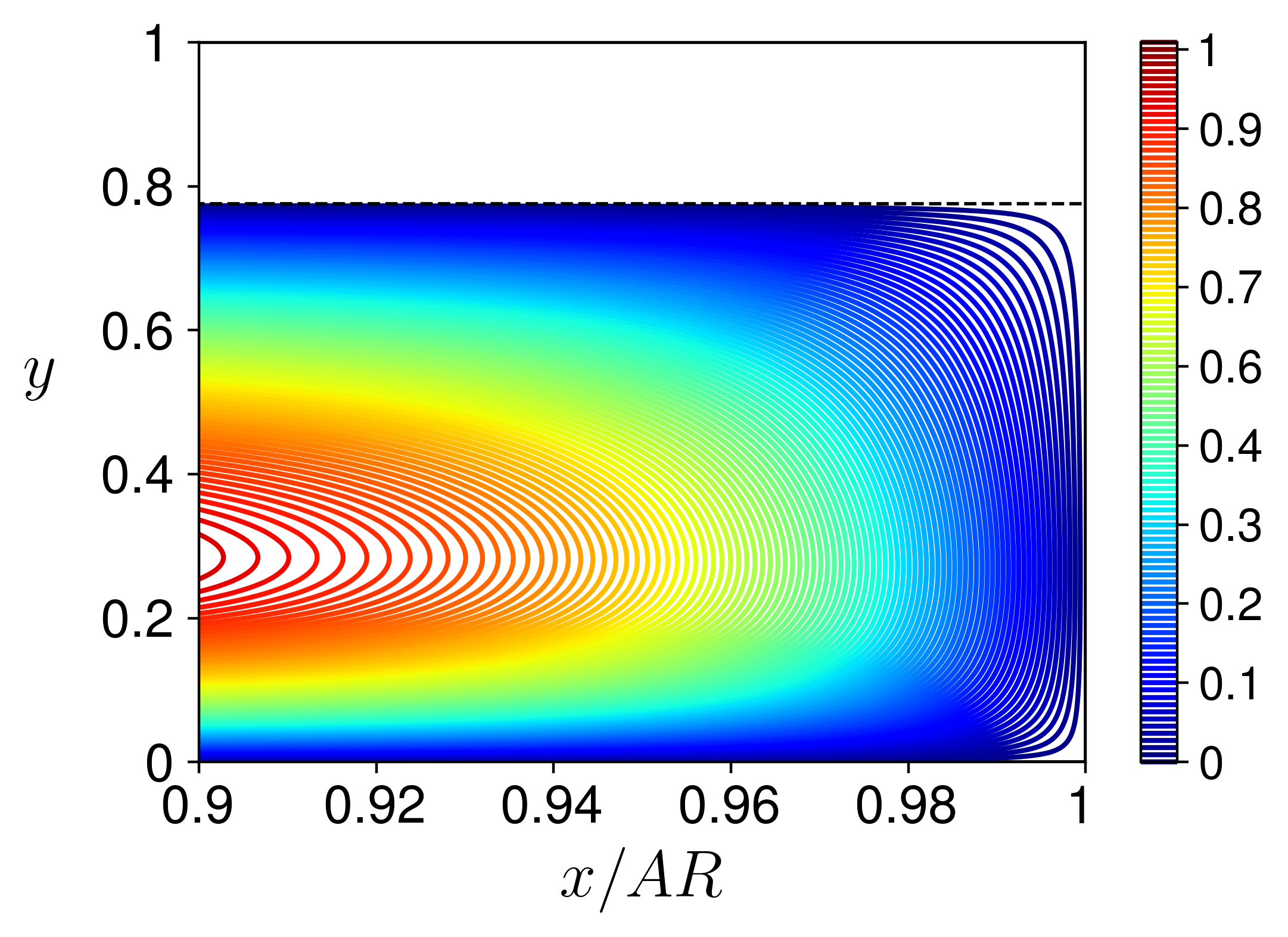}}
	\subfloat[$AR=100$, $b_\text{max}/\Ha=0.142$, $h=0.773$]{\includegraphics[width=0.32\textwidth,clip]{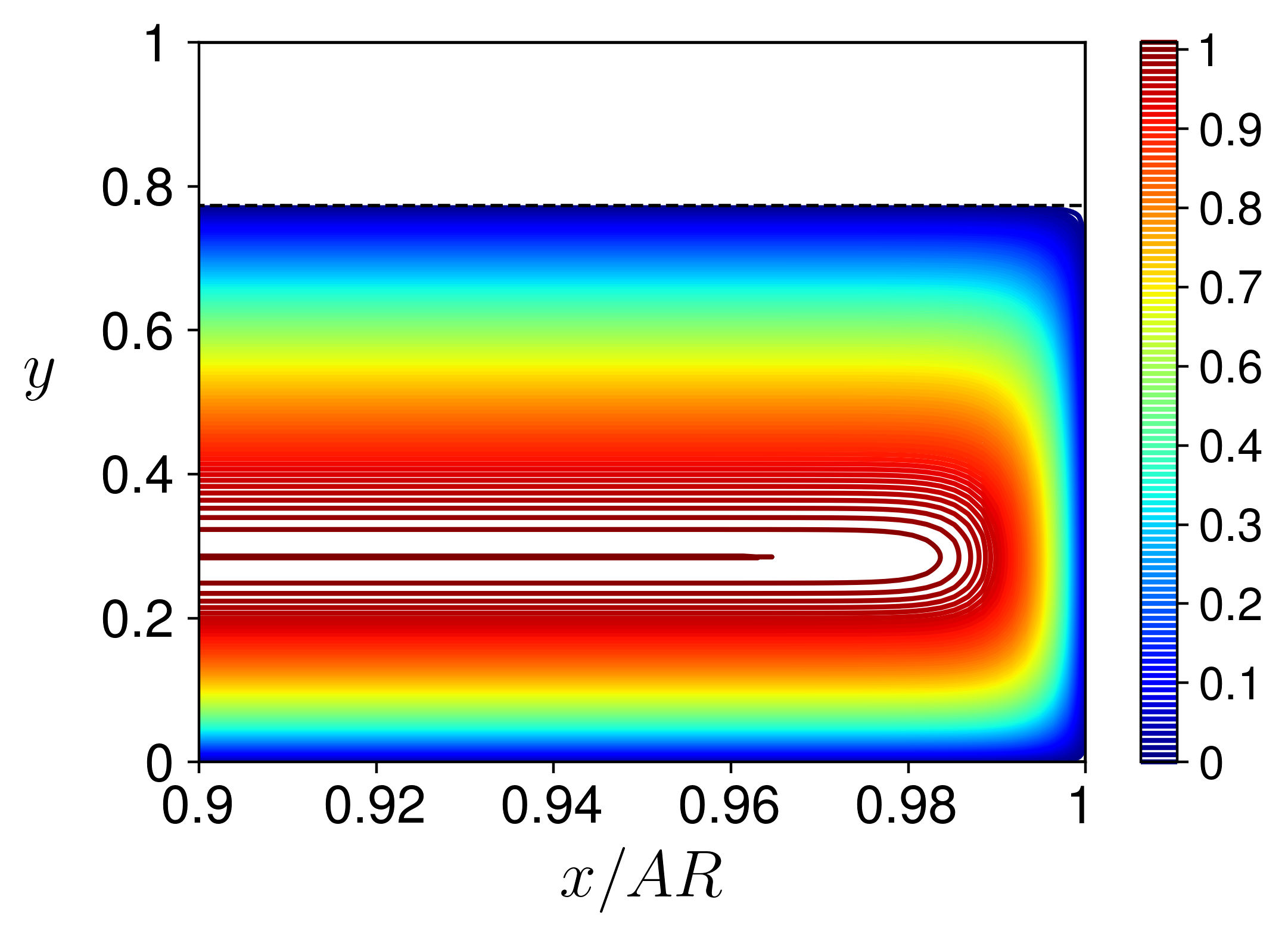}}
	\caption{\label{Fig: B_contours_Ha_5}Contours of $b/\Ha$ scaled by its maximum value, corresponding to the velocity contours shown in Fig.\ \ref{Fig: U_contours_Ha_5}.}	
\end{figure} 

\begin{figure}[!htb]
	\centering
	\subfloat[$b_\text{max}/\Ha=0.010$, $h=0.896$]{\includegraphics[width=0.32\textwidth,clip]{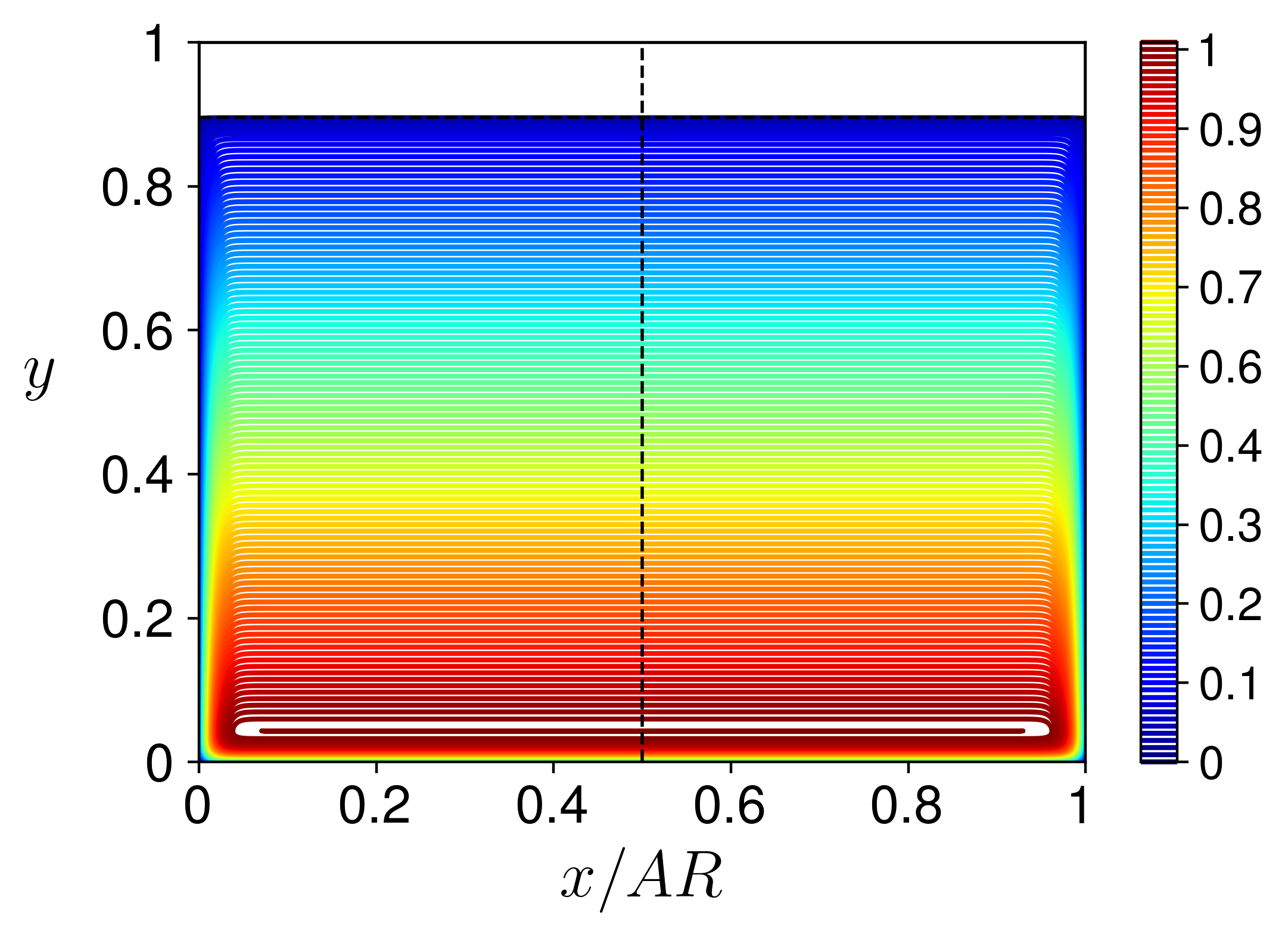}}
	\subfloat[$b_\text{max}/\Ha=1.000$, $h=0.976$]{\includegraphics[width=0.32\textwidth,clip]{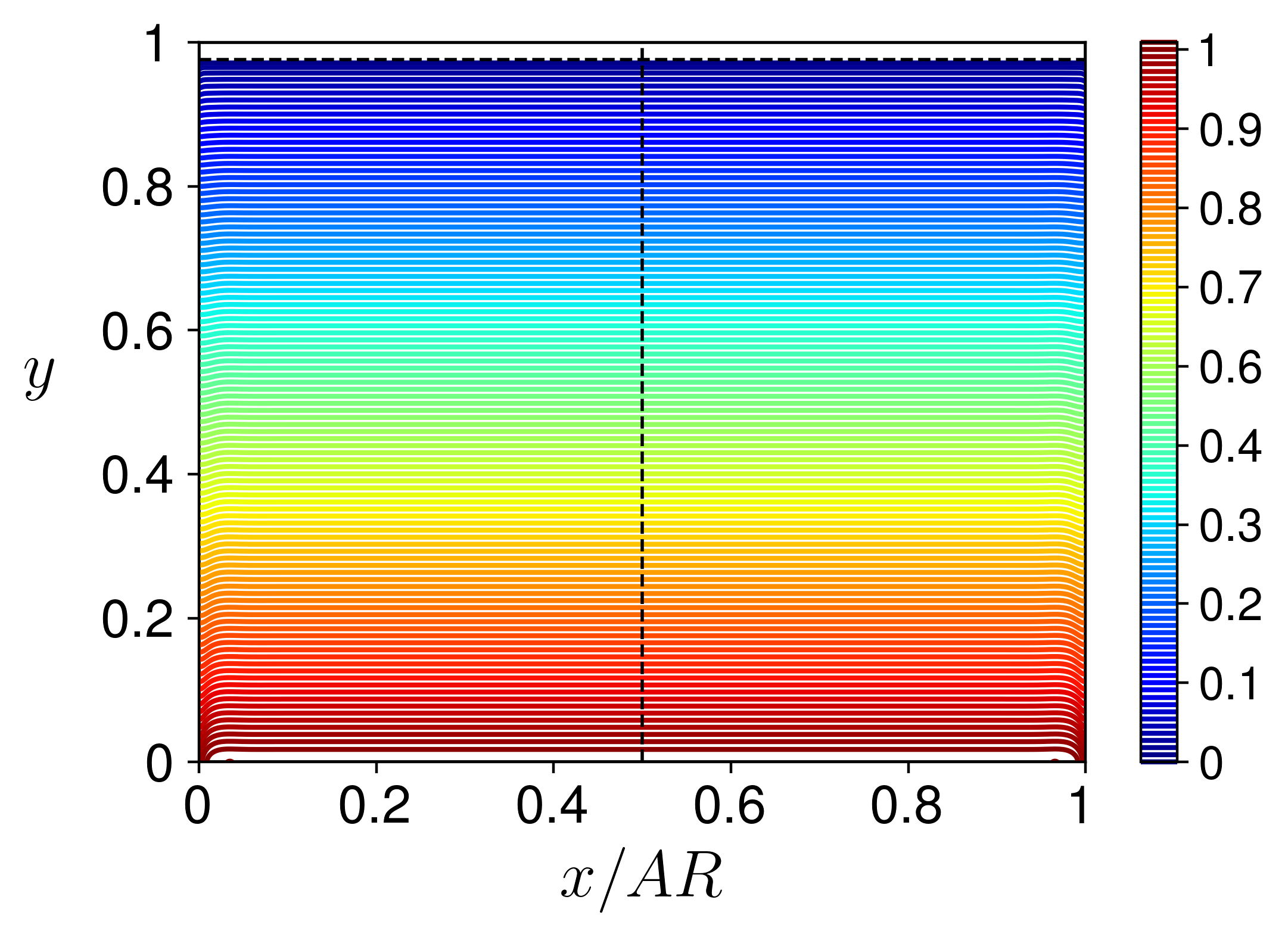}}
	\subfloat[$b_\text{max}/\Ha=0.665$, $h=0.973$]{\includegraphics[width=0.32\textwidth,clip]{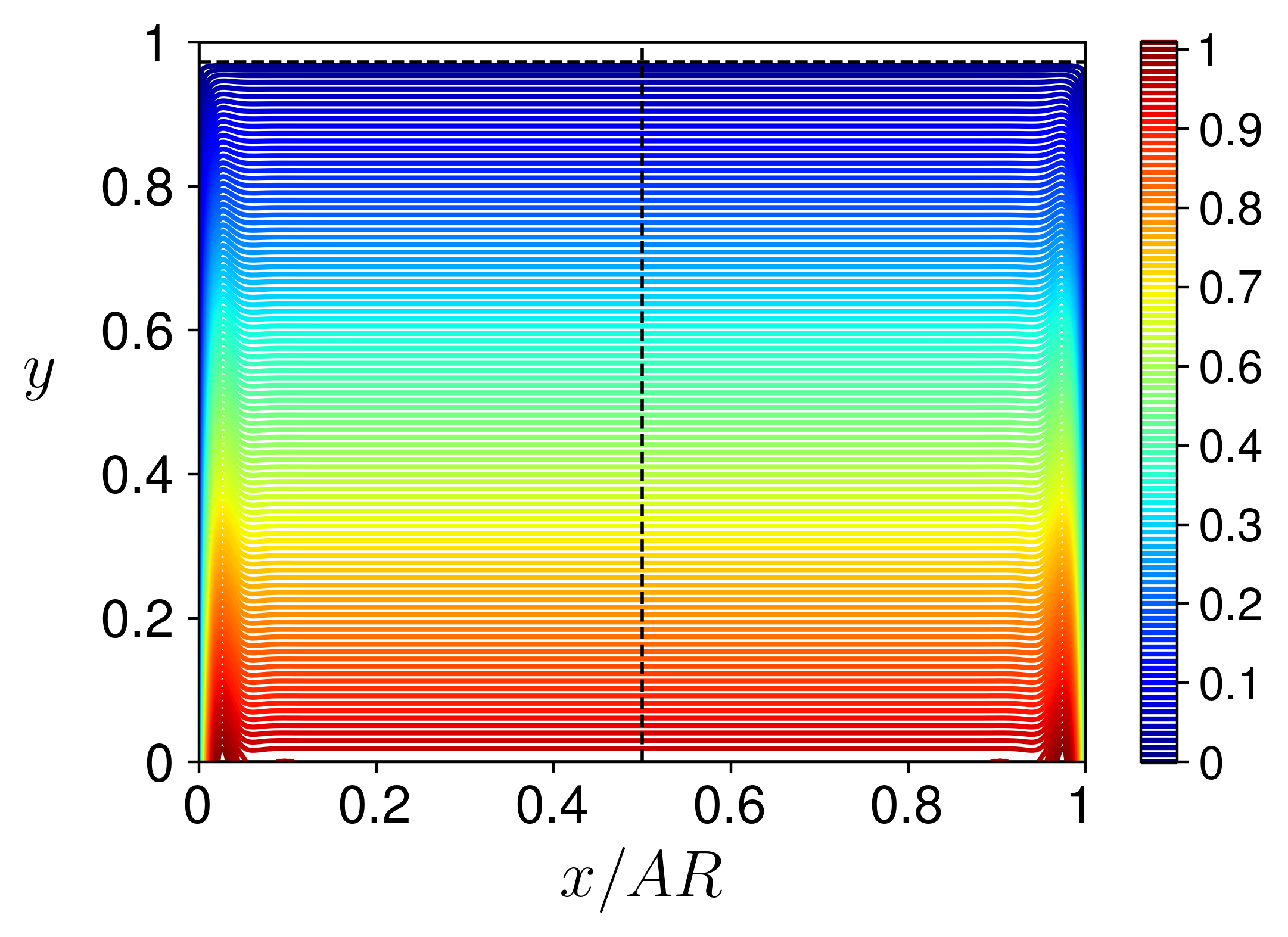}}
	\\
	\subfloat[$b_\text{max}/\Ha=0.010$, $h=0.896$]{\includegraphics[width=0.32\textwidth,clip]{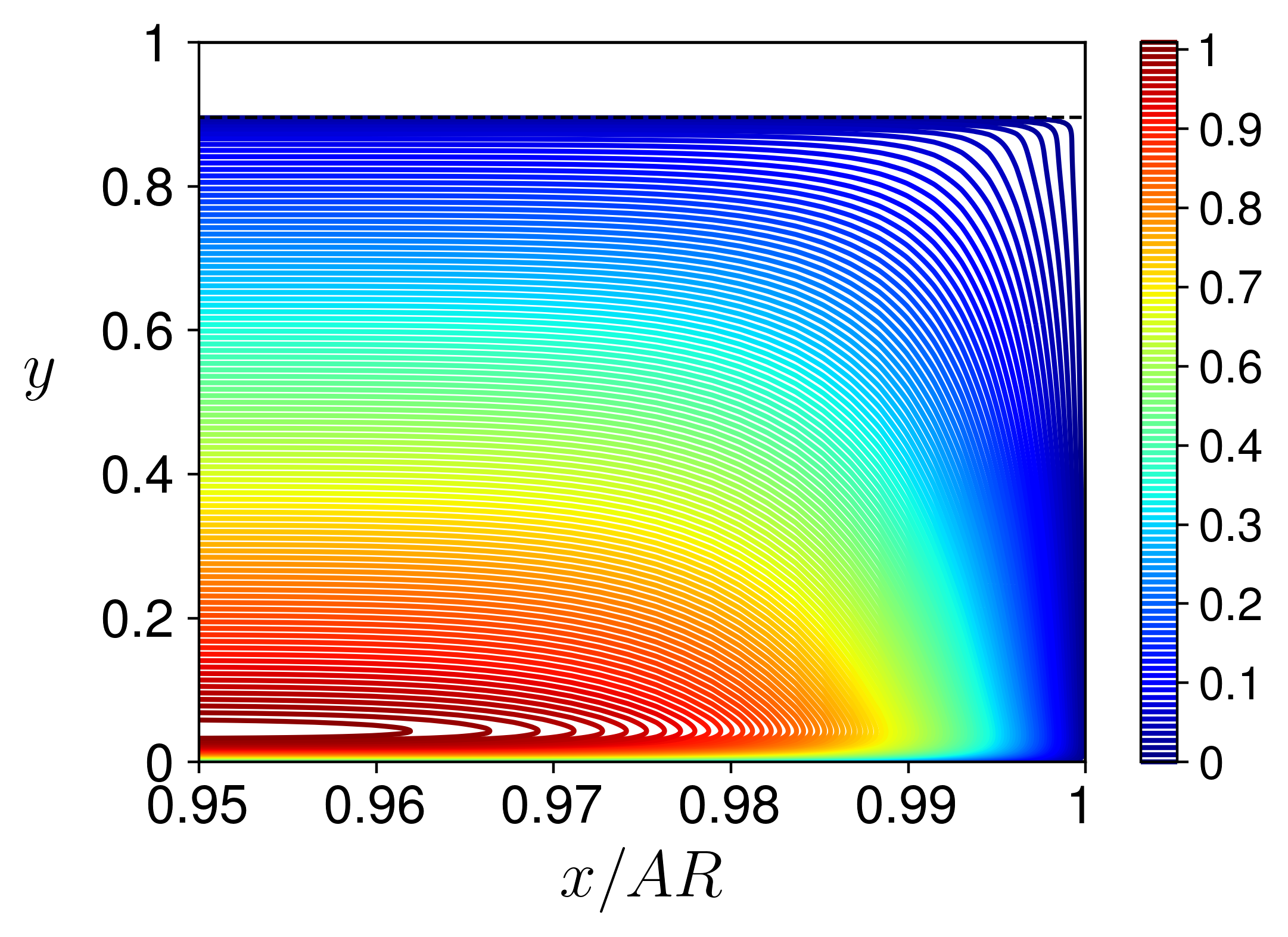}}
	\subfloat[$b_\text{max}/\Ha=1.000$, $h=0.976$]{\includegraphics[width=0.32\textwidth,clip]{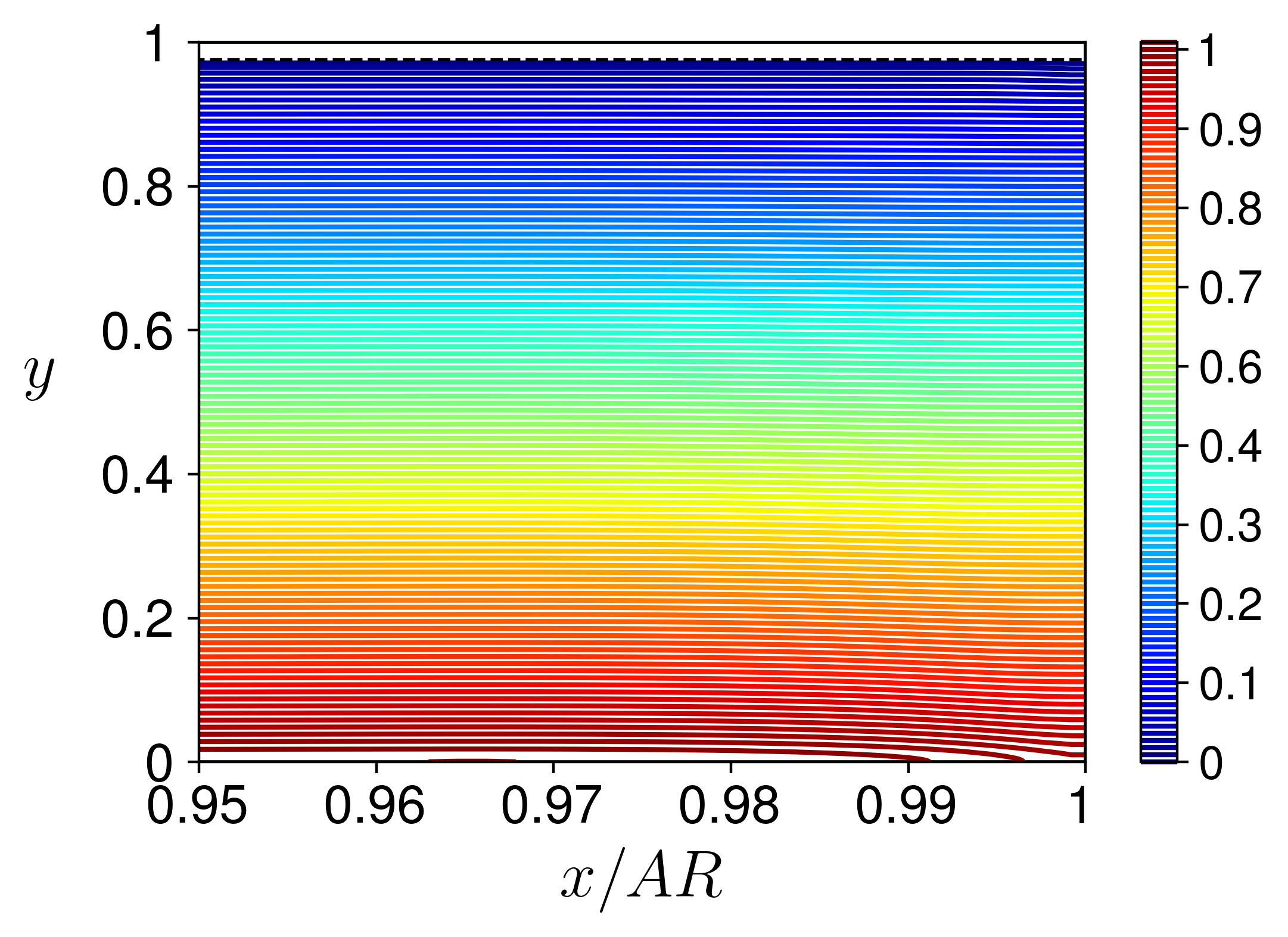}}
	\subfloat[$b_\text{max}/\Ha=0.665$, $h=0.973$]{\includegraphics[width=0.32\textwidth,clip]{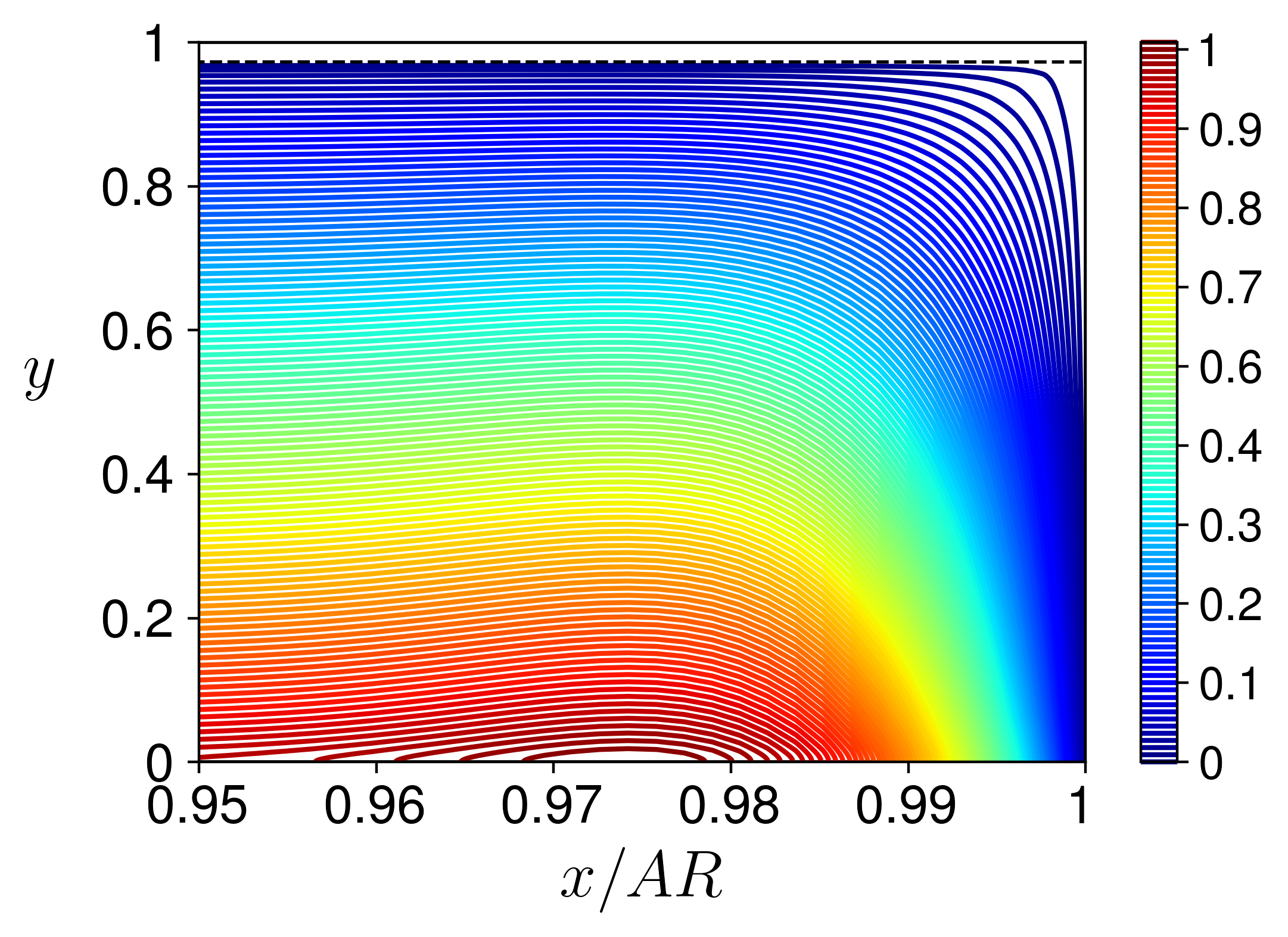}}
	\caption{\label{Fig: B_contours_Ha_103}Contours of $b/\Ha$ scaled by its maximum value, corresponding to the velocity contours shown in Fig.\ \ref{Fig: U_contours_Ha_103}.}	
\end{figure}

The isolines of the induced magnetic field, $b/\Ha=B_{ind}/\Rey_m$, corresponding to the velocity profiles shown in Fig.\ \ref{Fig: U_contours_Ha_5}, are depicted in Fig.\ \ref{Fig: B_contours_Ha_5}. These isolines are, in fact, the streamlines of the induced electrical current. Since the electrical current cannot penetrate the insulating walls or the interface, the currents induced in the bulk, and thus the $b$ isolines, are closed through the boundary layers (with $b=0$ on the walls and the interface). These contours resemble those obtained in single phase flow of a conductive fluid in a duct of a reduced height, as determined by the holdup. However, the location of maximum is not in the center of the conductive layer, but shifted towards the bottom wall. Similarly to the velocity field discussed above, the isolines of the induced magnetic field align with the bottom wall over most of the duct cross section, and the variation of b in the y-direction matches that predicted by the TP model. The maximal value of $\displaystyle \partial b/\partial y$ ($>0$) is reached at the bottom wall, while the minimal vertical gradient ($<0$) is observed at the interface. Near the side walls, on the other hand, $\displaystyle \partial b/\partial y$ is close to zero across the whole conducting layer. Hence, the Lorentz force, which facilitates the flow over most of the bottom wall, becomes ineffective near the side walls, and the viscous shear in the boundary layers is balanced by the pressure gradient. This region, however, becomes thinner with increasing $\Ha$ (compare Fig.\ \ref{Fig: B_contours_Ha_5}a with Fig.\ \ref{Fig: B_contours_Ha_103}a below).
\begin{figure}[h!]
	\centering
	\subfloat[$AR=10$, $\text{max}(U)=9.784$, $h=0.851$]{\includegraphics[width=0.32\textwidth,clip]{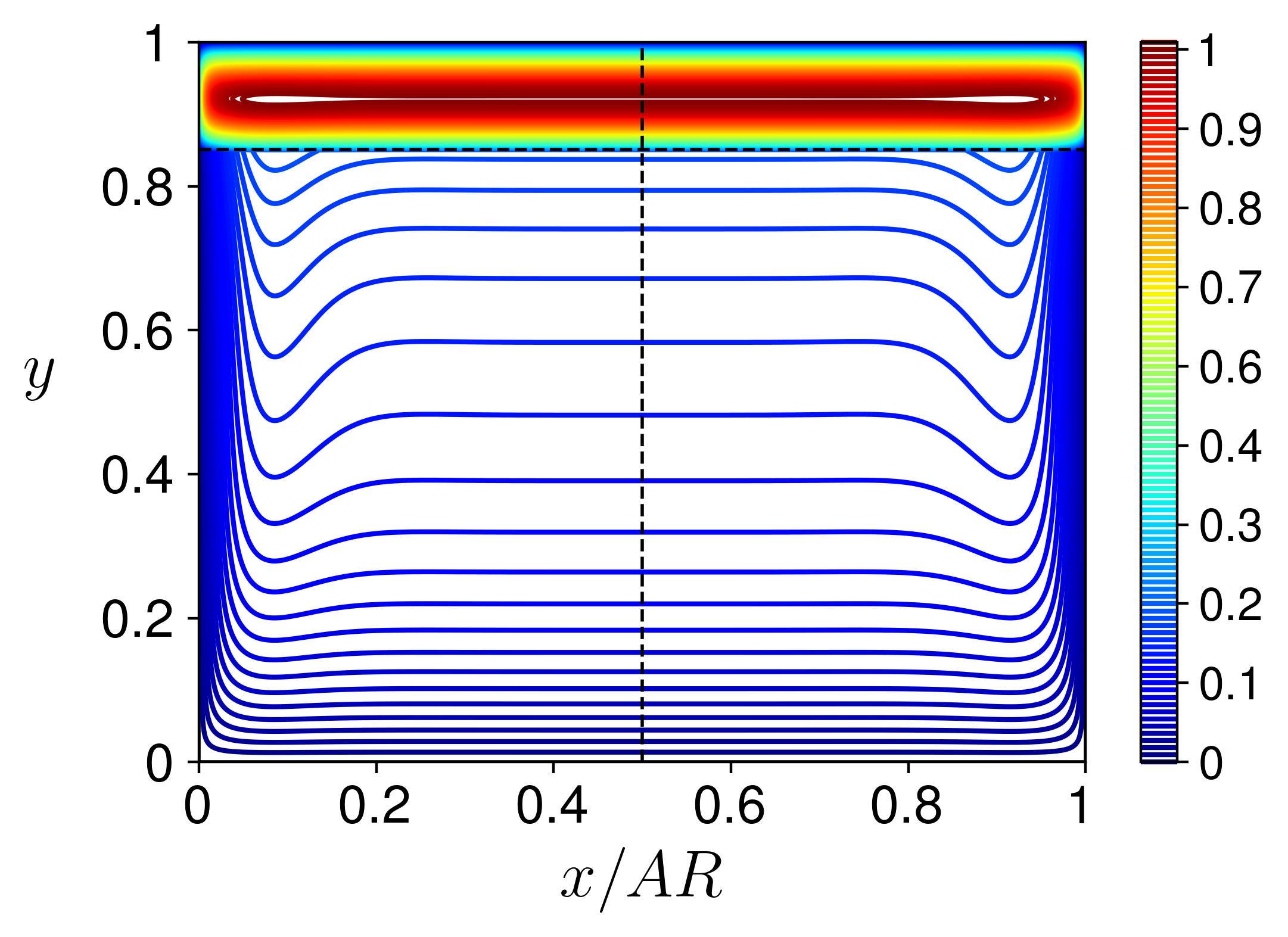}}
	\subfloat[$AR=10$]{\includegraphics[width=0.32\textwidth,clip]{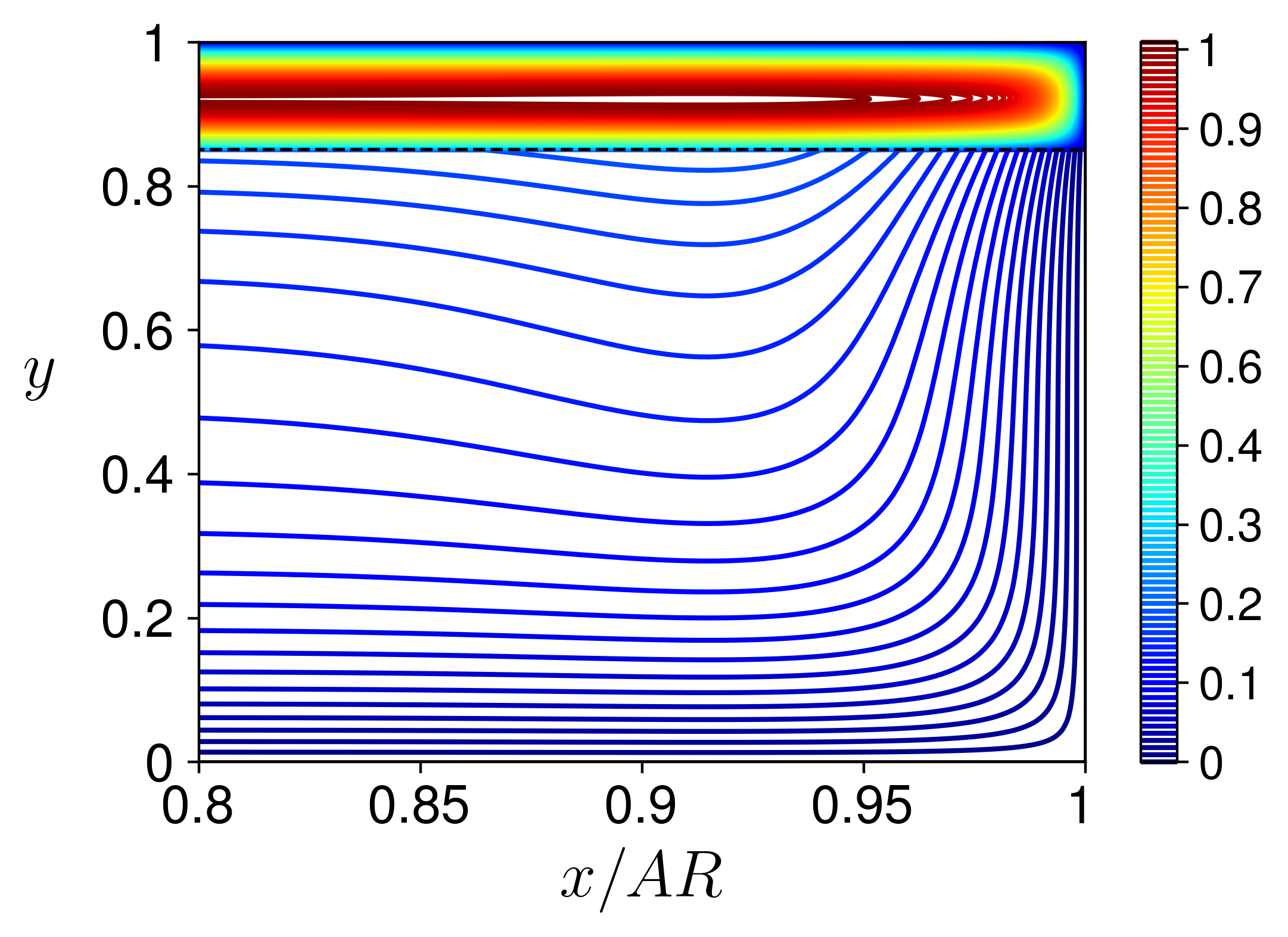}}
	\subfloat[$AR=100$, $\text{max}(U)=9.632$, $h=0.850$]{\includegraphics[width=0.32\textwidth,clip]{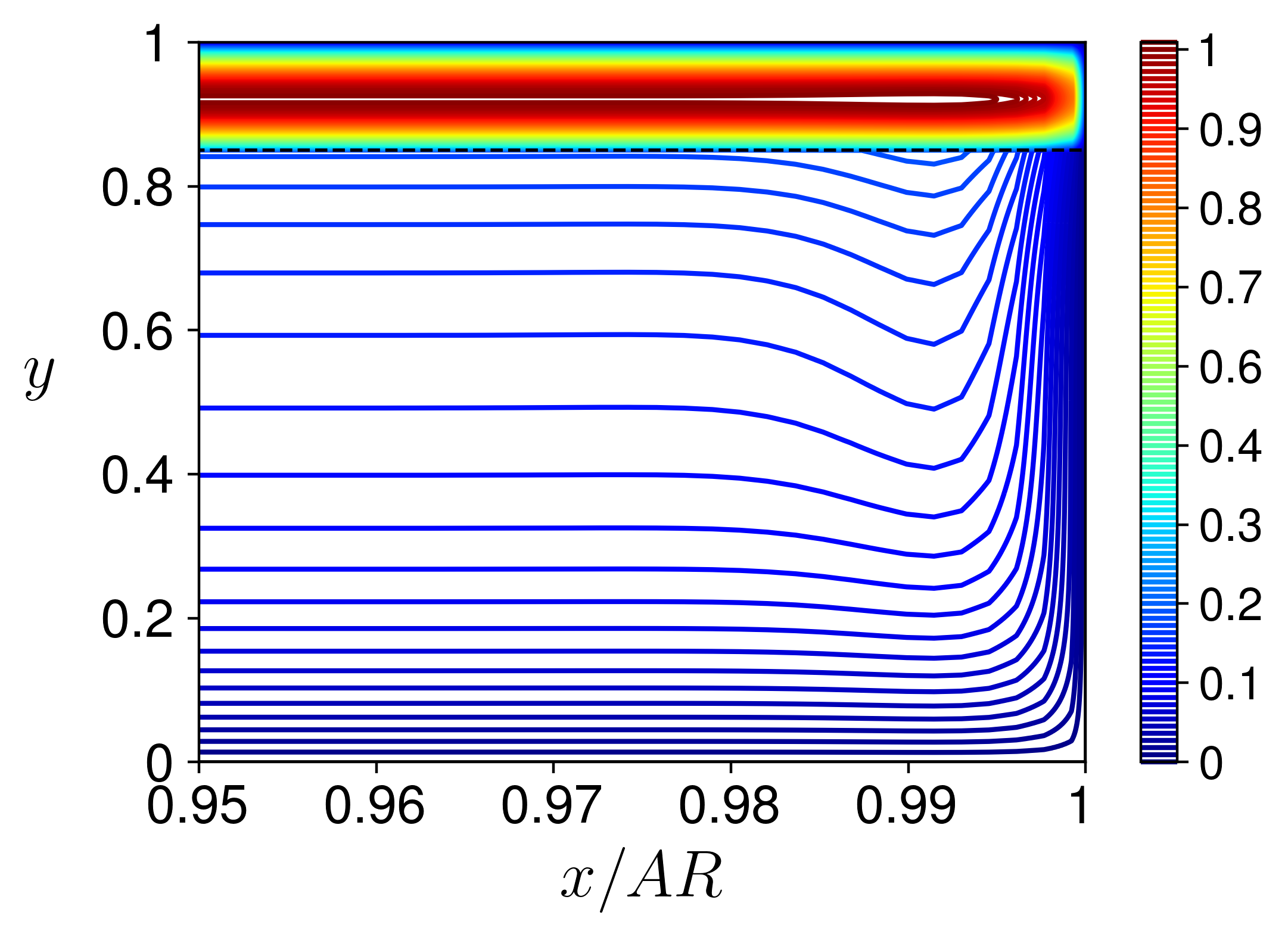}}
	\caption{\label{Fig: U_contours_Ha_5_cond}Contours of $U/\text{max}(U)$. Perfectly conducting walls, $Q_{2 1}=1$, $\Ha=5.181$. (a) The entire cross section; (b) and (c): enlargement of the near-side-wall region of ducts of $AR=10$ and $100$, respectively. The unperturbed interface is denoted by horizontal dashed black line; the cross-section centerline -- vertical dash-dot black line.}	
\end{figure} 
\begin{figure}[h!]
	\centering
	\subfloat[$AR=10$, $b_\text{max}/\Ha=1.028$, $h=0.851$]{\includegraphics[width=0.32\textwidth,clip]{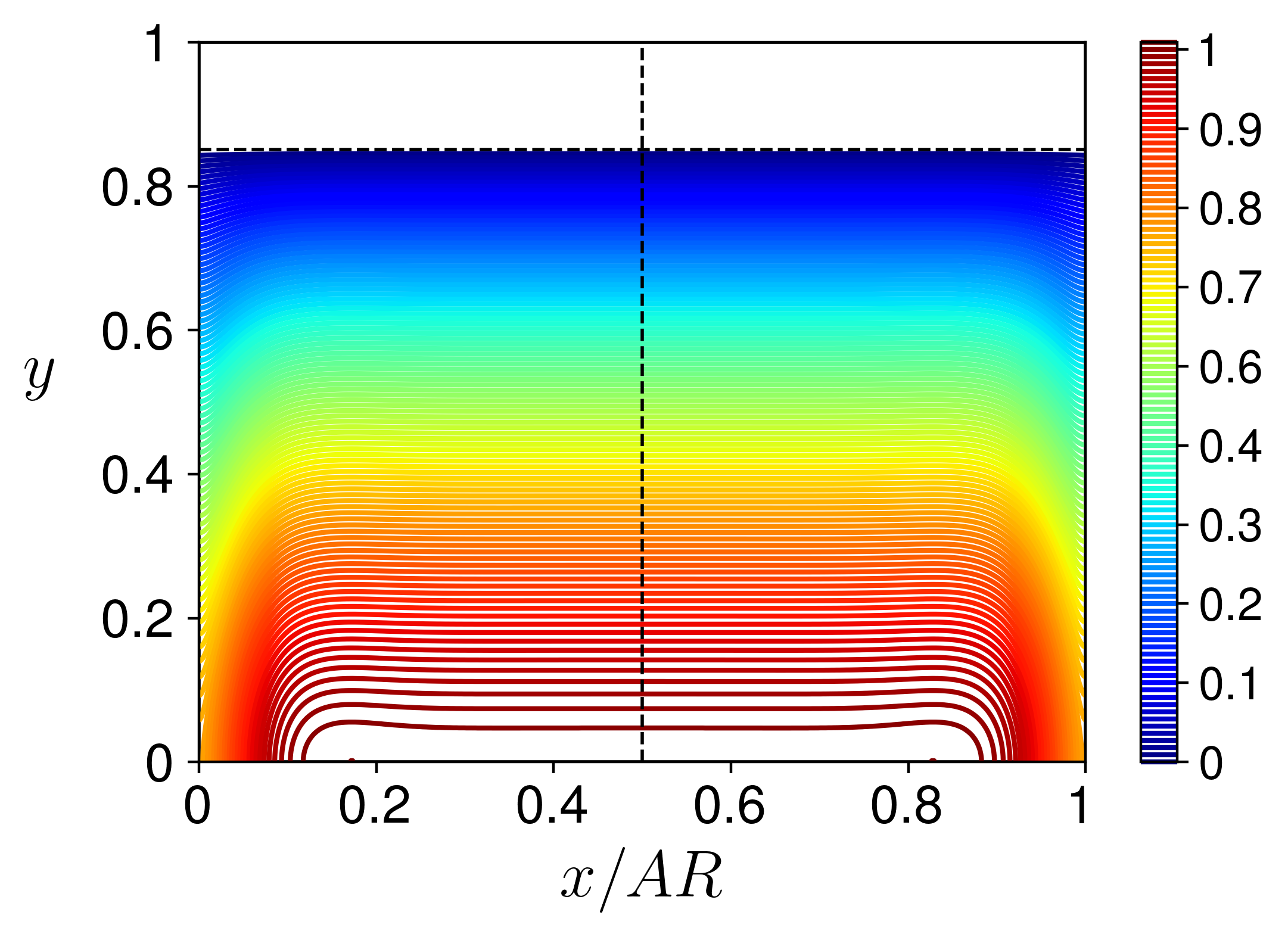}}
	\subfloat[$AR=10$]{\includegraphics[width=0.32\textwidth,clip]{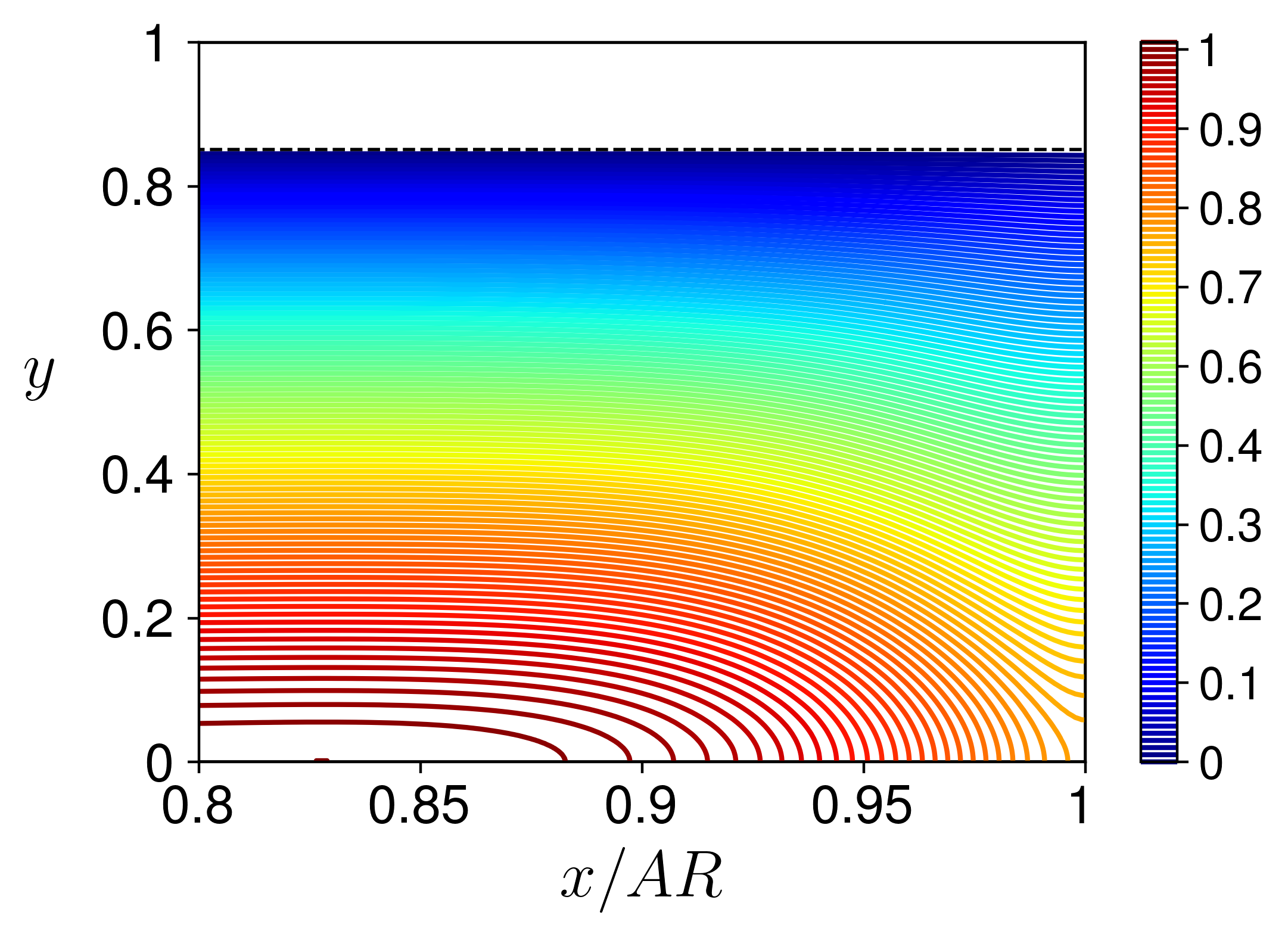}}
	\subfloat[$AR=100$,$b_\text{max}/\Ha=1.004$, $h=0.850$]{\includegraphics[width=0.32\textwidth,clip]{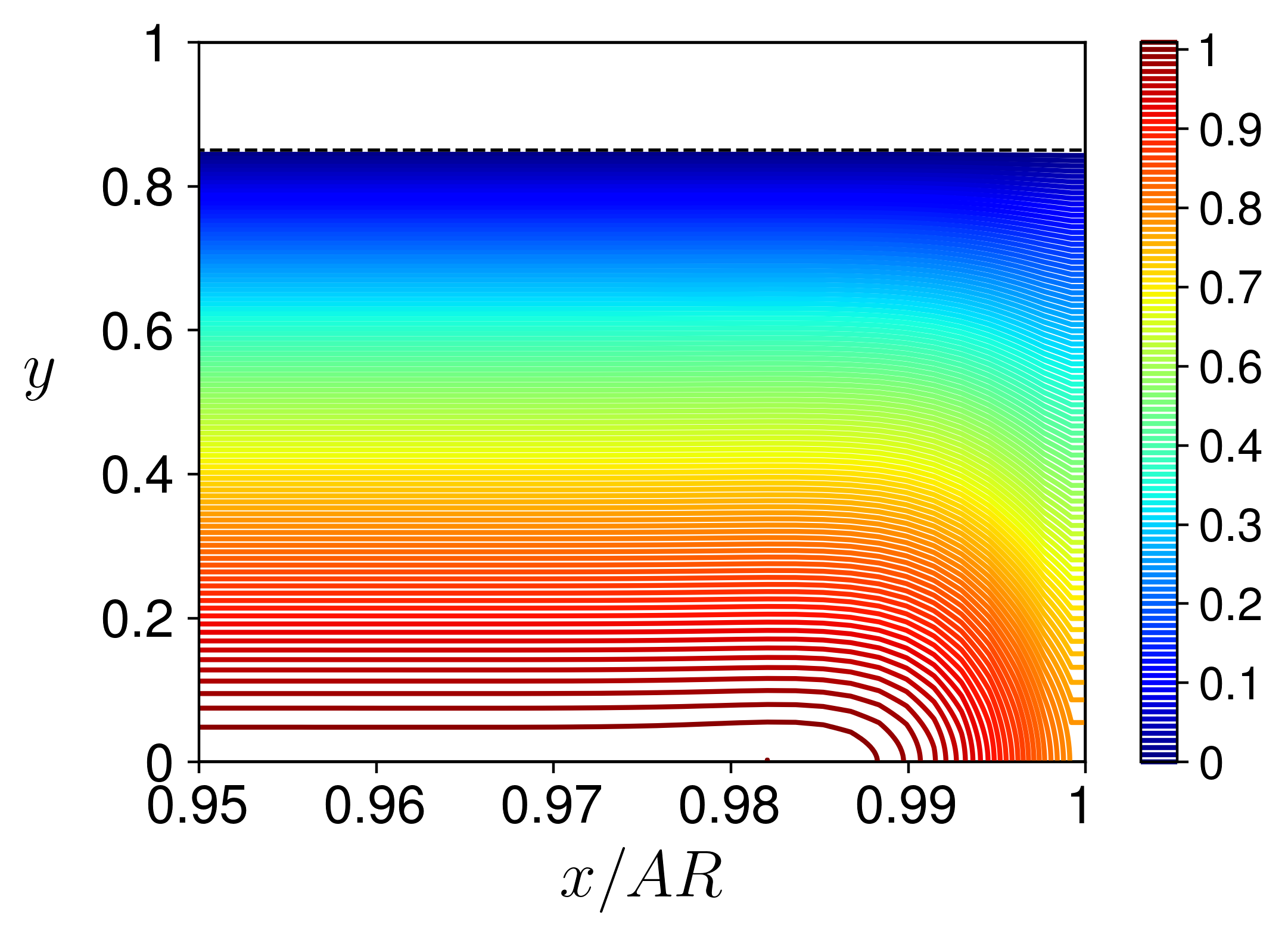}}
	\caption{\label{Fig: B_contours_Ha_5_cond}Contours of $b/\Ha$ scaled by its maximum value, corresponding to the velocity contours shown in Fig.\ \ref{Fig: U_contours_Ha_5_cond}.}	
\end{figure} 

The velocity and induced magnetic field contours for wide rectangular ducts ($AR=10$ and $100$) with perfectly conducting walls are shown in Figs.\ \ref{Fig: U_contours_Ha_5_cond} and \ref{Fig: B_contours_Ha_5_cond}, respectively. The flow rate ratio and the Hartmann number are the same as in the previous contour plots. As the currents induced in the conductive fluid cannot penetrate the interface, they are closed through the side and bottom walls, where the resistance to the current flow is zero. Thereby the isolines of the induced magnetic field are perpendicular to the wall (Fig.\ \ref{Fig: B_contours_Ha_5_cond}). The pattern of these isolines resembles that obtained in single phase flow of the conductive fluid in the lower half of a duct with perfectly conducting walls, where, due to symmetry, no currents can cross the duct (horizontal) centerline and $b=0$ as at the mercury--air interface. As shown in Fig.\ \ref{Fig: B_contours_Ha_5_cond}, the $b$ isolines are nearly parallel to the bottom wall, except near its corners formed with the side walls. The combined effects of the side walls, as reflected by increasing $AR$ and $\Ha$ on the boundary layers and the velocity profiles in the conductive fluid core, are similar to those in the case of the insulating walls (compare Fig.\ \ref{Fig: U_contours_Ha_5_cond}b with Fig.\ \ref{Fig: U_contours_Ha_5_cond}c and with Fig.\ \ref{Fig: U_contours_Ha_103}e). However, a noticeable difference is the regions of higher velocity observed near the side walls, where the local velocity exceeds the velocity in the core. This is clearly observed in Fig.\ \ref{Fig: U_i_duct}b, where the spanwise variation of the interface velocity normalized by the corresponding TP model value is shown for different $\Ha$. The velocity peak is located approximately on the verge of the Shercliff B. L., hence it approaches the side wall at high $\Ha$. As seen in the figure, this (normalized) peak value is not very sensitive to the Hartmann number and exceeds the centerline interfacial velocity by about $10\%$. The higher velocity near the side walls is a result of weaker (negative) $\displaystyle \partial b/\partial y$, hence a reduced retarding Lorentz force in this region, which enables higher local velocity under the pressure gradient set in the flow. In the case of perfectly conducting side walls, those localized small velocity peak does not significantly affect the core velocity and the pressure gradient required to drive the specified flow rate of the conductive fluid. This effect is, however, much more pronounced in the case of insulating side walls (e.g., Fig.\ \ref{Fig: U1_vs_x_bot_cond}), which is discussed below. 
\begin{figure}[h!]
	\centering
	\subfloat[$AR=10$, $\text{max}(U)=9.807$, $h=0.847$]{\includegraphics[width=0.32\textwidth,clip]{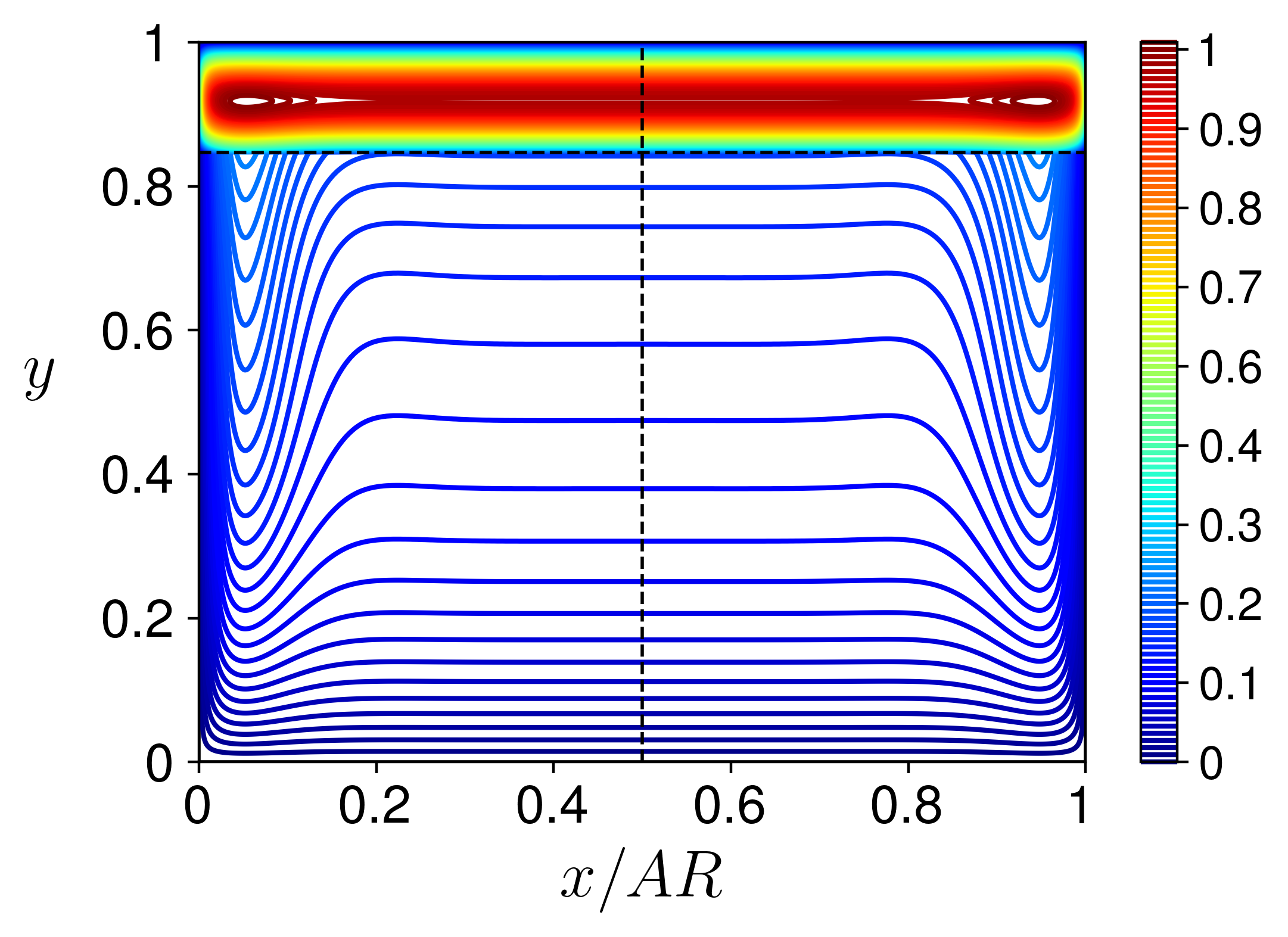}}
	\subfloat[$AR=10$]{\includegraphics[width=0.32\textwidth,clip]{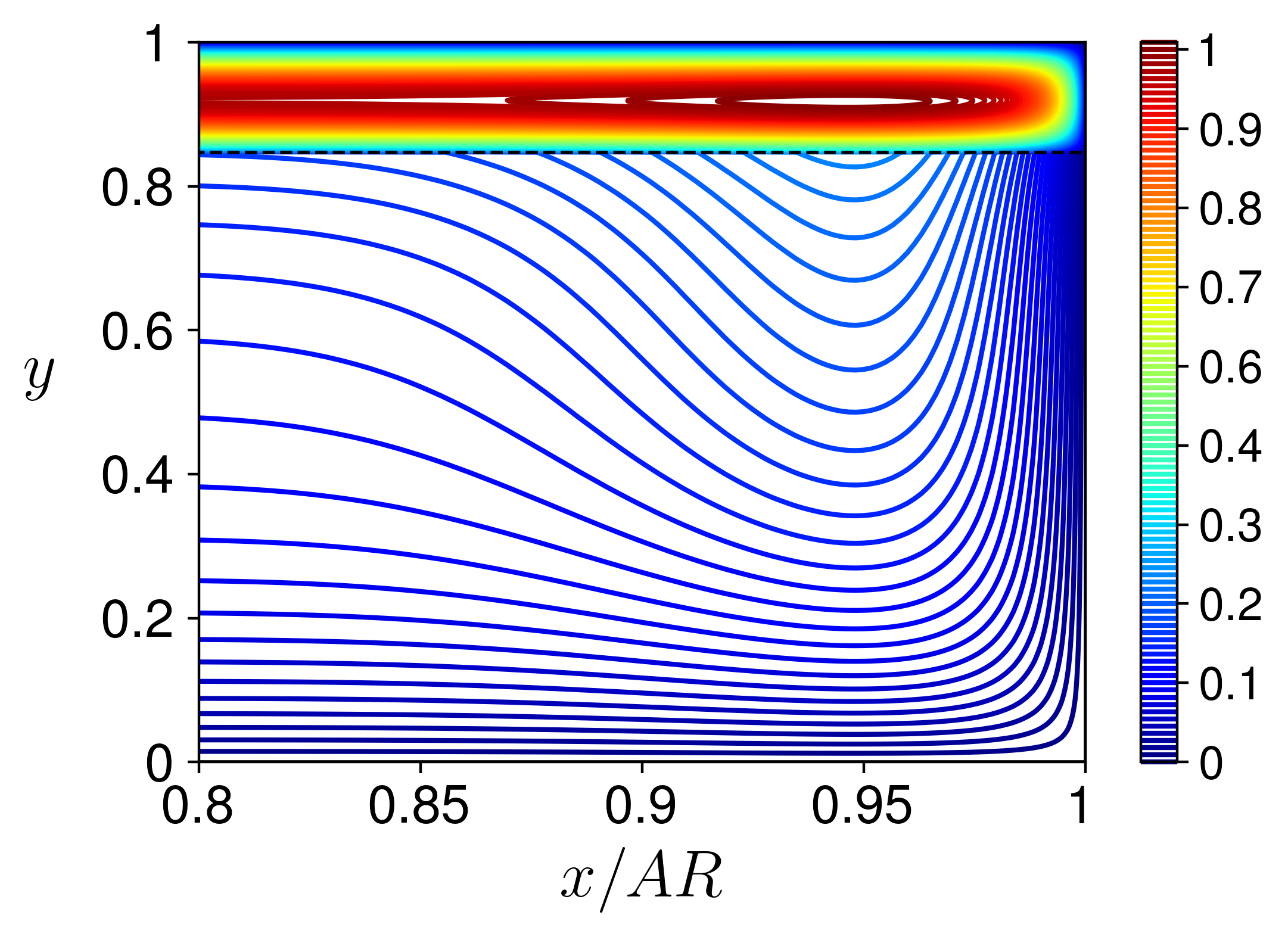}}
	\subfloat[$AR=100$, $\text{max}(U)=9.928$, $h=0.849$]{\includegraphics[width=0.32\textwidth,clip]{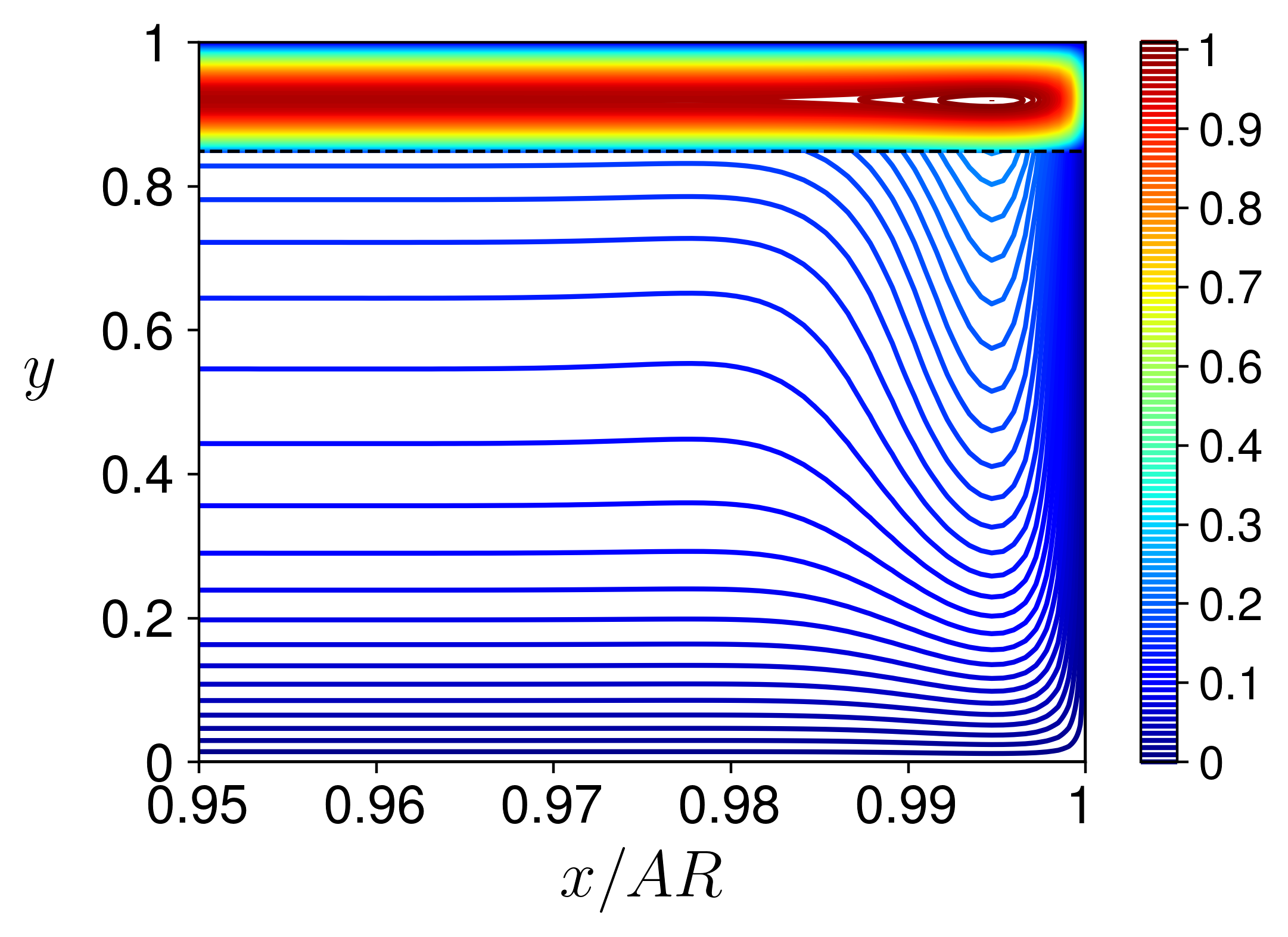}}
	\caption{\label{Fig: U_contours_Ha_5_bot_cond}Contours of $U/\text{max}(U)$. Perfectly conducting bottom wall and insulating side walls, $Q_{2 1}=1$, $\Ha=5.181$. (a) The entire cross section; (b) and (c): enlargement of the near-side-wall region of ducts of $AR=10$ and $100$, respectively. The unperturbed interface is denoted by horizontal dashed black line; the cross-section centerline -- vertical dash-dot black line.}	
\end{figure} 
\begin{figure}[h!]
	\centering
	\subfloat[$AR=10$, $b_\text{max}/\Ha=0.970$, $h=0.847$]{\includegraphics[width=0.32\textwidth,clip]{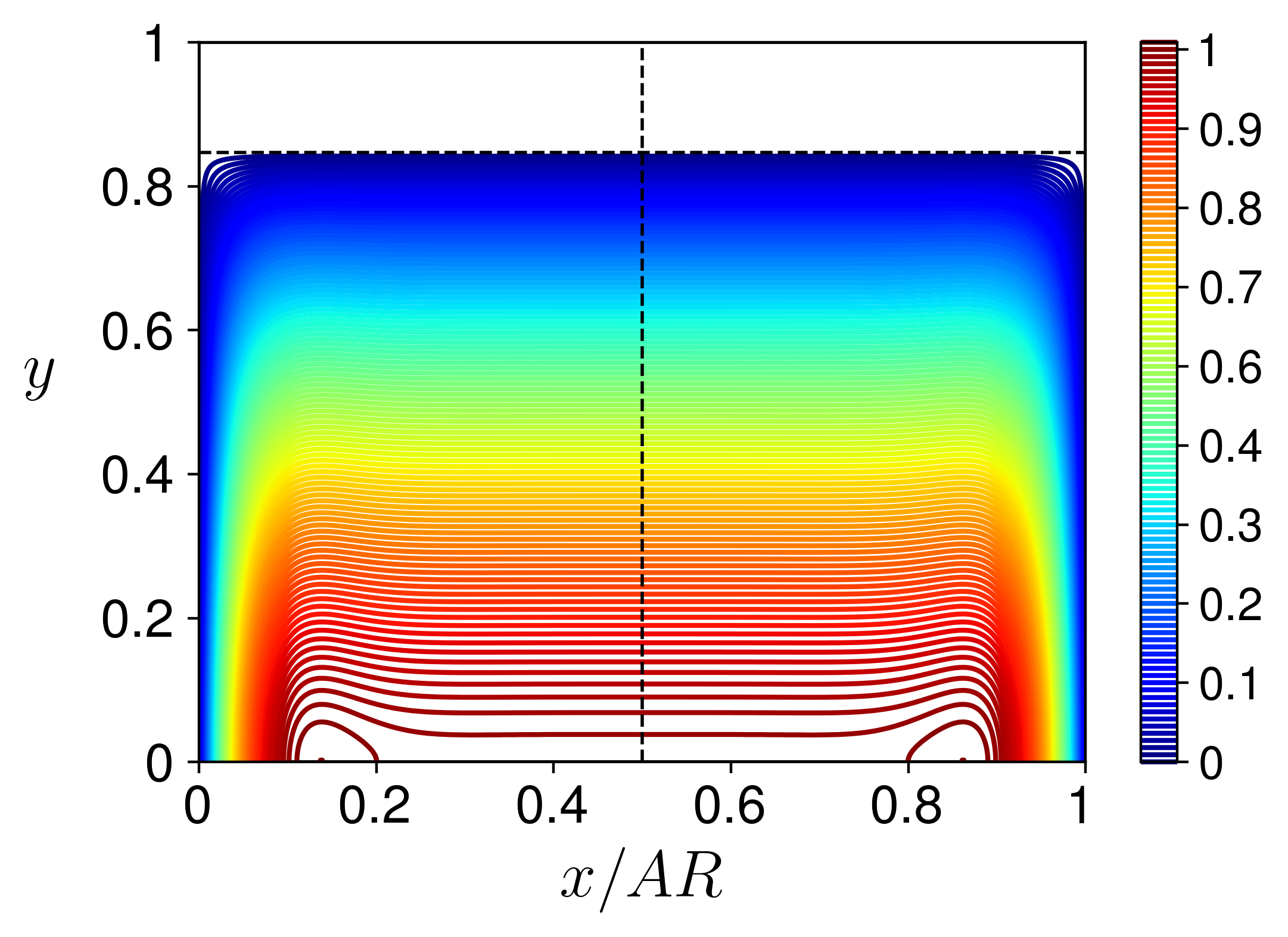}}
	\subfloat[$AR=10$]{\includegraphics[width=0.32\textwidth,clip]{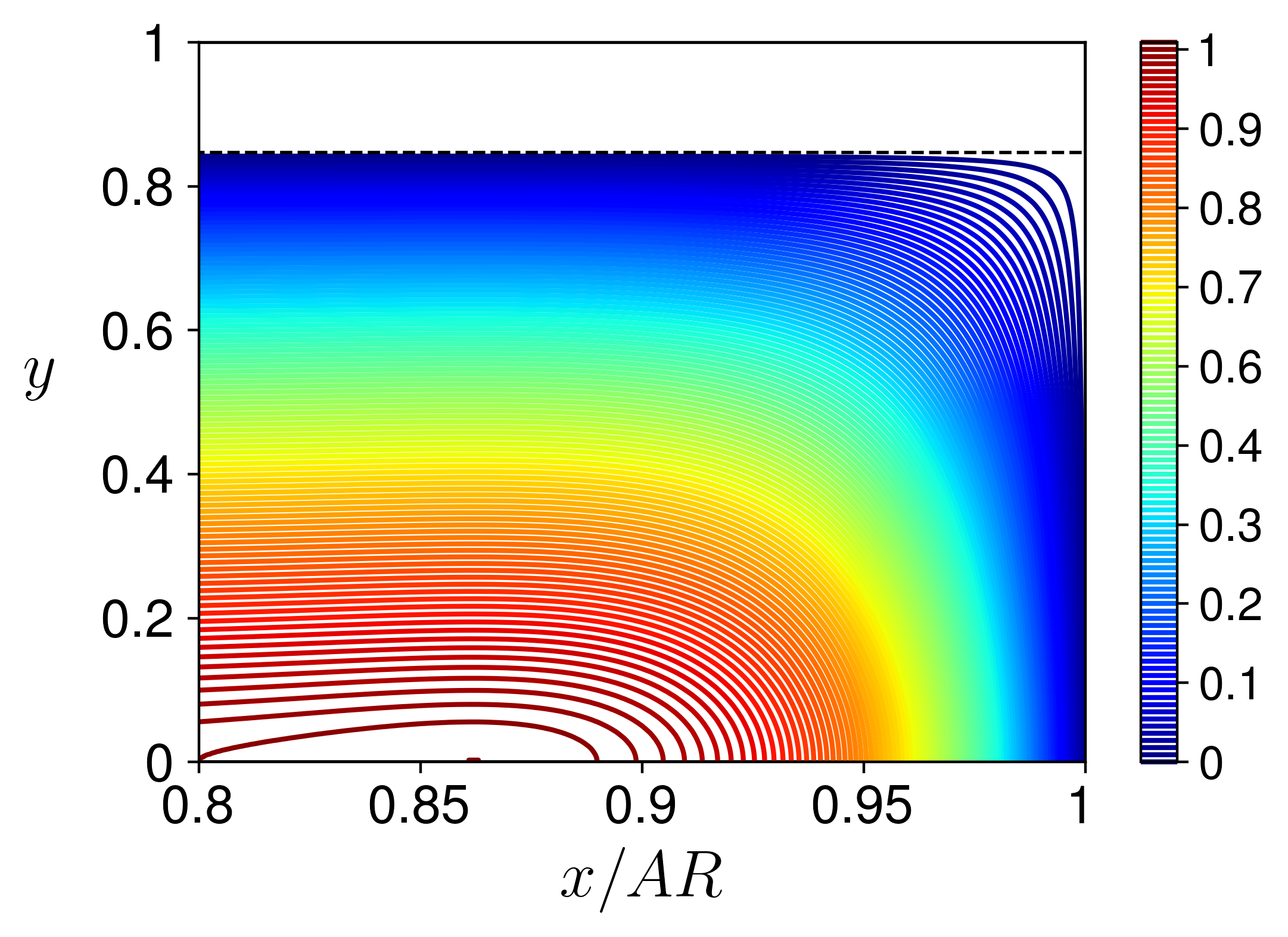}}
	\subfloat[$AR=100$,$b_\text{max}/\Ha=1.008$, $h=0.849$]{\includegraphics[width=0.32\textwidth,clip]{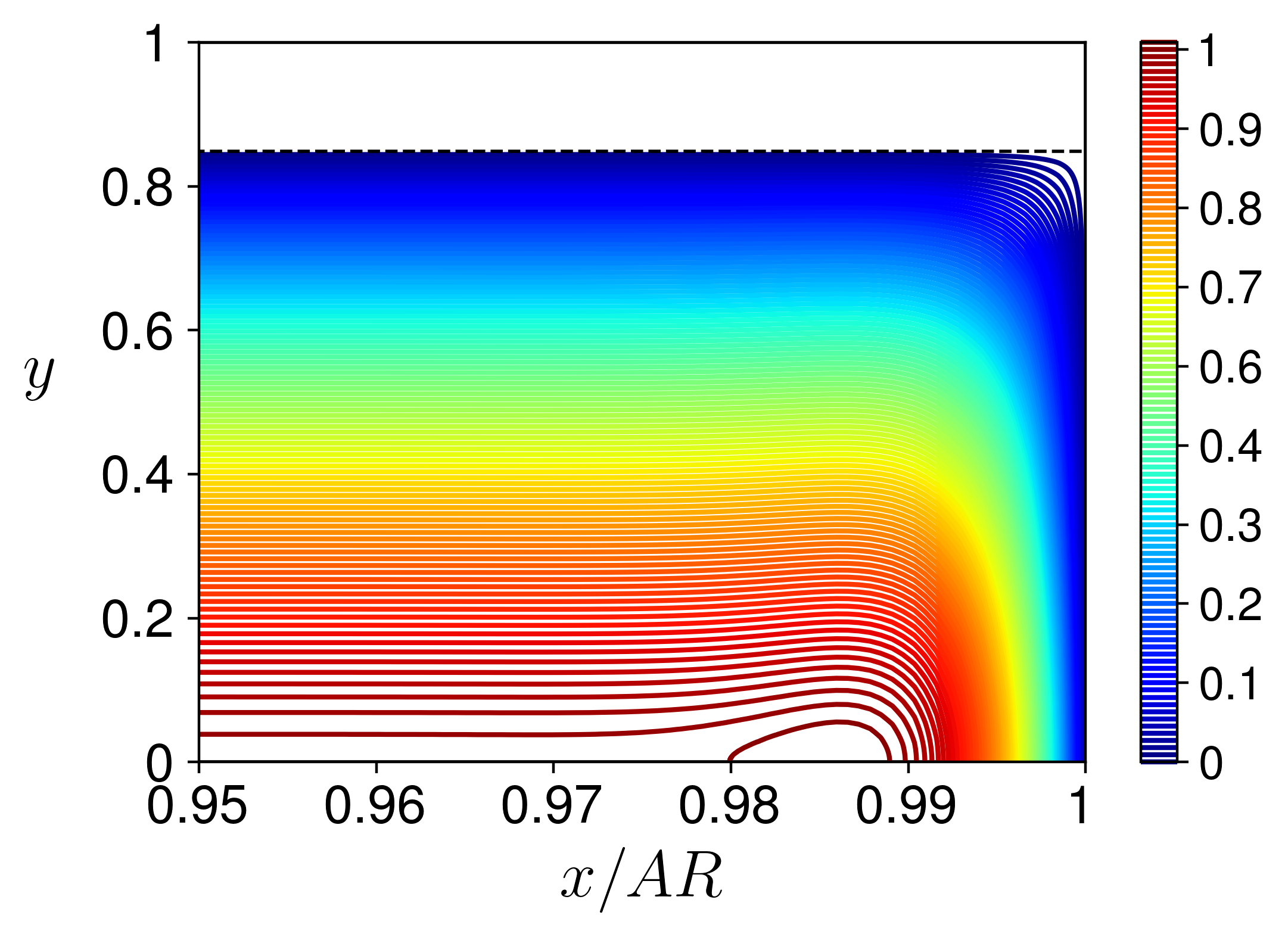}}
	\caption{\label{Fig: B_contours_Ha_5_bot_cond}Contours of $b/\Ha$ scaled by its maximum value, corresponding to the velocity contours shown in Fig.\ \ref{Fig: U_contours_Ha_5_bot_cond}.}	
\end{figure} 
\begin{figure}[h!]
	\centering
	\subfloat[$\Ha=5.181$, $h=0.847$]{\includegraphics[width=0.48\textwidth,clip]{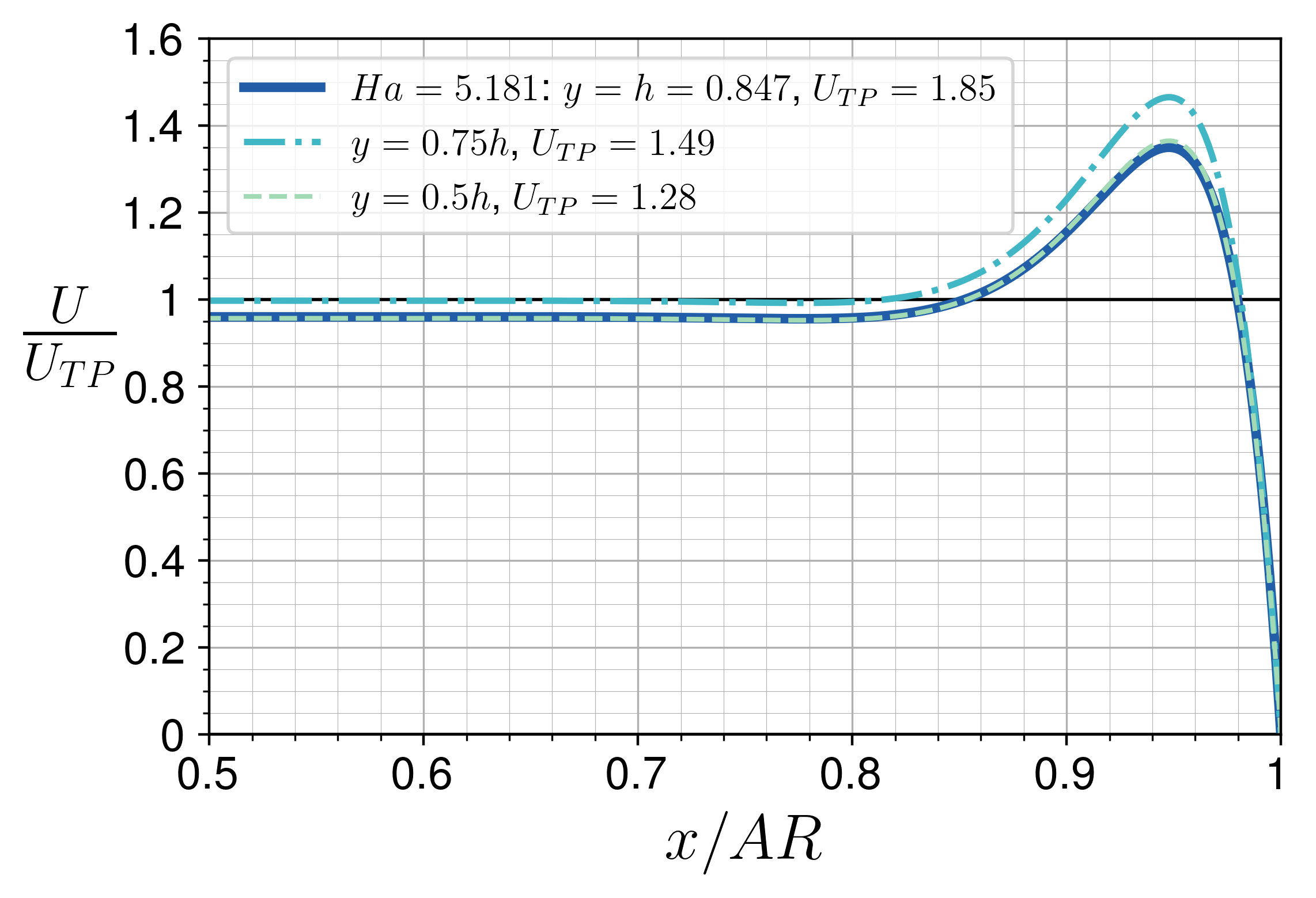}}
	\subfloat[$\Ha=103.62$, $h=0.973$]{\includegraphics[width=0.48\textwidth,clip]{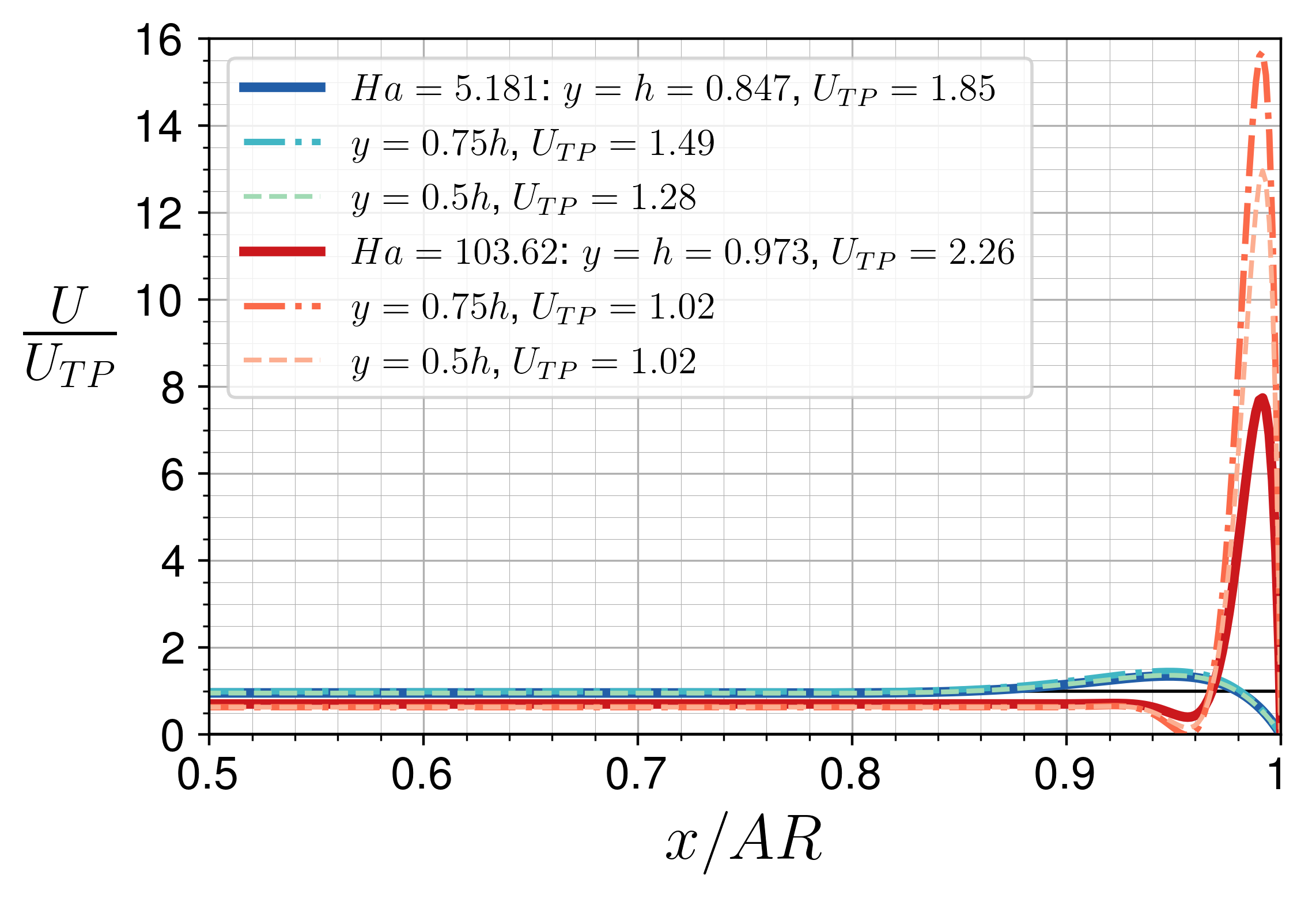}}
	\caption{\label{Fig: U1_vs_x_bot_cond}Spanwise velocity profiles at the interface and in the conducting phase. Perfectly conducting bottom wall and insulating side walls, $Q_{2 1}=1$, $AR = 10$. The corresponding velocity contour plots are shown in Fig.\ \ref{Fig: U_contours_Ha_5_bot_cond} and Fig.\ \ref{Fig: U_contours_Ha_103} for $\Ha=5.181$ and $\Ha=103.62$, respectively.}	
\end{figure} 

Two additional cases of wall conductivities can be considered. The first case is when the bottom wall is insulating, while the side walls are perfectly conducting. The MHD flow behavior in this case is, in fact, very similar to the case of all insulating walls with the only difference being in the induced magnetic field behavior near the side walls. Its isolines are perpendicular to the perfectly conducting side wall, instead of being parallel (currents are closed through the mercury B. L. in the case of insulating wall). This, however, almost does not affect the velocity profiles as well as the holdup and pressure gradient values for $AR\in[10,1000]$ as already demonstrated above (see discussion with reference to Figs.\ \ref{Fig: AR_convergence}(a)-(d)). Therefore, we do not show the contour plots for this case.

\begin{figure}[!htb]
	\centering
	\subfloat[$\Ha=5.181$: $U$]{\includegraphics[width=0.25\textwidth,clip]{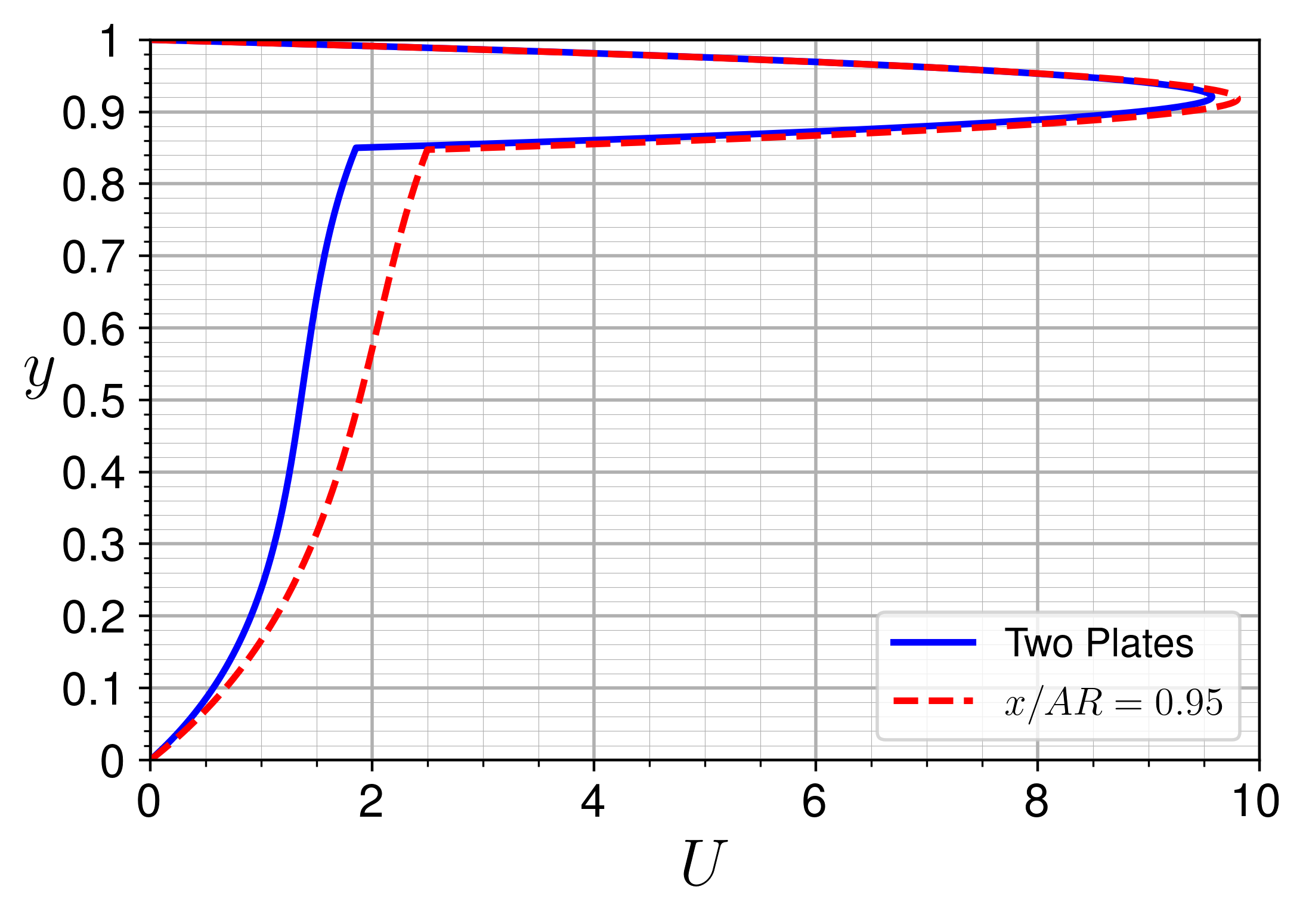}}
	\subfloat[$b/\Ha$]{\includegraphics[width=0.25\textwidth,clip]{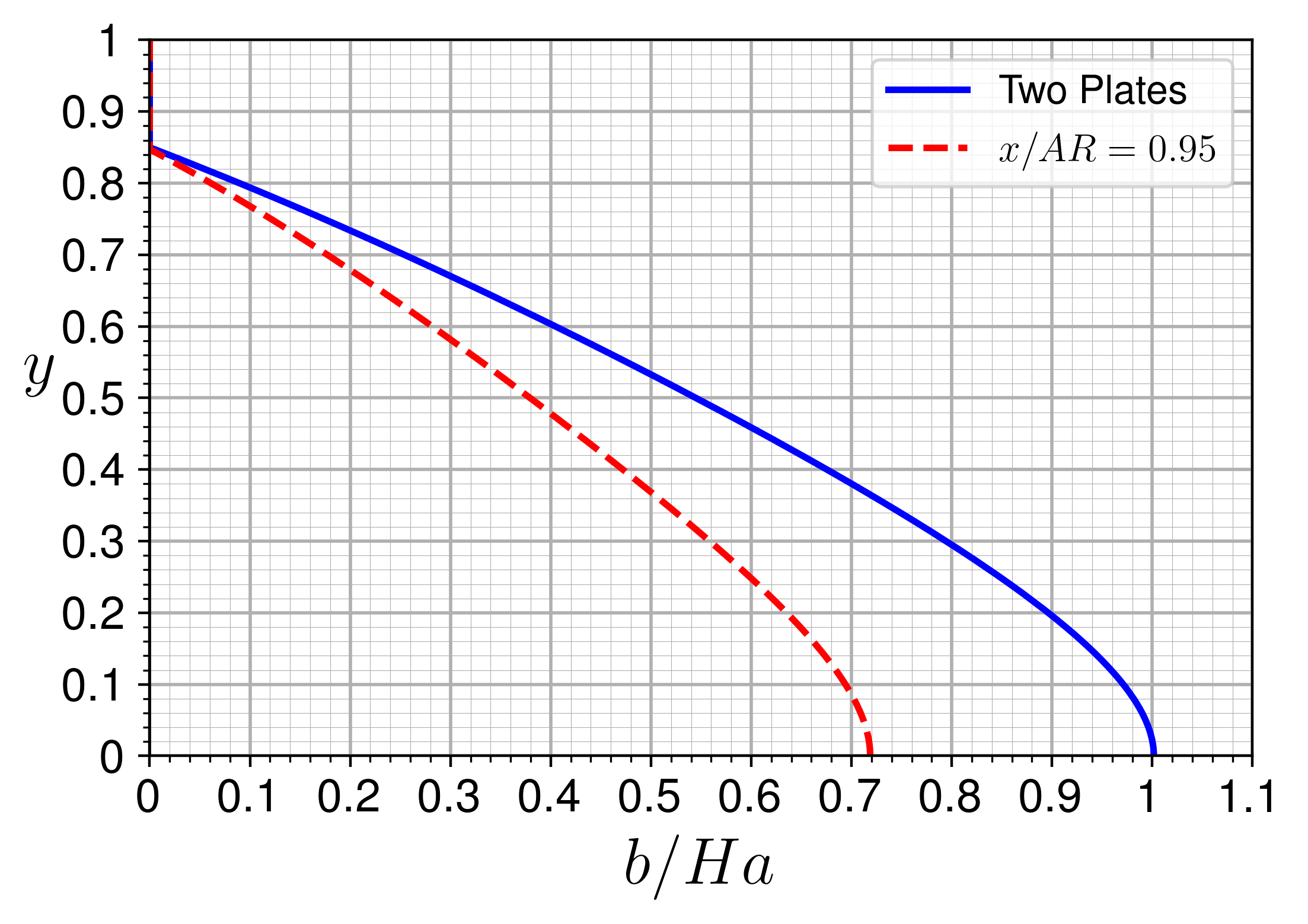}}
	\subfloat[$\Ha=103.62$: $U$]{\includegraphics[width=0.25\textwidth,clip]{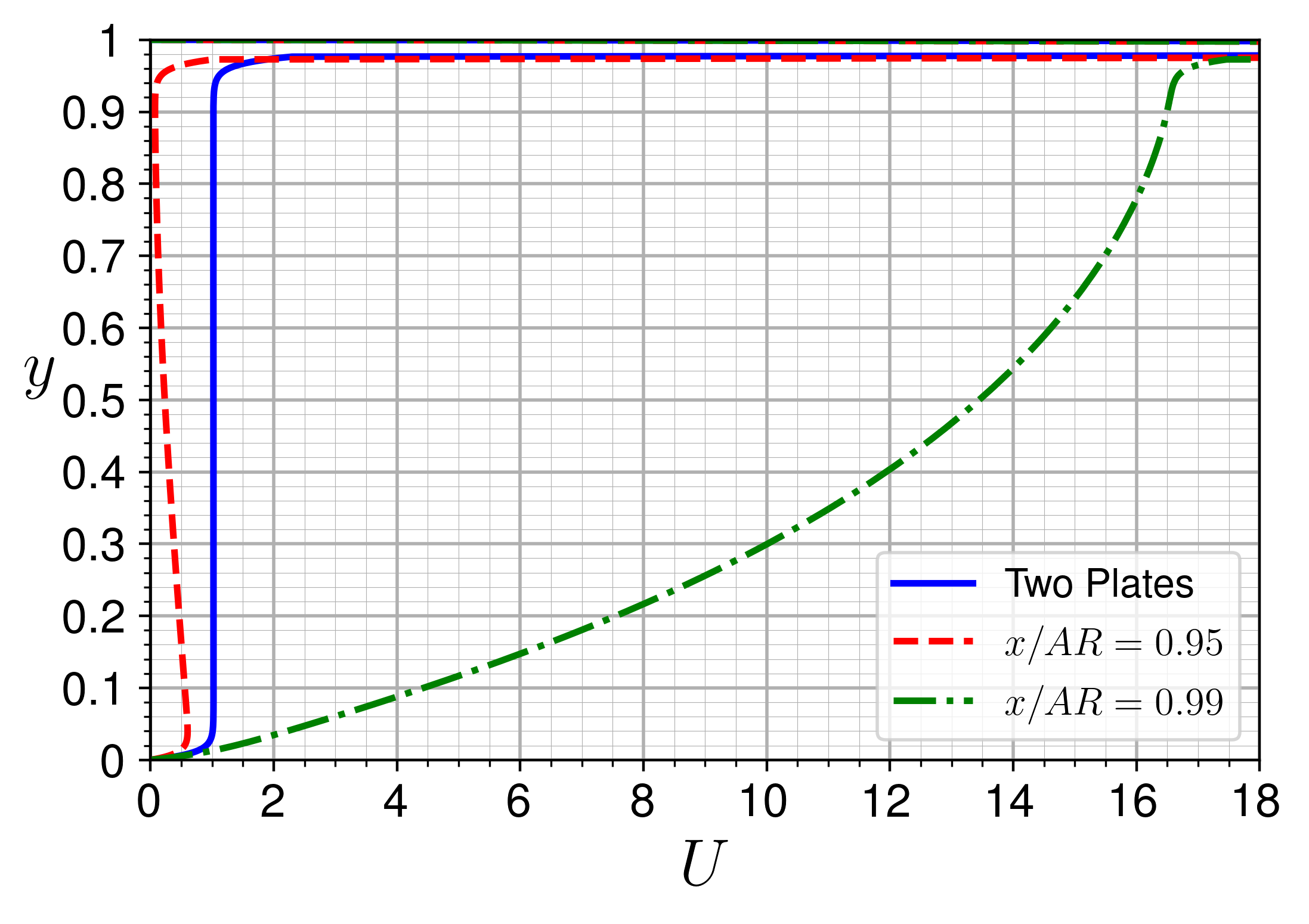}}
	\subfloat[$b/\Ha$]{\includegraphics[width=0.25\textwidth,clip]{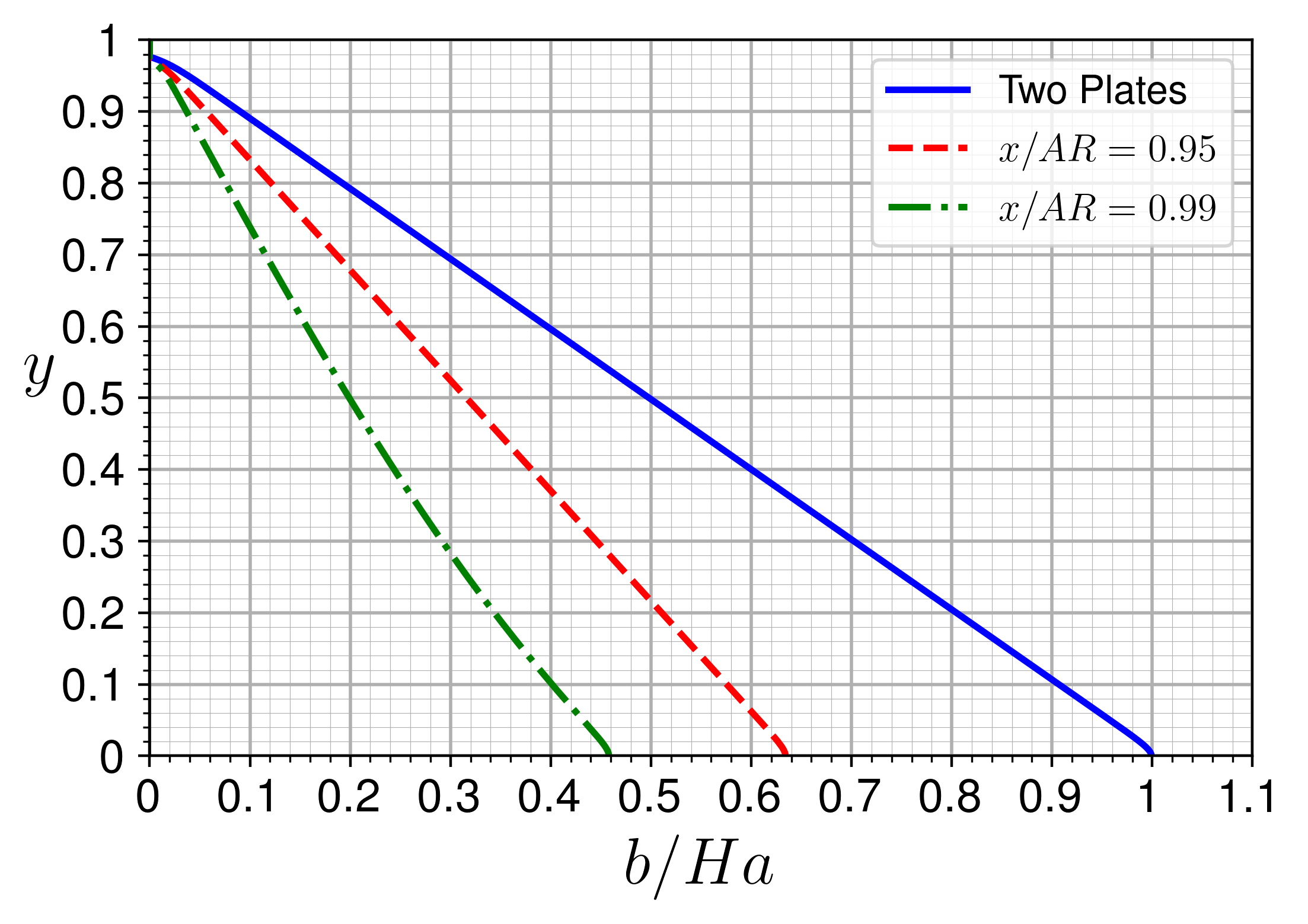}}
	\caption{\label{Fig: U_vs_y_comp_TP_vs_side}Comparison of the vertical profiles of $U$ and $b/\Ha$ predicted by the Two-Plate model (solid blue lines) with those obtained near an insulating side wall in a rectangular duct of $AR = 10$ with perfectly conducting bottom wall. $Q_{2 1}=1$.}	
\end{figure}

On the other hand, the combination of a conducting bottom wall and insulating side walls results in a dramatic change in the velocity field. The velocity and induced magnetic field contours for this case are demonstrated in Figs.\ \ref{Fig: U_contours_Ha_5_bot_cond} and \ref{Fig: B_contours_Ha_5_bot_cond}, respectively, for $Q_{21} = 1$ and $\Ha=5.181$. In this case, the induced currents in the fluid can only penetrate the perfectly conducting bottom wall. As a result, all of the induced magnetic field isolines cross the bottom wall perpendicularly. At the insulating side walls, however, $b=0$, and similarly to the case of fully insulating duct, the gradient $\displaystyle \partial b/\partial y$, and consequently the local Lorentz force, is significantly reduced. However, since the pressure gradient in the flow with a conducting bottom wall is higher than in a fully insulating duct, extended regions of high velocity develop near the side walls, and the velocity in the conductive fluid core is significantly lower. This is further elaborated in light of Figs.\ \ref{Fig: U1_vs_x_bot_cond}-\ref{Fig: U_vs_y_comp_TP_vs_side}.


Fig.\ \ref{Fig: U1_vs_x_bot_cond} shows the spanwise velocity profiles at the interface and at the middle and $75\%$ of the interface height ($AR=10$). The velocity at each height is scaled by the corresponding TP value. For $\Ha=5.181$ (Fig.\ \ref{Fig: U1_vs_x_bot_cond}a), at the distance of about $0.05AR$ from the side wall, the interfacial velocity (solid line) becomes $\approx 35\%$ larger than that obtained by the TP model. It gets even larger below the interface (see dashed-dot line) before decreasing toward the bottom wall. The velocity across most of the cross section is nevertheless similar to the TP one ($U/U_{TP}\approx 1$ in the core). The effect of the side wall is much more dramatic for high Hartmann number (e.g., $\Ha=103.62$, red lines in Fig.\ \ref{Fig: U1_vs_x_bot_cond}b). In this case, the velocity field (Fig.\ \ref{Fig: U_contours_Ha_103}) changes drastically as a significant part of the conductive fluid flow is carried in those high velocity regions (jets) near the side walls (similar structures were observed in single-phase MHD duct flows, see \cite{Muller01}, section 3.1 therein). For the higher Hartmann numbers, the presence of the conducting side walls (Figs.\ \ref{Fig: U_contours_Ha_103}b and \ref{Fig: U_contours_Ha_103}e) results in a core velocity similar to the TP model. On the other hand, a $35\%$ lower core velocity is obtained when the side walls are insulating (Fig.\ \ref{Fig: U_contours_Ha_103}c). The side jet occupies a thinner layer (Fig.\ \ref{Fig: U_contours_Ha_103}f) and the maximal velocity reaches the values more than $10$ times larger than the TP values at the distance about $0.01AR$ from the side wall (Fig.\ \ref{Fig: U1_vs_x_bot_cond}b), along with a small region of weak backflow ($min(U_1)\approx -0.03$) a bit further away (see closed blue contour line in Fig.\ \ref{Fig: U_contours_Ha_103}f). The pressure gradient is mainly set by the core velocity, so that for high $\Ha$ and $AR=10$ it is significantly lower than that predicted by the TP model ($-38\%$ deviation for $AR=10$, see Figs.\ \ref{Fig: AR_convergence}f,h).

As mentioned above, the side wall jet originates from the significant reduction of the Lorentz force near the insulating side walls in a duct with a conductive bottom wall. This can be clearly observed in Figs.\ \ref{Fig: B_contours_Ha_103}c and \ref{Fig: B_contours_Ha_103}f, showing the weak (negative) $\partial b/\partial y$ near the side insulating walls as compared to those gradients in the case of conductive side walls (Figs.\ \ref{Fig: B_contours_Ha_103}b and \ref{Fig: B_contours_Ha_103}e), as well as to the gradients observed in the duct core region. The impact of insulating side walls on reduced  values is also clearly observed when comparing the variation of  vertical profiles in their vicinity to the corresponding TP profiles in the case of conductive bottom wall (Figs.\ \ref{Fig: U1_vs_x_bot_cond}b,d). When compared to the TP model velocity profile for high Ha (Fig.\ \ref{Fig: U1_vs_x_bot_cond}c), the conductive fluid velocities are much higher close to the side walls (e.g., green line for $x=0.99AR$), and lower velocities in an adjacent region (e.g., red line for $x=0.95AR$). However, the significance of the flow phenomena near the side walls diminishes when  is further increased, and for  the pressure gradient over prediction by the TP model is about $7\%$ (see Figs.\ \ref{Fig: AR_convergence}f,h). 

\section{Conclusions} \label{Sec: Conclusions} 

A study has been conducted on fully developed two-phase stratified magnetohydrodynamic (MHD) flow of a conducting liquid and a non-conducting gas in wide horizontal channels under a vertical external magnetic field.

We present both an analytical solution based on the Two-Plate (TP) model and a numerical solution for ducts with high width-to-height aspect ratios ($AR>10$). The holdup (i.e., the dimensionless flow area occupied by the conducting liquid) and the dimensionless pressure gradient are found to be governed by the Hartmann number ($Ha$), flow rate ratio of the fluids ($Q_{2 1}$), their viscosity ratio ($\eta_{1 2}$), and the conductivity of the bottom wall ($c_{bw}$). The latter influences the induced magnetic field and thereby the Lorentz force acting on the conductive fluid. Their effects are however independent on the magnetic Reynolds number, $\Rey_m$, making the results for the flow characteristics applicable for any $\Rey_m$ (providing the flow is laminar). To explore the effect of the bottom wall conductivity we referred to the two limiting cases: an insulating and a perfectly conducting bottom wall. As the problem formulation is independent of the fluids densities, the results obtained for horizontal channel are valid also for zero-gravity conditions.

Unlike in single-phase flow of a conductive fluid, where the shape of the dimensionless velocity profile (scaled by the average, or maximal, velocity) is unaffected by the walls conductivity, in the two-phase flow, the velocity profiles differ in shape when the bottom wall is insulating or conducting. This variation arises from the influence of the bottom wall conductivity on the holdup of the conducting liquid. With an insulating wall, the retarding Lorentz force is weaker, leading to a higher bulk velocity of the conductive liquid. As a result, a smaller flow area is sufficient to maintain the same flow rate, compared to the case with a conducting wall. For the same gas-to-liquid flow rate ratio and Hartmann number, the pressure gradient is consistently higher when the bottom wall is conducting, primarily due to the stronger Lorentz force. 

The results were used to investigate the air flow lubrication effect, as evidenced by a reduction in pressure gradient and pumping power compared to single-phase flow of the conducting liquid in a channel with the same Ha and bottom wall conductivity. For tested case of air-mercury flow with a conducting bottom wall, air lubrication is limited to low $\Ha$, where some pumping power savings can be achieved by introducing a low air flow rate  ($Q_{2 1} < 0.1$). The lubrication effect is more significant when the bottom wall is insulating. In this case, pumping power reduction becomes largely insensitive to the Hartmann number and persists across a broader range of the air-to-mercury flow rate ratios ($Q_{2 1} < 1$). Notably, for $Q_{2 1} \approx 0.1$, introducing air into the channel can lead to a power saving of approximately $50\%$. These findings highlight the importance of accounting for the bottom wall conductivity and the associated induced magnetic field in the two-phase flow analysis, even in the limit of low $\Rey_m$.

Since the simplified TP model cannot be realized in practice, we investigated the influence of side walls in wide ducts on the two-phase flow characteristics and the induced magnetic field. The effect of duct aspect ratio ($AR$) was examined under four configurations: (1) all walls insulating, (2) all walls conducting, (3) insulating bottom wall with conducting side walls, and (4) conducting bottom wall with insulating side walls. In all cases, the influence of the side walls diminishes as the aspect ratio increases. However, the rate of convergence to the TP model predictions differs depending on the bottom wall conductivity. The effect of the side wall has been elaborated by examining the contours of the velocity and induced magnetic field in the various combinations of the bottom wall and side walls conductivities.   

When the bottom wall is insulating, the conductivity of the side walls has only a minor effect on the holdup and pressure gradient for aspect ratios greater than $10$. In this case, the presence of side walls results in slightly higher holdup and pressure gradient compared to the TP model. In contrast, when the bottom wall is perfectly conducting, the side walls can exert a much more significant influence. Specifically, insulating side walls reduce the pressure gradient, leading to a substantial overprediction by the TP model even in wide ducts. For example, at $AR=10$ and $\Ha\approx100$, the pressure gradient can be overestimated by as much as $\approx40\%$ for the test case of mercury--air flow a wide range of $Q_{21}$. This unexpected result is attributed to the reduced induced magnetic field and the associated Lorentz force near the insulating side walls. The weaker Lorentz force allows significantly higher velocities of the conductive liquid, forming a jet-like flow near the side walls. This in turn reduces the core velocity and results in a lower pressure gradient than that predicted by the TP model. 

The current work and the presented results provide a foundation for future studies on the stability of two-phase stratified MHD flows.


\bmhead{Acknowledgments}

This research was supported by Israel Science Foundation (ISF) grant No 1363/23.

\begin{appendices}
	
	\section{Boundary conditions for induced magnetic field} \label{Sec: BC_induced_B}
	\numberwithin{equation}{section}
	\setcounter{equation}{0}
	
	On the border between two different materials with no surface charges, i.e., between fluid and solid wall or between two fluids, the tangential component of the electrical field is continuous \cite{Muller01}. In the following, without loss of generality, we present the derivation based on example of the bottom wall:
	\begin{equation} \label{Eq: BC_bottom_E}
		\bigl(\vec{n} \times \vec{E}\bigr)_1\Biggr\rvert_{x, y=0} = \bigl(\vec{n} \times \vec{E}\bigr)_w\Biggr\rvert_{x, y=0}
	\end{equation}
	
	One can rewrite Eq.\ \ref{Eq: BC_bottom_E} using Amp\`ere's law, Eq.\ \ref{Eq: Maxwell's equations}c, and Ohm's law, Eq.\ \ref{Eq: Ohm's law}, to obtain:
	\begin{equation} \label{Eq: BC_bottom_rotB}
		\frac{1}{\sigma_{e1}} \bigl[\vec{n} \times \bigl(\nabla \times \vec{B}\bigr)\bigr]_1\Biggr\rvert_{x, y=0}
		= \frac{1}{\sigma_{w}} \bigl[\vec{n} \times \bigl(\nabla \times \vec{B}\bigr)\bigr]_w\Biggr\rvert_{x, y=0}
	\end{equation}
	
	Taking into account that the total magnetic field $\displaystyle\vec{B}=\bigl(0,B_0,\hat{B}_{ind}(x,y)\bigr)$, Eq.\ \ref{Eq: BC_bottom_rotB} becomes:
	\begin{equation}
		\frac{1}{\sigma_{e1}} \biggl(\frac{\partial \hat{B}_{ind}}{\partial \hat{y}}\biggr)_1\Biggr\rvert_{x, y=0}
		= \frac{1}{\sigma_{w}} \biggl(\frac{\partial \hat{B}_{ind}}{\partial \hat{y}}\biggr)_w \Biggr\rvert_{x, y=0}
	\end{equation}
	
	Furthermore, the channel wall is assumed thin $\displaystyle t_w \ll H$, and the induced magnetic field vanishes in the insulating domain outside the channel walls (i.e., $\hat{B}_{ind\rvert out}=0$), so that:
	\begin{equation}
		\frac{1}{\sigma_{e1}} \biggl(\frac{\partial \hat{B}_{ind}}{\partial \hat{y}}\biggr)\Biggr\rvert_{x, y=0}
		= \frac{1}{\sigma_{w} } \frac{\hat{B}_{ind}(x,y=0) - \hat{B}_{ind\rvert out}}{t_w}
		= \frac{1}{\sigma_{w} t_w} \hat{B}_{ind}(x,y=0)
	\end{equation}
	
	\section{Single-phase flow of conductive fluid between two plates} \label{Sec: Single_phase_TP}
	
	The two-plate model analytical solution for a single-phase flow of the conductive fluid is presented in order to explore the possible benefit (or penalty) of the flow of the non-conductive phase (e.g., air). The pressure gradient required for driving a specified flow rate of the conductive phase is derived. The corresponding lubrication and pumping power factors resulting from adding a non-conductive gas phase can then be obtained.
	
	In the case of single-phase flow, the flow and induced magnetic fields are symmetrical about the horizontal centerline of the channel (assuming the same conductivity of its walls, $c_{bw}$). Therefore, it is convenient to use vertical coordinate symmetrical around the channel centerline, i.e., $\displaystyle y_{sp} = y - 1/2$, so that $y_{sp}=0$ at the center. The solution for the dimensionless (same scaling as above) velocity (for consistency, it is denoted as $U_1$ also in single-phase flow) and induced magnetic field in the case of insulating channel walls are:
	\begin{subequations} \label{Eq: SingleP_TP_insul}
		\begin{align}
			U_1(y_{sp}, c_{bw}=0) &= -\frac{G_1}{2 \Ha} \dfrac{\cosh\Bigl(\dfrac{\Ha}{2}\Bigr)}{\sinh\biggl(\dfrac{\Ha}{2}\biggr)}
			\Biggl[1 
			- \dfrac{\cosh(\Ha y_{sp})}{\cosh\Bigl(\dfrac{\Ha}{2}\Bigr)}
			\Biggr]
			\\
			b(y_{sp}, c_{bw}=0) &= \frac{G_1}{\Ha}
			\Biggl[y_{sp}
			- \dfrac{\sinh(\Ha y_{sp})}{2 \sinh\biggl(\dfrac{\Ha}{2}\biggr)}
			\Biggr]
		\end{align}
	\end{subequations}
	and in the case of the perfectly conducting walls:
	\begin{subequations} \label{Eq: SingleP_TP_cond}
		\begin{align}
			U_1(y_{sp}, c_{bw}\to\infty) &= -\frac{G_1}{\Ha^2}
			\Biggl[1 
			- \dfrac{\cosh(\Ha y_{sp})}{\cosh\biggl(\dfrac{\Ha}{2}\biggr)}
			\Biggr]
			\\
			b(y_{sp}, c_{bw}\to\infty) &= \frac{G_1}{\Ha}
			\Biggl[y_{sp}
			- \dfrac{\sinh(\Ha y_{sp})}{\Ha \cosh\biggl(\dfrac{\Ha}{2}\biggr)}
			\Biggr]
		\end{align}
	\end{subequations}
	
	The pressure gradient corresponding to a specified flow rate can be found by integration of the velocity profiles:
	\begin{align} \label{Eq: SinglePhase_G}
		G_1 (c_{bw}=0) &= - \dfrac{2 \Ha \tanh \Bigl(\dfrac{\Ha}{2}\Bigr)}{
			1-\dfrac{2}{\Ha}\tanh\biggl(\dfrac{\Ha}{2}\biggr)
		}
		\\
		G_1 (c_{bw}\to\infty) &= - \dfrac{\Ha^2}{
			1-\dfrac{2}{\Ha}\tanh\biggl(\dfrac{\Ha}{2}\biggr)
		}
	\end{align}
	
	The solution for the pressure gradient should then be substituted into the corresponding profiles, Eqs.\ \ref{Eq: SingleP_TP_insul} and \ref{Eq: SingleP_TP_cond}, in order to observe the solution dependence on the Hartmann number only. 
	
	The dimensionless wall shear stress (scaled by a value for the case of $\Ha=0$), can be found by calculating the velocity gradient at the wall. Although the velocity profile is different in the two cases of wall conductivity (compare Eqs.\ \ref{Eq: SingleP_TP_insul}a and \ref{Eq: SingleP_TP_cond}a), the wall shear stress is found to be the same for these cases and depends on the Hartmann number only:
	\begin{align} \label{Eq: Shear_stress_factor_SP}
		\tau_{w\rvert1}^0 = \dfrac{\tau_{w\rvert1}}{\tau_{w\rvert1,\Ha=0}}
		= \dfrac{\Ha \tanh \Bigl(\dfrac{\Ha}{2}\Bigr)}{
			6 \Bigl[1-\dfrac{2}{\Ha}\tanh\biggl(\dfrac{\Ha}{2}\biggr)\Bigr]
		}
	\end{align}
	
	Referring to Eq.\ \ref{Eq: BC_Cb} for the case of insulating walls, $c_{bw}=0$, the vertical derivative of the induced magnetic field at the bottom wall is:
	\begin{equation}
		\frac{\partial b}{\partial y_{sp}}\Biggr\rvert_{y_{sp}=-1/2} = C_b 
		= \frac{G_1}{\Ha} \biggl[1 - \Ha \coth\biggl(\frac{\Ha}{2}\biggr)\biggr]
	\end{equation}
	
	The momentum equation for the conductive fluid (Eq.\ \ref{Eq: TP_U_equation}) then reads:
	\begin{equation}
		\frac{\partial^2 U_{1}}{\partial y_{sp}^2}
		-\Ha^2 U_1 
		= G_1 \frac{\Ha}{2} \coth\biggl(\frac{\Ha}{2}\biggr)
	\end{equation}
	Hence, for $c_{bw}=0$ the effective force (per unit volume) driving the flow is $\displaystyle G_1 \bigl(\Ha/2\bigr) \coth\bigl(\Ha/2\bigr)$ compared to $G_1$ for the case of $c_{bw}\to\infty$ (Eq.\ \ref{Eq: TP_U_equation_perf_cond}). Indeed, the ratio of the corresponding pressure gradients can be found from Eqs.\ \ref{Eq: SinglePhase_G} for specified flow rate (superficial velocity):
	\begin{equation}
		\frac{G_1(c_{bw}=0)}{G_1(c_{bw}\to\infty)}\Biggr\rvert_{U_{1S}=const} = \frac{2}{\Ha} \tanh\biggl(\frac{\Ha}{2}\biggr)
	\end{equation}
	
	On the other hand, for the same pressure gradient the ratio of superficial velocities and the of the maximal (centerline) velocities is as following:
	\begin{equation}
		\frac{U_{1S}(c_{bw}=0)}{U_{1S}(c_{bw}\to\infty)}\Biggr\rvert_{G_1=const}
		= \frac{U_{max}(c_{bw}=0)}{U_{max}(c_{bw}\to\infty)}\Biggr\rvert_{G_1=const}
		= \frac{\Ha}{2} \coth\biggl(\frac{\Ha}{2}\biggr)
	\end{equation}
	
	These results agree with the observation made in \cite{Muller01}, that for a given Hartmann number and pressure gradient, the flow rate is the largest in a channel with electrically insulating walls and the smallest for perfectly conducting walls. On the other hand, for the same flow rate, the pressure gradient is the smallest for perfectly insulating walls. Note that in \cite{Muller01} the pressure gradient was fixed at a value of $1$, which corresponds to $G_1=4$ in the present notation.
	
\end{appendices}

\bibliography{Barmak_MHD_2phase_arxiv}

\end{document}

%% file: figures/Fig1.pdf_tex
\begingroup%
  \makeatletter%
  \providecommand\color[2][]{%
    \errmessage{(Inkscape) Color is used for the text in Inkscape, but the package 'color.sty' is not loaded}%
    \renewcommand\color[2][]{}%
  }%
  \providecommand\transparent[1]{%
    \errmessage{(Inkscape) Transparency is used (non-zero) for the text in Inkscape, but the package 'transparent.sty' is not loaded}%
    \renewcommand\transparent[1]{}%
  }%
  \providecommand\rotatebox[2]{#2}%
  \newcommand*\fsize{\dimexpr\f@size pt\relax}%
  \newcommand*\lineheight[1]{\fontsize{\fsize}{#1\fsize}\selectfont}%
  \ifx\svgwidth\undefined%
    \setlength{\unitlength}{333.77252053bp}%
    \ifx\svgscale\undefined%
      \relax%
    \else%
      \setlength{\unitlength}{\unitlength * \real{\svgscale}}%
    \fi%
  \else%
    \setlength{\unitlength}{\svgwidth}%
  \fi%
  \global\let\svgwidth\undefined%
  \global\let\svgscale\undefined%
  \makeatother%
  \begin{picture}(1,0.41496348)%
    \lineheight{1}%
    \setlength\tabcolsep{0pt}%
    \put(0,0){\includegraphics[width=\unitlength,page=1]{Fig1.pdf}}%
    \put(0.34508358,0.19927497){\color[rgb]{0,0,0}\makebox(0,0)[lt]{\lineheight{1.25}\smash{\begin{tabular}[t]{l}$Q_2$\end{tabular}}}}%
    \put(0.34508358,0.06801799){\color[rgb]{0,0,0}\makebox(0,0)[lt]{\lineheight{1.25}\smash{\begin{tabular}[t]{l}$Q_1$\end{tabular}}}}%
    \put(0,0){\includegraphics[width=\unitlength,page=2]{Fig1.pdf}}%
    \put(0.23296678,0.06804637){\color[rgb]{0,0,0}\makebox(0,0)[lt]{\lineheight{1.25}\smash{\begin{tabular}[t]{l}$h_1$\end{tabular}}}}%
    \put(0.23296678,0.17857493){\color[rgb]{0,0,0}\makebox(0,0)[lt]{\lineheight{1.25}\smash{\begin{tabular}[t]{l}$h_2$\end{tabular}}}}%
    \put(0.85340378,0.01958499){\color[rgb]{0,0,0}\makebox(0,0)[lt]{\lineheight{1.25}\smash{\begin{tabular}[t]{l}$z$\end{tabular}}}}%
    \put(0.92528592,0.15747201){\color[rgb]{0,0,0}\makebox(0,0)[lt]{\lineheight{1.25}\smash{\begin{tabular}[t]{l}$x$\end{tabular}}}}%
    \put(0.75527404,0.38309534){\color[rgb]{0,0,0}\makebox(0,0)[lt]{\lineheight{1.25}\smash{\begin{tabular}[t]{l}$y$\end{tabular}}}}%
    \put(0.54105169,0.02942559){\color[rgb]{0,0,0}\makebox(0,0)[lt]{\lineheight{1.25}\smash{\begin{tabular}[t]{l}$\rho_1, \eta_1, \sigma_{e1}$\end{tabular}}}}%
    \put(0.54105169,0.31221998){\color[rgb]{0,0,0}\makebox(0,0)[lt]{\lineheight{1.25}\smash{\begin{tabular}[t]{l}$\rho_2, \eta_2$\end{tabular}}}}%
    \put(0.01476216,0.14013416){\color[rgb]{0,0,0}\makebox(0,0)[lt]{\lineheight{1.25}\smash{\begin{tabular}[t]{l}$H$\end{tabular}}}}%
    \put(0,0){\includegraphics[width=\unitlength,page=3]{Fig1.pdf}}%
    \put(0.08144722,0.3183301){\color[rgb]{0,0,0}\makebox(0,0)[lt]{\lineheight{1.25}\smash{\begin{tabular}[t]{l}$W$\end{tabular}}}}%
    \put(0,0){\includegraphics[width=\unitlength,page=4]{Fig1.pdf}}%
    \put(0.91582903,0.29777767){\color[rgb]{0,0,0}\makebox(0,0)[lt]{\lineheight{1.25}\smash{\begin{tabular}[t]{l}$\textbf{g}$\end{tabular}}}}%
    \put(0,0){\includegraphics[width=\unitlength,page=5]{Fig1.pdf}}%
    \put(0.3339296,0.31738423){\color[rgb]{0,0,0}\makebox(0,0)[lt]{\lineheight{1.25}\smash{\begin{tabular}[t]{l}$\textbf{B}_0$\end{tabular}}}}%
  \end{picture}%
\endgroup%